\documentclass[twocolumn,floats,floatfix,
               showpacs,prd,superscriptaddress,
               longbibliography,
               nofootinbib]{revtex4-2}
\usepackage[utf8]{inputenc}

\usepackage{graphicx,epsfig}
\usepackage{epstopdf}
\usepackage{amssymb,amsmath,amsthm,amsfonts}
\usepackage{bm}

\usepackage[caption=false]{subfig}
\usepackage[usenames,dvipsnames]{xcolor}
\usepackage{url}
\usepackage[inline]{enumitem}
\usepackage{dsfont}
\usepackage{float}
\usepackage{placeins}
\usepackage[T1]{fontenc}
\usepackage{mathrsfs}
\usepackage{anyfontsize}
\usepackage{xspace}
\usepackage{tensor}

\usepackage[linktocpage,breaklinks]{hyperref}
\hypersetup{colorlinks=true,
            citecolor=NavyBlue,
            linkcolor=NavyBlue,
            urlcolor=NavyBlue}

\renewcommand{\vec}[1]{\boldsymbol{#1}}

\newcommand{\bracket}[1]{\left( #1 \right)}
\newcommand{\Bracket}[1]{\left[ #1 \right]}

\def\ETK{\textsc{Einstein Toolkit\,}}
\def\cactus{\textsc{Cactus}\,}
\def\carpet{\textsc{Carpet}\,}
\def\canuda{\textsc{Canuda}\,}
\def\lean{\textsc{Lean}\,}

\def\TP{\textsc{TwoPunctures}\,}
\def\TPBBHSF{\textsc{TwoPunctures\_BBHSF}\,}
\def\newacronym#1#2#3{\gdef#1{\gdef#1{#2\xspace}#3 (#2)\xspace}}
\newacronym{\amr}{AMR}{adaptive mesh refinement}
\newacronym{\adm}{ADM-York}{Arnowitt-Deser-Misner-York}
\newacronym{\ode}{ODE}{ordinary differential equation}
\newacronym{\pde}{PDE}{partial differential equation}
\def\bh#1{black hole#1 (BH#1)\gdef\bh{BH}}
\newacronym{\bssn}{BSSN}{Baumgarte-Shapiro-Shibata-Nakamura}
\newacronym{\gr}{GR}{General Relativity}
\newacronym{\gw}{GW}{gravitational wave}
\newacronym{\emri}{EMRI}{extreme mass-ratio inspiral}

\def\dif{\textrm{d}}
\def\p{\partial}

\def\R4D{\,^{(4)}R}
\def\muS{\mu_{\rm S}}
\def\psiBL{\psi_{\mathrm{BL}}}

\def\bgamma{\bar{\gamma}}
\def\bA{\bar{A}}
\def\tgamma{\tilde{\gamma}}
\def\tA{\tilde{A}}
\def\tGamma{\tilde{\Gamma}}
\newcommand{\tracefree}[1]{\left[{#1}\right]^{\rm TF}}

\def\rex{r_{\rm ex}}
\def\H{\mathcal{H}}
\def\Lie{\mathcal{L}}
\def\M{\mathcal{M}}

\begin{document}

\title{
Dephasing in binary black hole mergers surrounded by scalar wave dark matter clouds
}

\author{Cheng-Hsin Cheng}
\email{chcheng3@illinois.edu}
\affiliation{The Grainger College of Engineering,
Department of Physics and Illinois Center for Advanced Studies of the Universe, University of Illinois Urbana-Champaign, Urbana, Illinois 61801, USA}

\author{Giuseppe Ficarra}
\email{giuseppe.ficarra@unical.it}
\affiliation{Dipartimento di Fisica, Università della Calabria, Arcavacata di Rende (CS), 87036, IT}

\author{Helvi Witek}
\email{hwitek@illinois.edu}
\affiliation{The Grainger College of Engineering,
Department of Physics and Illinois Center for Advanced Studies of the Universe, University of Illinois Urbana-Champaign, Urbana, Illinois 61801, USA}
\affiliation{Center for AstroPhysical Surveys, National Center for Supercomputing Applications, University of Illinois Urbana-Champaign, Urbana, IL, 61801, USA}

\begin{abstract}
    Scalar fields of masses between $10^{-21}$ and $10^{-11} \rm{eV}/c^2$ can exhibit enhanced gravitational interactions with black holes,
    and form scalar clouds around them.
    Such a cloud modifies the dynamics of a coalescing black-hole binary,
    and the resulting gravitational waves
    may provide a new channel to detect light scalar fields, such as axion-like particles or wave-like dark matter candidates.
    In this work we simulate a series of black-hole mergers with mass ratios $q=1$ and $q=1/2$, immersed in an scalar field overdensity with masses in the range $M\muS\in[0,1.0]$.
    To do so, we implemented a constraint-satisfying initial data solver based on the puncture method,
    we improved the accuracy of our open-source software {\textsc{Canuda}} to eighth order finite differences,
    and we reduced the initial orbital eccentricity.
    We investigate the impact of the scalar mass on the gravitational and scalar radiation.
    We find that binaries can undergo a delayed or an accelerated merger
    with respect to the vacuum.
    Our study highlights the challenge and importance of accurately modeling 
    black-hole binaries in dark matter environments.
\end{abstract}

\date{{\today}}

\maketitle

\tableofcontents

\section{Introduction}
In 2015, the direct detection of a \gw{} signal emitted from a \bh{} binary coalescence \cite{LIGOScientific:2016aoc} by the LIGO/Virgo Collaboration marked the dawn of \gw astronomy, 
and paved the way for a new era of breakthroughs in fundamental physics.
To this day, over $300$ direct \gw{} detections have been confirmed~\cite{LIGOScientific:2025slb}.
\gw observations provide an exquisite window to gather insights on the properties of astrophysical \bh{s} and their surrounding environments.
Upcoming ground-based \gw detectors such as the Einstein Telescope \cite{ET:2019dnz} or Cosmic Explorer \cite{Reitze:2019iox}, and space-based missions such as LISA \cite{LISA:2022kgy}, will enable us to infer properties about these \bh{} binaries and their environments with even higher sensitivity.

One of the most tantalizing prospects of probing fundamental physics through \gw{s} 
is that of shedding
light on the nature of dark matter, a key open problem in modern physics.
In recent years, scalar fields with masses below the eV scale arose as compelling dark matter candidates, given their influence on the growth of cosmic structure \cite{Hui:2016ltb}.
Excitingly,
\bh{s} can be used as probes for massive scalar fields
when the (reduced) Compton wavelength of the scalar is comparable to the gravitational radius of the \bh{}. This condition is commonly expressed in terms of the product of \bh{} and scalar masses as~\cite{Dolan:2007mj,Baumann:2019eav}
\begin{align}
\label{eq:SRMassInPhysicalMasses}
\frac{G M_{\rm BH} }{c^2}
\left[\frac{\hbar}{m_{\rm S}c} \right]^{-1}
 & \sim
  10^{10}
  \left[\frac{M_{\rm BH}}{M_\odot}\right] 
  \left[\frac{m_{\rm S}}{{\rm eV}/c^2}\right]
  \sim \mathcal{O}(1)
\,,
\end{align}
where $M_{\rm BH}$ is the \bh{'s} mass and $m_{\rm {S}} = \hbar \muS$ is the scalar's physical mass.
For astrophysical \bh{s} 
ranging from stellar-mass to supermassive ones,
$5\lesssim M_{\rm{BH}}/M_{\odot} \lesssim10^{10}$,
the criterion in Eq.~\eqref{eq:SRMassInPhysicalMasses} implies that \bh{s} can be sensitive probes of ultralight scalars with masses between $10^{-21}$ eV and $10^{-11}$ eV.
This range encompasses 
axion-like particles as well as  fuzzy dark matter candidates.

The average density of galactic dark matter is very low
with
$\rho_{\rm DM} \sim 0.01 M_\odot/{\rm pc}^3$~\cite{Pato:2015dua}.
For the effects of dark matter to have a sizable impact on the dynamics of astrophysical \bh{s}, the density of dark matter has to be enhanced, which can be achieved if a portion of dark matter consists of massive scalar fields.
Field overdensities
up to $10^{18} M_\odot/{\rm pc}^3$
can arise from the adiabatic growth of dark matter minispikes~\cite{Quinlan:1994ed,Gondolo:1999ef,Ullio:2001fb,Kavanagh:2020cfn,Coogan:2021uqv,Kim:2022mdj},
up to $10^8 M_\odot/{\rm pc}^3$ from accretion in a solitonic core~\cite{Hui:2019aqm,Hancock:2025ois},
depending on the cloud's angular momentum as well as parameters describing the \bh{} \cite{Clough:2019jpm,Bamber:2020bpu}.
Another example of such an enhancement process is the superradiant instability, where a massive bosonic field extracts energy and angular momentum from a rotating \bh{}, favoring the macroscopic growth of a bosonic cloud in its surroundings \cite{Brito:2015oca, Detweiler:1980uk,Dolan:2007mj,Witek:2012tr,Shlapentokh-Rothman:2013ysa,East:2017ovw,East:2023nsk},
potentially yielding
synchronised \bh{} hair~\cite{Herdeiro:2014goa,Nicoules:2025bhh}.
Such (quasi-)stationary bosonic clouds can grow up to a few percent of the \bh{} mass~\cite{Brito:2015oca,East:2017ovw}, with densities reaching $10^{-5}$ in geometric units, corresponding to
$10^{23} M_\odot/{\rm pc}^3$ for a \bh{} of mass $10^6M_\odot$.
As the superradiant cloud grows and the \bh{} spins down, the depletion of \bh{} populations in the superradiant region of the \bh{} mass-spin plane can provide indirect evidence for \bh{} superradiance or exclude fields of a given mass parameter~\cite{Brito:2014wla,Ficarra:2018rfu,Hui:2022sri}.

In the presence of a sufficiently large scalar field overdensity, the dynamics of a \bh{} binary can be altered,
potentially leading to observable signatures in their gravitational waveforms~\cite{Macedo:2013qea,Brito:2017zvb,Ghosh:2018gaw,Zhang:2019eid,Siemonsen:2022yyf}.
Therefore, when modeling such \gw{} sources, a non-negligible environment will need to be carefully considered~\cite{Barausse:2014tra,Baumann:2018vus,Baumann:2022pkl,Tomaselli:2024dbw,Boskovic:2024fga}.
With the scalar field accreted onto individual \bh{s}~\cite{Hui:2019aqm,Clough:2019jpm}, the overdensities act as orbiting scalar ``charges'' which radiate energy from the system \cite{Yang:2017lpm}.
The formation of a scalar ``wake'' trailing the \bh{s} has also been investigated numerically and analytically, which results in a dynamical friction force dragging the \bh{s} as they move through the medium \cite{Traykova:2021dua,Vicente:2022ivh,Traykova:2023qyv,Xin:2025ymm}.
Additional features are also expected to arise due to tidal deformations, where the scalar clouds may be depleted in some circumstances \cite{Zhang:2018kib,Cardoso:2020hca,Takahashi:2021eso}.
Moreover, the presence of self-gravitating field configurations, such as dark matter cores and boson stars, can further affect the dynamics of inspiraling \bh{} binaries \cite{Annulli:2020ilw,Annulli:2020lyc,Takahashi:2024fyq}.

If the \gw source is an \emri, where a primary, supermassive \bh{} is perturbed by a secondary, stellar-mass \bh{}, 
the scalar field overdensity is expected to persist around the secondary \bh{} during the \emri evolution~\cite{Maselli:2020zgv,Maselli:2021men,Barsanti:2022vvl,Barsanti:2022ana,Cardoso:2022whc,Duque:2023seg}.
Recent developments using perturbation theory have enabled the calculation of fully-relativistic waveform templates for \emri{s} in a scalar field environment~\cite{Brito:2023pyl,DellaRocca:2024pnm,Dyson:2025dlj}.
In \emri{s} as well as intermediate mass-ratio inspirals,
the presence of a dark matter environment can be detectable by future \gw observations \cite{Cole:2022yzw}, significantly enhance the merger event rate~\cite{Yue:2018vtk}, and cause dephasing in the \gw signal
from energy dissipation due to dynamical friction~\cite{Kavanagh:2020cfn,Kim:2022mdj}.

\begin{figure}[htp!]
    \centering    \includegraphics[width=.23\textwidth]{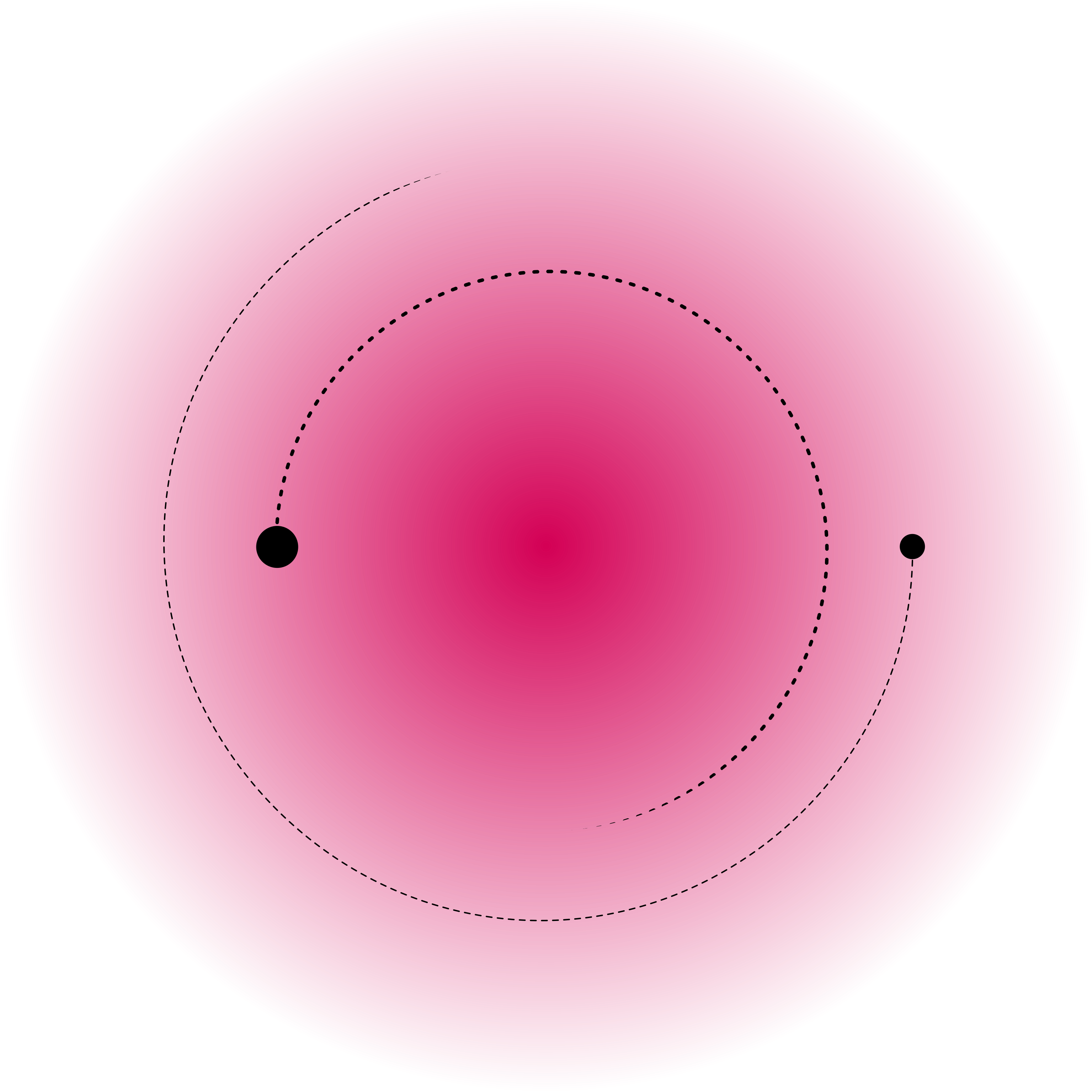}
    \centering    \includegraphics[width=.23\textwidth]{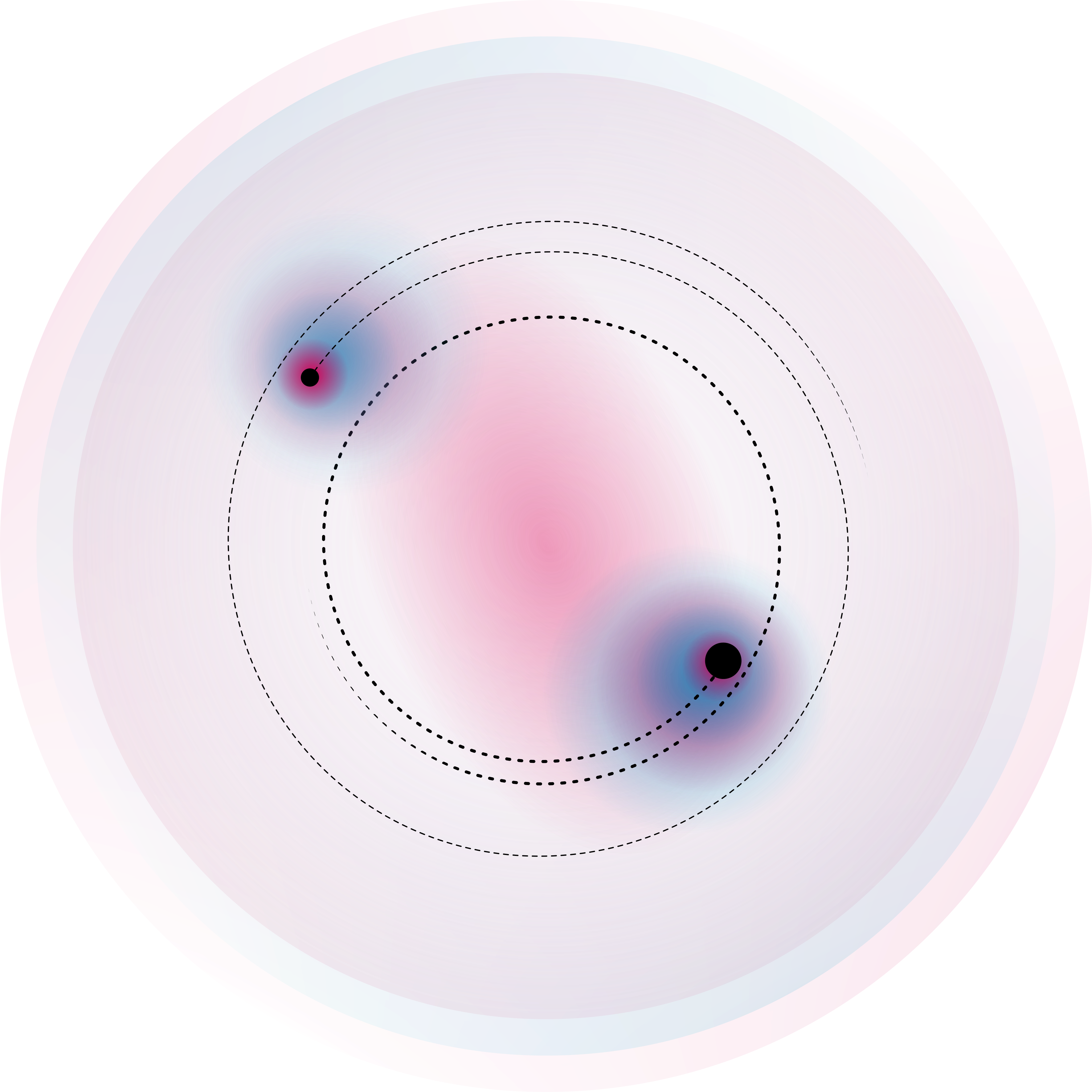}
    \caption{Sketch of an unequal-mass \bh{} binary interacting with a massive scalar cloud. Solid black circles indicating the \bh{s}, their trajectories as dotted and dashed lines, and the oscillating scalar field represented by color.
    Left panel: Initial configuration of the binary and scalar cloud that is set up as spherically symmetric Gaussian profile at the \bh{s}' center-of-mass.
    Right panel: The system one orbit into the inspiral. Around the \bh{s}, the scalar field forms a pair of scalar overdensities that generate scalar radiation.
    }
    \label{fig:sketch}
\end{figure}

In this paper, we focus on the impact of a scalar cloud on comparable-mass \bh{} binaries, as illustrated in Fig.~\ref{fig:sketch}.
For the early inspiral, earlier works have characterized modicfications to the \gw{} signals \cite{Berti:2019wnn, Zhang:2019eid, Guo:2024iye, Guo:2025ckp}.
On the other hand, the nonlinear regime 
that unfolds during the merger is less well-understood, 
and a consistent solution of both the inital data and the evolution 
of the Einstein--Klein-Gordon equations is needed.
Taking into account the scalar cloud back-reaction in the evolution alone, simulations have shown that the presence of the massive scalar cloud can affect the \gw ringdown~\cite{Choudhary:2020pxy}.
Recently, the construction of constraint-satisfying initial data has been made possible by a modification of the conformal transverse-traceless formalism called the ``CTTK'' method~\cite{Aurrekoetxea:2022mpw}, which has been used to show that a scalar environment can lead to strong dephasing of the \gw signal~\cite{Aurrekoetxea:2023jwk,Aurrekoetxea:2024cqd}.
A similar approach helped identify kicks in simulations of spinning and unequal-mass binaries evolved for a few orbits before merger~\cite{Zhang:2022rex}.

In the present work, we adopt the puncture method for conformally flat \bh{} binaries \cite{Brandt:1997tf},
and prescribe the profile the 
(conformally rescaled) 
scalar field itself.
With massive scalar field back-reaction in the initial data, evolution, and waveform extraction, the primary goal of this work is to explore the impact of the scalar field mass on the gravitational waveform.
We consider equal-mass and unequal-mass \bh{} binaries with mass ratios $q=1$ and $q=1/2$, respectively.
We vary the scalar's mass parameter in the range $M\muS\in[0.0,1.0]$.
Our simulations found that quasi-circular binaries can undergo an accelerated or a delayed merger,
depending on the scalar's and \bh{s'} properties.
Moving beyond proof-of-principle simulations of quasi-circular inspirals, we make several improvements to our software
to produce gravitational waveforms that are suitable for template building and parameter estimation studes.
In particular,
we have implemented a new, constraint-satisfying initial data solver \TPBBHSF~\cite{TPBBHSFGitWeb}, that is publicly available in the \ETK~\cite{EinsteinToolkitWeb,roland_haas_2024_14193969,Loffler:2011ay}.
We performed careful eccentricity reduction to ensure that the binary is quasi-circular with
a sufficiently small eccentricity comparable to that used for \gw{} analysis of vacuum \bh{} binaries.
We upgraded our open-source software~\canuda{}~\cite{CanudaGitWeb},
to provide up to eighth order finite differences
in the 
spacetime and scalar evolution, and in the wave extraction.
With the enhanced waveform quality, our work establishes the first step towards constructing waveform templates of \bh{} binaries in non-vacuum environments, suitable for next-generation gravitational wave detectors.
As a closing remark, we comment on the possibility of finding imprints of a massive scalar cloud in realistic astrophysical configurations, and the implications towards probing dark matter and ultralight scalars through strong gravity.

The paper is organized as follows:
In Sec.~\ref{sec:setup} we describe the theoretical framework
and initial value formulation.
In Sec.~\ref{sec:initialdata} we discuss our approach 
construct constraint-satisfying initial data for the Einstein--Klein-Gordon equations.
We summarize the numerical relativity framework in Sec.~\ref{sec:NRframework}.
In Sec.~\ref{sec:results} we present results for our simulations of quasi-circular \bh{} binaries embedded in a dark matter environment.
We conclude in Sec~\ref{sec:conclusions}.

Throughout this study, we use natural units where $G=c=1$ and $\hbar=1$, and we adopt the $(-,+,+,+)$ signature for the spacetime metric.
We denote symmetrization and anti-symmetrization of a tensor by
$A_{(\mu\nu)} = \frac{1}{2}\left(A_{\mu\nu} + A_{\nu\mu} \right)$
and
$A_{[\mu\nu]} = \frac{1}{2}\left(A_{\mu\nu} - A_{\nu\mu} \right)$, respectively.

\section{General relativity and massive scalars}\label{sec:setup}

\subsection{Action and equations of motion}
We consider a complex, massive scalar field $\Phi$ minimally coupled to gravity,
described by the action
\begin{align}
  S = \int\dif^{4}x\sqrt{-g}
  \left(
    \frac{\R4D}{16\pi}
    - \frac{1}{2} g^{\mu\nu} \nabla_\mu \Phi^\dagger  \nabla_\nu\Phi
    - \frac{\muS^2}{2} \Phi^\dagger \Phi
    \right)
\,,
  \label{eq:action}
\end{align}
where $\Phi^\dagger$ is the complex conjugate of the scalar,
and $\muS=m_{S}/\hbar$ is its mass parameter.
We denote the four-dimensional spacetime metric, $g_{\mu\nu}$,
its determinant $g=\det(g_{\mu\nu})$,
the covariant derivative, $\nabla_{\mu}$, 
and Ricci scalar, $\R4D$,
associated to the metric.
Varying the action, Eq.~\eqref{eq:action}, yields the Einstein--Klein--Gordon equations
\begin{subequations}
\label{eq:EinsteinKleinGordon}
\begin{align}
    \label{eq:Einstein}
    \R4D_{\mu\nu} - \frac{1}{2} g_{\mu\nu} \R4D & =  8\pi T_{\mu\nu}
    \,,\\
    \label{eq:Klein-Gordon}
    \left(g^{\mu\nu} \nabla_{\mu} \nabla_{\nu} - \muS^2\right) \Phi &  = 0
    \,,
\end{align}
\end{subequations}
where $\R4D_{\mu\nu}$ is the Ricci tensor
of the spacetime metric,
and the scalar field energy-momentum tensor is
\begin{align}
    \label{eq:Tmunu}
    T_{\mu\nu} & = \nabla_{(\mu}\Phi^\dagger \nabla_{\nu)}\Phi
    - \frac{1}{2} g_{\mu\nu} \left[
        \nabla^{\alpha}\Phi^\dagger  \nabla_{\alpha}\Phi
        + \muS^2 \Phi^\dagger \Phi
    \right]
\,.
\end{align}
We solve Eqs.~\eqref{eq:EinsteinKleinGordon} and~\eqref{eq:Tmunu}
consistently
to simulate the dynamical evolution of a coalescing \bh{} binary and the surrounding scalar field condensate.

Note that we provide the action, field equations and time evolution equations (below) for a complex scalar field.
In \canuda, the complex scalar field is implemented as
two real fields representing its real and imaginary part.
Since their field equations decouple, \canuda has the capability to evolve both a real or a complex scalar field.
In this paper, we focus on a single real scalar.
This corresponds to evolving the real part of $\Phi$, while setting its imaginary part to zero throughout the simulations.

\subsection{Time evolution formulation}
To evolve Eqs.~\eqref{eq:EinsteinKleinGordon},
we formulate them as an initial value problem.
We first foliate the spacetime manifold, $\mathcal{M}$, into a family of spacelike hypersurfaces indexed by a time coordinate $t$.
The geometry on each hypersurface is determined by the $3$-dimensional, spatial metric, $\gamma_{ij}$.
We introduce the unit timelike vector, $n^{\mu}$, that is normal to the hypersurface.
The spacetime coordinates are determined by the lapse function, $\alpha$, and shift vector, $\beta^{i}$, that are collectively referred to as gauge functions.
Putting all together, the spacetime metric can be expressed as 
\begin{align}
\dif s^{2} & =  g_{\mu\nu} \dif x^{\mu} \dif x^{\nu}
\nonumber \\ & =
- \left( \alpha^2 - \beta_i \beta^i\right) \dif t^{2}
+ 2\beta_i\, \dif t\, \dif x^i
+ \gamma_{ij}\, \dif x^i\, \dif x^j.
\end{align}
In this coordinate system, the 
unit vector normal to each hypersurface has components
$n^\mu = (1/\alpha, -\beta^i/\alpha)$ and $n_\mu = (-\alpha, \vec{0}).$

Using the $3+1$ decomposition,
the Einstein field equations can be rewritten in terms of 3-dimensional quantities, in
the \adm formulation \cite{Arnowitt:1962hi,York:1972sj,York:1978gql}.
In this formulation, we introduce the extrinsic curvature,
\begin{align}
\label{eq:DefKij}
K_{ij} & = - \frac{1}{2} \mathcal{L}_n \gamma_{ij}
    = - \frac{1}{2\alpha} (\partial_t - \mathcal{L}_\beta) \gamma_{ij}
\,,
\end{align}
where $\mathcal{L}_{n}$ and $\mathcal{L}_{\beta}$ denote the Lie derivatives along the normal vector $n^{\mu}$ and the shift vector $\beta^\mu$, respectively.
We note that the extrinsic curvature can be related to the conjugated momentum of the metric in the original Hamiltonian formulation of Einstein's equations introduced by Arnowitt, Deser and Misner~\cite{Arnowitt:1962hi}.
Similarly, we introduce the
``momentum'' of the scalar field $\Phi$,
\begin{align}
\label{eq:DefPi}
  \Pi & =  - \frac{1}{2} \mathcal{L}_n \Phi
  = - \frac{1}{2\alpha} (\partial_t - \mathcal{L}_\beta) \Phi
\,,
\end{align}
and analogously for its complex conjugate.
With these definitions, the time evolution of the spatial metric and of the scalar field is determined by
\begin{subequations}
\label{eq:KinematicEvolution}
\begin{align}
\label{eq:Evolgammaij}
\left(\p_t - \mathcal{L}_\beta \right) \gamma_{ij}
    & = -2\alpha K_{ij}
\,,\\
\label{eq:EvolPhi}
\left(\p_t - \mathcal{L}_\beta \right) \Phi & =
    -2\alpha \Pi
\,,
\end{align}
\end{subequations}
and similarly for the scalar's complex conjugate.
We obtain evolution equations for the extrinsic curvature and the scalar field ``momentum''
by decomposing the Einstein--Klein-Gordon equations~\eqref{eq:EinsteinKleinGordon},
and taking their spatial projection.
They are given by
\begin{subequations}
\label{eq:DynamicEvolution}
\begin{align}
\label{eq:EvolKij}
\left(\p_t - \mathcal{L}_\beta \right) K_{ij} & =
    - D_i D_j \alpha
    + \alpha \left[R_{ij} + K K_{ij} - 2K_{ik} K^k_j \right]
\nonumber \\ &
    + 4\pi\alpha \left[ \gamma_{ij} (S-\rho) - 2S_{ij} \right]
\,,\\
\label{eq:EvolPi}
\left(\p_t - \mathcal{L}_\beta \right) \Pi & =
    \alpha \left[ K\Pi - \frac{1}{2} D^i D_i \Phi + \frac{1}{2} \muS^2 \Phi \right]
\nonumber \\ &
    - \frac{1}{2} D^i\alpha\, D_i\Phi
\,,
\end{align}
\end{subequations}
and analogously for its complex conjugate.
Here, $D_{i}$ and $R_{ij}$ are the covariant derivative and Ricci tensor with respect to the spatial metric,
$\gamma_{ij}$,
and $K=\gamma^{ij}K_{ij}$ is the trace of the extrinsic curvature.
The matter contribution is encoded in the energy density $\rho$,
energy-momentum flux $j_{i}$ and the spatial stress tensor $S_{ij}$ with its trace $S=\gamma^{ij} S_{ij}$.
They are determined by projections of the energy-momentum tensor.
For a massive (complex) scalar field, it is given by Eq.~\eqref{eq:Tmunu}, and we obtain
\begin{subequations}
\label{eq:TmnScalarProjections}
\begin{align}
\label{eq:RhoSF}
\rho & = n^\mu n^\nu T_{\mu\nu}
        = 2\Pi^\dagger \Pi
        + \frac{\muS^2}{2} \Phi^\dagger \Phi
        + \frac{1}{2} D_k\Phi^\dagger  D^k\Phi
\,,\\
\label{eq:JiSF}
j_i & = -\gamma_i^\mu n^\nu T_{\mu\nu}
    = \Pi^\dagger D_i\Phi
    + \Pi D_i \Phi^\dagger
\,,\\
\label{eq:SijSF}
S_{ij} & = \gamma_i^\mu \gamma_j^\nu T_{\mu\nu}
        =  D_{(i}\Phi^\dagger D_{j)}\Phi
        - \frac{1}{2}\gamma_{ij} D^k\Phi^\dagger D_k\Phi
\nonumber\\ & \qquad\qquad\qquad
        + \gamma_{ij} \left( 2\Pi^\dagger \Pi
            - \frac{\muS^2}{2}\Phi^\dagger \Phi  \right)
\,.
\end{align}
\end{subequations}

The evolution equations of the extrinsic curvature determine six of the ten independent components of Einstein's equations~\eqref{eq:Einstein}.
The remaining components give the
Hamiltonian and momentum constraints,
\begin{subequations}
\label{eq:constraints}
\begin{align}
\label{eq:Hamiltonian}
\H & = R - K_{ij} K^{ij} + K^2 - 16\pi\rho
    = 0
\,,\\
\label{eq:momentum}
\M_{i} & = D^{k} K_{ik} - D_{i} K - 8\pi j_{i}
    = 0
\,,
\end{align}
\end{subequations}
computed by contracting Eq.~\eqref{eq:Einstein} with the normal vector.
They are coupled elliptic partial differential equations that have to be satisfied at all times in the continuum limit of the Einstein equations.
Rather than enforcing the constraints at each time step,
we adopt the free-evolution approach, i.e., we solve the constraints only to obtain the initial data, and monitor the constraint violation throughout the evolution.
We will describe our approach to solving the constraints in detail in Sec.~\ref{sec:initialdata}.

Although the \adm formulation provides a simple description of the Einstein's equations as a Cauchy problem, they are only weakly hyperbolic~\cite{Sarbach:2012pr},
and simulations become numerically unstable.
Instead, we use the \bssn formulation~\cite{Shibata:1995we,Baumgarte:1998te}, which is rendered strongly hyperbolic
by introducing auxiliary variables and using the constraints.

We adopt the ``$W$'' version of the \bssn formulation~\cite{Marronetti:2007wz}, i.e.,
we decompose the physical metric into
the conformal factor, $W\equiv \gamma^{-1/6}$,
and the conformal metric
\begin{align}
\label{eq:WBSSNMetric}
  \tgamma_{ij} & = W^2 \gamma_{ij}
\,.
\end{align}
The extrinsic curvature $K_{ij}$ is split into
its trace $K$ and its conformal trace-free part $\tA_{ij}$,
\begin{align}
\label{eq:WBSSNCurvature}
  K & = \gamma^{ij} K_{ij},
  \quad
  \tA_{ij} = W^2 \left( K_{ij} - \frac{1}{3}\gamma_{ij} K \right)
\,.
\end{align}
We introduce the conformal connection function,
\begin{align}
\label{eq:WBSSNGamma}
  \tGamma^i & = \tgamma^{jk} \tGamma^{i}{}_{jk}
    = - \p_{k} \tgamma^{ik}
\,,
\end{align}
where $\tGamma^{i}{}_{jk}$ is the
Levi-Civita connection of the conformal metric $\tgamma_{ij}$.
We summarize the evolution equations
of the \bssn variables and of the scalar field in the \bssn formulation in App.~\ref{appendix:BSSN}.

We close the system of evolution by prescribing
the moving puncture gauge~\cite{Campanelli:2005dd,Baker:2005vv} for the gauge functions.
That is, we employ the $1+\log$ slicing condition
and the $\Gamma$-driver shift condition~\cite{Alcubierre:2002kk, vanMeter:2006vi},
\begin{subequations}
\label{eq:movingpuncturegauge}
\begin{align}
\label{eq:1LogSlicing}
\p_t \alpha & = \beta^k \partial_k \alpha - 2\alpha K
\,,\\
\label{eq:GammaDriver}
\p_t \beta^i &= \beta^k \partial_k \beta^i - \eta \beta^i + \frac{3}{4} \tGamma^{i}
\,,
\end{align}
\end{subequations}
where we choose $\eta=1$ in our simulations.

\subsection{Gravitational wave extraction}
\label{ssec:WaveExtraction}
We extract \gw{s} from our numerical simulations
by computing the Newman-Penrose scalar~\footnote{We follow the notation of Ref.~\cite{alcubierre2008introduction} throughout.}
\begin{align}
\label{eq:DefPsi4inWeyl}
\Psi_4 & = C_{\mu\nu\rho\sigma}\, k^{\mu}\bar{m}^{\nu} k^{\rho} \bar{m}^{\sigma}
\,,
\end{align}
that represents outgoing gravitational radiation.
Here, $C_{\mu\nu\rho\sigma}$ is the Weyl curvature tensor and $k^{\mu}$ and $\bar{m}^{\mu}$ are vectors of a null tetrad $\{l^{\mu},k^{\mu},m^{\mu},\bar{m}^{\mu}\}$ with vanishing inner products except for
$-l^{\mu} k_{\mu} = m^{\mu} \bar{m}_{\mu} = 1$.

We construct the vectors of the null tetrad from
\begin{subequations}
\label{eq:DefineTetradVectors}
\begin{align}
l^{\mu} & = \frac{1}{\sqrt{2}} \left( n^{\mu} + u^{\mu} \right)
\,,\quad
k^{\mu}   = \frac{1}{\sqrt{2}} \left( n^{\mu} - u^{\mu} \right)
\,,\\
m^{\mu} & = \frac{1}{\sqrt{2}} \left( v^{\mu} + \imath w^{\mu} \right)
\,,\quad
\bar{m}^{\mu} = \frac{1}{\sqrt{2}} \left( v^{\mu} - \imath w^{\mu} \right)
\,,
\end{align}
\end{subequations}
where $n^{\mu}$ is the timelike unit normal vector, and
$\{u^{\mu},v^{\mu},w^{\mu}\}$ are spatial vectors forming a Cartesian orthonormal basis~\footnote{In the code they are constructed following App. C of Ref.~\cite{Sperhake:2006cy}.}.
Asymptotically, they correspond to radial, polar and azimuthal normal vectors.

In practice, we use the gravito-electric and gravito-magnetic decomposition of the Weyl tensor;
see standard textbooks, e.g., Refs.~\cite{alcubierre2008introduction,baumgarte_shapiro_2010,shibata2015numerical} for details.
The gravito-electric and -magnetic fields are defined as
\begin{align}
\label{eq:DefEijBijInWeyl}
E_{\mu \nu} & = C_{\mu \lambda \nu \rho}\, n^{\lambda} n^{\rho}
\,,\quad
B_{\mu\nu}  = \,^{\ast} C_{\mu \lambda \nu \rho}\, n^{\lambda} n^{\rho}
\,,
\end{align}
where $\,^{\ast} C_{\mu \nu \rho\sigma} = \frac{1}{2} \epsilon_{\rho\sigma}{}^{\kappa\lambda} C_{\mu\nu\kappa\lambda }$ is the dual Weyl tensor.
The symmetries of the Weyl tensor imply that the gravito-electric and -magnetic fields are
traceless, spatial and symmetric, i.e.,
$g^{\mu\nu} E_{\mu\nu} = 0 = g^{\mu\nu}B_{\mu\nu}$,
$E_{\mu\nu}n^{\nu} = 0 = B_{\mu\nu}n^{\nu}$
and $E_{[\mu\nu]}= 0 = B_{[\mu\nu]}$.
Conversely, the Weyl tensor can be reconstructed from~\cite{alcubierre2008introduction}
\begin{align}
\label{eq:WeylinEijBij}
C_{\mu\nu \rho \sigma} & =
    2 \left( l_{\mu[\rho } E_{\sigma ]\nu} - l_{\nu [\rho} E_{\sigma]\mu} \right)
\\ & \quad
    - 2 \left( n_{[\rho} B_{\sigma ] \kappa } \epsilon^{\kappa}{}_{\mu\nu} + n_{[\mu} B_{\nu ] \kappa } \epsilon^{\kappa}{}_{\rho \sigma}
    \right)
\,,\nonumber
\end{align}
where $l_{\mu\nu}= \gamma_{\mu\nu} + n_{\mu} n_{\nu}$.
Putting everything together and making spatial tensor components explicit,
the desired
Newman-Penrose scalar can be written as
\begin{align}
\label{eq:Psi4InEBmbar}
\Psi_{4} & = \left(E_{ij} - \imath B_{ij} \right) \bar{m}^{i} \bar{m}^{j}
\,.
\end{align}

Finally, we express the gravito-electric and -magnetic fields in terms of the $3$-metric, extrinsic curvature and projections of the energy-momentum tensor.
Performing the 3+1 decomposition of Eqs.~\eqref{eq:DefEijBijInWeyl} and
inserting
Eqs.~\eqref{eq:DynamicEvolution}~and~\eqref{eq:constraints},
we find
\begin{subequations}
\label{eq:EijBijIn3p1}
\begin{align}
\label{eq:EijIn3p1}
E_{ij} & = \left[ R_{ij} + K K_{ij} - K_{ik} K^{k}{}_{j} - 4\pi S_{ij} \right]^{\rm{TF}}
\,,\\
\label{eq:BijIn3p1}
B_{ij} & = \left[ \epsilon_{i}{}^{lm}
    \left( D_{l} K_{m j} -4 \pi \gamma_{j l} j_{m} \right) \right]^{\rm{TF}}
\,,
\end{align}
\end{subequations}
where
$\epsilon_{\mu\nu\rho} = n^{\sigma} \epsilon_{\sigma\mu\nu\rho}$ is the three-dimensional Levi-Civita tensor,
and $\left[\dots\right]^{\rm{TF}}$ denotes the tracefree part of a tensor with respect to the 3-metric.

In practice, we implement Eq.~\eqref{eq:Psi4InEBmbar} after inserting Eqs.~\eqref{eq:DefineTetradVectors} and evaluating Eqs.~\eqref{eq:EijBijIn3p1},
where the explicit expressions are given in
Sec.~4 of Ref.~\cite{Zilhao:2015tya}.
We extract the multipoles of the scalar field and of the Newman-Penrose scalar  on spheres of fixed radii $\rex$.
We interpolate the fields $f\in\{\Psi_{4},\Phi\}$, on spheres of radii $\rex$, and expand them as
\begin{align}
f(t,\rex,\theta,\varphi) & =
    \sum_{\ell=0}^\infty \sum_{m=-\ell}^{\ell}
        f_{\ell m}(t,\rex)\, {_s}Y_{\ell m}(\theta, \varphi)
\,,
\end{align}
where $_{s}Y_{\ell m}(\theta, \varphi)$ are the spin-$s$-weighted spherical harmonics.
The spin-weight is $s=-2$ for the Newman-Penrose scalar $\Psi_{4}$ and  $s=0$ for the scalar field $\Phi$.
The $(\ell,m)$ multipole is computed by the projection
\begin{align}
\label{eq:Multipoles}
f_{\ell m}(t, \rex) & =
    \int \dif\Omega\,
        f(t, \rex, \theta, \varphi)\, {_s}Y_{\ell m}^{\ast}(\theta, \varphi)
\,.
\end{align}

Finally, we compute the luminosity
(i.e., energy flux)
of the gravitational radiation using
\begin{align}
\label{eq:RadiatedEnergyGW}
\frac{{\rm d}E_{\rm{GW}}}{{\rm d}t}
&=
\lim_{\rex\to\infty}
\frac{\rex^2}{16\pi}
\sum_{\ell=2}^\infty
\sum_{m=-\ell}^\ell
\left|
\int_{-\infty}^t
{\rm d}t'
\,
\Psi_{4, \ell m}
\right|^2.
\end{align}
In practice, we sum over the first $\ell=2,\ldots,8$ multipoles
and compute the energy flux at $r_{\rm ex}=100M$.

\section{Initial data for black-hole binaries
immersed in a scalar cloud}\label{sec:initialdata}

Constraint-satisfying initial data is key for a
consistent time evolution.
In fact, a simple superposition of scalar fields onto the metric introduces constraint violations that can
yield incorrect physical results~\cite{Okawa:2014nda}.
We derive the formulation of the initial data problem for binary \bh{s} immersed in a scalar cloud in Sec.~\ref{subsec:id_procedure},
and discuss its well-posedness in Sec.~\ref{ssec:Wellposedness}.
The numerical implementation is presented in Sec.~\ref{sec:NRframework}.

\subsection{Constraints with scalar field contributions}\label{subsec:id_procedure}
We first need to construct constraint-satisfying initial data for the metric in the presence of a scalar cloud.
Consistent initial data requires one to solve simultaneously for
(i) constraint-satisfying metric data in the presence of the scalar
and (ii) a quasi-equilibrium state of the scalar field.
In this paper
we focus on the first task,
and solve the constraint equations
for the metric
using the conformal, transverse, traceless decomposition~\cite{Lichnerowicz:1944zz,York:1971hw,York:1972sj}.
The scalar field source is set up via an
approximate, analytic prescription.
In particular, we leverage the result that
simple scalar field profiles reach their equilibrium configuration around a \bh{} binary within a few orbits~\cite{Bamber:2022pbs}.

Our method allows for a general scalar field profile,
and only assumes that the field is momentarily at rest,
\begin{align}
\label{eq:ID_SF_PiZero}
\Pi(t=0) & = 0
\,.
\end{align}
Then, the energy density and flux of the scalar field, Eqs.~\eqref{eq:TmnScalarProjections},
reduce to
\begin{subequations}
\begin{align}
\rho & = \frac{\muS^2}{2} \Phi^{\dagger}\Phi + \frac{1}{2} D^{k}\Phi^{\dagger} D_{k}\Phi
\,,\\
j_{i} & = 0
\,.
\end{align}
\end{subequations}
With this choice, the momentum constraint~\eqref{eq:momentum} reduces to that of vacuum 
General Relativity.

Next, we write the constraints, Eqs.~\eqref{eq:constraints}, in the conformal, transverse, traceless  decomposition~\cite{Lichnerowicz:1944zz,York:1971hw,York:1972sj},
which prescribes a conformal rescaling of
the metric and extrinsic curvature
\begin{align}
\label{eq:CTT_gammaij_Kij}
\gamma_{ij} & = \psi^4 \bgamma_{ij}
\,,\quad
K_{ij} = \psi^{-2}\bA_{ij}
        + \frac{1}{3} \psi^4 \bgamma_{ij} K
\,.
\end{align}
Here,
$\psi$ is the conformal factor, $\bgamma_{ij}$ is the conformal metric, and $\bA_{ij}$ is the traceless part of the conformal extrinsic curvature.
We simulate asymptotically flat spacetimes,
and we therefore impose the boundary condition
$\lim_{r\to\infty} \psi = 1$.

We further simplify the constraint equations by imposing
maximal slicing, $K=0$,
and conformal flatness,
$\bgamma_{ij} = \eta_{ij}$
where $\eta_{ij}=\rm{Diag}(1,1,1)$ is the flat spatial metric.
With these assumptions and the conformal decomposition of Eq.~\eqref{eq:CTT_gammaij_Kij},
the constraint equations decouple
and become
\begin{subequations}
\begin{align}
\label{eq:CTT_HC_confflat}
\bar{D}^{i} \bar{D}_{i} \psi
+\frac{1}{8} \psi^{-7} \bar{A}_{i j} \bar{A}^{i j}
+ 2 \pi \psi^{5} \rho & =0
\,, \\
\label{eq:CTT_MC_final}
\bar{D}_{j} \bar{A}^{i j} & = 0
\,,
\end{align}
\end{subequations}
where $\bar{D}$ denotes the covariant derivative with respect to the conformal (flat) metric,
and the density is now
\begin{equation}
\label{eq:CTT_rhoSF}
\rho =
    \frac{\muS^2}{2} \Phi^{\dagger}\Phi
    + \frac{1}{2} \psi^{-4}\bgamma^{kl} \bar{D}_{k}\Phi^{\dagger} \bar{D}_{l}\Phi
\,.
\end{equation}

The momentum constraint, Eq.~\eqref{eq:CTT_MC_final},
can be solved exactly by the Bowen-York extrinsic curvature~\cite{Bowen:1980yu}.
For a single \bh{} it is given by
\begin{align}
\label{eq:BY_solution}
\bA^{\rm{BY}}_{ij} &=
    \frac{3}{2r^2}\left[q_{i} P_{j} + q_{j} P_{i} + q^{k} P_{k} \left(q_{i}q_{j} - \eta_{ij} \right)  \right]
\\ &\quad
    - \frac{3}{r^3} \left(\epsilon_{ikl} q_{j} + \epsilon_{jkl} q_{i} \right)  q^{k} S^{l}
\,,\nonumber
\end{align}
where
$q^{i}$ is an outward-pointing unit radial vector,
$\epsilon_{\mu\nu\rho} = n^{\sigma} \epsilon_{\sigma\mu\nu\rho}$ is the Levi-Civita tensor in three dimensions,
$r$ the \bh{'s} position,
and $P^{i}$ and $S^{i}$ are the ADM
linear and angular momenta, respectively.
Because the momentum constraint, Eq.~\eqref{eq:CTT_MC_final},
is a linear differential equation,
we obtain solutions for multiple \bh{s}
via a superposition of individual Bowen-York solutions as shown by Brandt \& Br\"{u}gmann~\cite{Brandt:1997tf}.

With the momentum constraint solved analytically,
all that remains is solving the Hamiltonian constraint, Eq.~\eqref{eq:CTT_HC_confflat}.
To this end, we adopt the puncture decomposition of the conformal factor~\cite{Brandt:1997tf},
\begin{align}
\label{eq:CTT_puncture_decomposition}
\psi & = \psi_{\mathrm{BL}} + u
\,,\,\, \textrm{with}\quad
\psi_{\mathrm{BL}} =
    1 + \sum_{a = 1}^{N} \frac{m_{(a)}}{2 |r - r_{(a)}|},
\end{align}
where
$m_{(a)}$ and $r_{(a)}$ are the mass parameter and position of the $a$-th \bh{,}
$\psi_{\rm{BL}}$ denotes the Brill-Lindquist solution
(for time symmetric \bh{} solutions in vacuum),
and $u$ is a $C^2$
function determining the correction to that
solution.

Applying Eq.~\eqref{eq:CTT_puncture_decomposition}
and using $\bgamma_{ij}=\eta_{ij}$ explicitly,
the Hamiltonian constraint becomes
\begin{align}
\label{eq:CTT_HC_final}
0 & = \Delta_{\eta} u
    + \frac{1}{8}\left(u + \psi_{\mathrm{BL}}\right)^{-7} \bar{A}_{ij}\bar{A}^{ij}
\\ & \quad
    + \pi  \left(u + \psi_{\mathrm{BL}}\right)^5\muS^2 \Phi^\dagger \Phi
    + \pi \left(u + \psi_{\mathrm{BL}}\right)
    \eta^{kl} \partial_k \Phi^\dagger \partial_{l} \Phi
\,, \notag
\end{align}
where
$\Delta_{\eta}=\eta^{ij}\partial_i\partial_j$
is the flat space Laplacian,
and we insert
the scalar's energy density.

Our strategy has reduced the initial data problem
to solving one elliptic \pde for the function $u$.
This approach captures a large class of interesting problems involving the interaction between \bh{s} and scalar fields, as long as the scalar can be treated as (approximately)
momententarily at rest.
More demanding setups,
such as superradiant quasi-bound states~\cite{Detweiler:1980uk,Dolan:2007mj}
which are time-dependent,
or highly spinning \bh{s}
(exceeding the spin bound of Bowen-York data~\cite{Dain:2008ck,Lovelace:2008tw})
would entail solving the coupled momentum constraint equations~\cite{East:2012zn,Aurrekoetxea:2022mpw}.
They are beyond the scope of the current work.

\subsection{Well-posedness of the Hamiltonian constraint}
\label{ssec:Wellposedness}

With the formalism in hand, we implemented Eq.~\eqref{eq:CTT_HC_final} in \TPBBHSF{} (see Sec.\ref{ssec:initialdatacode} for details).
The initial numerical solutions of \bh{s}
in the presence of a massive scalar field,
appeared  physical and reasonably well behaved.
However, they failed to converge as we increased the resolution of the spectral solver.

This behavior is a strong indication for a numerical instability,
and we identified the
mathematical structure of Eq.~\eqref{eq:CTT_HC_final} as its root.
To see this, recall that
a necessary condition for numerical stability is the well-posedness of the underlying \pde{.}
Here, we are concerned with elliptic \pde{s}.
They are well-posed if their linearized version,
\begin{align}
\label{eq:EllipticPDELinearized}
\Delta \epsilon - h\epsilon & = 0
\,,
\end{align}
where $\epsilon$ is a small perturbation,
satisfies $h>0$ \cite{Choquet-Bruhat:1999gli,Gourgoulhon:2007ue}.

Analysing Eq.~\eqref{eq:CTT_HC_final},
we find that the mass term $\sim \muS^{2} (u+\psiBL)^5 \Phi^{\dagger}\Phi$ changes the sign of the function $h$ to $h<0$.
That is, the mass term changes the elliptic character of the constraint, thus spoiling the convergence properties of our code.
We cure this problem by conformally rescaling the scalar field as
\begin{align}
\label{eq:SF_conformal_rescaling}
\Phi & = \psi^{\delta} \bar\Phi
\,, \qquad
\Phi^\dagger = \psi^{\delta} \bar\Phi^\dagger
\,,
\end{align}
where $\psi$ is the conformal factor.
The key idea is to choose the exponent $\delta$ such that
 $h>0$ in the linearized equation.
This corresponds to choosing $\delta$ such that
all powers of the conformal factor in
the Hamiltonian constraint become negative.
After applying the rescaling, the Hamiltonian constraint, Eq.~\eqref{eq:CTT_HC_final}, becomes
\begin{widetext}
\begin{align}
\label{eq:CTT_HC_final_SF_rescaled}
0 = \Delta_{\eta} u
    + \frac{\bar{A}_{ij}\bar{A}^{ij}} {8(u + \psiBL)^{7}}
    & + \pi  \left(u + \psiBL \right)^{2\delta+5} \muS^2 \bar{\Phi}^\dagger \bar{\Phi}
    + \pi \delta \left( u + \psiBL \right)^{2 \delta}
      \left( \p^{i} u + \p^{i} \psiBL \right)
      \left( \bar{\Phi} \p_{i} \bar{\Phi}^{\dagger}
            + \bar{\Phi}^{\dagger} \p_{i} \bar{\Phi} \right)
\\ &
    + \pi \left( u + \psiBL \right)^{2\delta+1} (\p^{i}\bar{\Phi}) (\p_{i}\bar{\Phi}^{\dagger})
    + \pi \delta^2
        \left(u+\psiBL\right)^{2\delta-1}
        \left( \p^{i} u + \p^{i} \psiBL \right)
        \left( \p_{i} u + \p_{i} \psiBL \right)
        \bar{\Phi}^{\dagger} \bar{\Phi}
\,, \notag
\end{align}
\end{widetext}
where we raise indices using the flat metric $\eta^{ij}$.
The well-posedness of the elliptic equation can be shown by linearizing Eq.\eqref{eq:CTT_HC_final_SF_rescaled} around a known solution $u_0$.
We introduce $u = u_0 + \epsilon$,
with $\left|\epsilon\right| \ll |u_0|$,
and expand Eq.~\eqref{eq:CTT_HC_final_SF_rescaled} to first order in $\epsilon$
to find
\begin{widetext}
\begin{align}
\label{eq:CTT_HC_final_SF_rescaled_lin}
0 & = \Delta_{\eta} \epsilon
    - \frac{7\bar{A}_{ij}\bar{A}^{ij}} {8(u_0 + \psi_{\mathrm{BL}})^{8}} \epsilon
    + \pi (u_0 + \psi_{\mathrm{BL}})^{2 \delta}
    \left[
        (2\delta+1)
        \left(\partial^{i}\bar{\Phi}^{\dagger}\right)
        \left(\partial_{i}\bar{\Phi}\right)
        \epsilon
        + \delta \left( \bar{\Phi}\partial^{i}\bar{\Phi}^{\dagger}{}
                      + \bar{\Phi}^{\dagger}{}  \partial^{i}\bar{\Phi} \right)
        \left( \partial_{i} \epsilon \right)
    \right]
\\ &
    + \pi (2\delta+5) \muS^2 |\bar{\Phi}|^2
    (u_0 + \psi_{\mathrm{BL}})^{2\delta+4}
    \epsilon
    + 2\pi \delta^2 (u_0 + \psi_{\mathrm{BL}})^{2\delta-1}
    \left(\partial^{i}u_0 + \partial^{i}\psi_{\mathrm{BL}}\right)
    \left[
        \left( \bar{\Phi}\partial_{i}\bar{\Phi}^{\dagger}{} +  \bar{\Phi}^{\dagger}{}  \partial_{i}\bar{\Phi} \right) \epsilon
        +  |\bar{\Phi}|^2 \left(\partial_{i}\epsilon\right)
    \right]
\notag \\ &
    + \pi (2\delta-1) \delta^2 |\bar{\Phi}|^2
    (u_0 + \psi_{\mathrm{BL}})^{2\delta-2}
    \left(\partial^{i}u_0 + \partial^{i}\psi_{\mathrm{BL}}\right)
    \left(\partial_{i}u_0 + \partial_{i}\psi_{\mathrm{BL}}\right)
    \epsilon
\,. \notag
\end{align}
\end{widetext}
This can be brought (approximately) into the form of Eq.~\eqref{eq:EllipticPDELinearized}.
Then, the existence of a unique solution requires that $h > 0$~\cite{Choquet-Bruhat:1999gli,Gourgoulhon:2007ue}.
Since the function $h$ is rather complicated we
focus on the term with the largest power of the conformal factor.
This corresponds to the term $\sim \muS^{2}$,
and it becomes negative when $\delta<-5/2$.
In practice, we set $\delta=-3$ in our simulations.
We verified that different values $\delta\leq -3$ yield stable and unique solutions.
Furthermore, the conformal rescaling implies regularity at the punctures, since the (physical) scalar field vanishes when approaching the punctures' positions by virtue of Eq.~\eqref{eq:SF_conformal_rescaling}.

\subsection{Initial scalar field profile}
\label{ssec:ScalarInitialProfile}
We initialize the (conformally rescaled)
scalar field as a Gaussian shell
that is momentarily at rest,
\begin{subequations}
\label{eq:ID_SF_profile}
\begin{align}
\label{eq:ID_SF_Phi}
\bar{\Phi} (t=0) & = A_{\rm{SF}}\, Z(\theta,\phi) \exp\left[-\frac{(r - r_0)^2}{w^2}\right]
\,,\\
\label{eq:ID_SF_Pi}
\bar{\Pi}(t=0) & = 0
\,,
\end{align}
\end{subequations}
where $A_{\rm{SF}}$ is the amplitude,
$r_{0}$ and $w$ are the peak position and width of the Gaussian, and $Z(\theta,\phi)$ indicates the field's angular distribution.
In the code we implemented the options to set $Z(\theta,\phi)=Y_{00}(\theta,\phi)$ or $Z(\theta,\phi)=Y_{11}(\theta,\phi)$,
where $Y_{lm}$ are spherical harmonics.
In this paper we always set
$Z(\theta,\phi)=Y_{00}(\theta,\phi)=1/\sqrt{4\pi}$.

\section{Numerical relativity framework}
\label{sec:NRframework}
To simulate binary \bh{s} immersed in a scalar cloud,
we extend and use the \ETK{}~\cite{roland_haas_2024_14193969,EinsteinToolkitWeb} alongside the \canuda code~\cite{witek_helvi_2023_7791842,CanudaGitWeb}.
We summarize the software in Sec.~\ref{ssec:SoftwareETKCanuda},
and describe our new initial data solver in Sec.~\ref{ssec:initialdatacode}.
In Sec.~\ref{ssec:simulationsetup}
we list the simulations performed for this paper and summarize their setup.

\subsection{Software description}
\label{ssec:SoftwareETKCanuda}
We perform  simulations with the \ETK{,}
an open-source software for numerical relativity and computational astrophysics~\cite{roland_haas_2024_14193969,EinsteinToolkitWeb}.
The toolkit consists of a set of ``arrangements'' that, in turn, are a collection of ``thorns'' which implement specific physics or infrastucture tasks.
The \ETK{} is based on the \cactus{} computational toolkit~\cite{Goodale:2002a,Cactuscode:web}
and the \carpet{}
boxes-in-boxes
\amr driver~\cite{Schnetter:2003rb,CarpetCode:web}.
In addition to \amr{,} it also provides thorns for I/O and MPI/OpenMP hybrid parallelization.

The physics modules for this project are provided by our open-source
\canuda{} code
for numerical relativity in fundamental physics~\cite{witek_helvi_2023_7791842}.
We use the {\textsc{Scalar}} arrangement
with its initial data and scalar evolution thorns~\cite{Okawa:2014nda,Cunha:2017wao},
and the spacetime evolution and wave extraction thorns
in the {\textsc{lean\_public}}
arrangement.

We construct constraint-satisfying initial data of binary \bh{} immersed in a massive scalar cloud with our new initial data solver \TPBBHSF{}~\cite{TPBBHSFGitWeb}.
It is an extension of the \TP{} spectral solver~\cite{Ansorg:2004ds}.
We present a detailed description of \TPBBHSF{,} code validation and convergence tests in Sec.~\ref{ssec:initialdatacode} below.
We determine the \bh{s'} initial momenta needed for a quasi-circular inspiral with the \textsc{NRPyPN} script~\cite{Habib:2020dba,NRPyWeb}
We evolve the coupled scalar and metric equations in the \bssn{} formulation~\cite{Baumgarte:1998te,Shibata:1995we} together with the moving puncture gauge~\cite{Campanelli:2005dd,Baker:2005vv,Alcubierre:2002kk}
as summarized in
App.~\ref{appendix:BSSN}.
The metric evolution equations are implemented in the
{\textsc{LeanBSSNMoL}}
thorn,
while the scalar's evolution is carried out with the
{\textsc{ScalarEvolve}}
thorn.
The latter also computes the components of the scalar field's energy-momentum tensor that are stored in the {\textsc{TmunuBase}} thorn.
To couple the scalar field to the metric, {\textsc{LeanBSSNMoL}} reads the components of the energy-momentum tensor from {\textsc{TmunuBase}}. 

The equations are numerically integrated via the method of lines.
We calculate spatial derivatives using finite differences.
We upgraded both evolution thorns to provide up to
eighth order finite difference stencils
that are centered for all regular evolution variables
and lop-sided for advection derivatives $\sim\p_{k}\beta^{k}$.
We add artifical Kreiss-Oliger dissipation to reduce high-frequency noise that is generated at mesh refinement boundaries.
The simulations presented here use eighth order finite difference stencils for spatial derivatives together with ninth order stencils for the dissipation and a dissipation coefficient $\epsilon_{\rm diss} = 0.325$.
We adopt Sommerfeld, i.e., radiative boundary conditions for both the metric and scalar field \cite{Alcubierre:2002kk}. 
We integrate the evolution equations with a fourth order Runge-Kutta integrator.
The \ETK{} uses
subcycling in time, and the time step on each refinement level is determined by the Courant factor
which we typically set to
$\dif t/\dif x=0.45$.

We compute two types of observables, namely the gravitational and scalar radiation generated by the coalescence, and (local) properties of the \bh{s}.
Information on the gravitational radiation is contained in the Newman-Penrose scalar $\Psi_{4}$; see Sec.~\ref{ssec:WaveExtraction}.
To compute the Newman-Penrose scalar in the presence of
the scalar field,
we extend \canuda{'s} wave extraction thorn {\textsc{NPScalars}}.
The updated {\textsc{NPScalars}} thorn reads the energy-momentum tensor
from the {\textsc{TmunuBase}} thorn, 
constructs the energy density,
energy-momentum flux and spatial stress,
and incorporates them via Eqs.~\eqref{eq:EijBijIn3p1}.
We further upgraded {\textsc{NPScalars}} by implementing up to eighth order finite difference stencils to compute $\Psi_{4}$.
We perform a multipolar decomposition of the Newman-Penrose scalar and of the scalar field (c.f. Eq.~\eqref{eq:Multipoles})
using the \textsc{Multipole} thorn.

We compute (local) properties of the individual and final \bh{s} using the \textsc{QuasiLocalMeasures} thorn \cite{Dreyer:2002mx}, and their apparent horizons with the \textsc{AHFinderDirect} thorn \cite{Thornburg:2003sf}.
These include the apparent horizons'
area $A_{\rm AH}$,
equatorial circumference $C_{\rm e}$,
and irreducible mass $M_{\rm irr}=\sqrt{A_{\rm AH}/(16\pi)}$.
The latter are used
to compute the \bh{s'} dimensionless spin
\begin{align}
\label{eq:AHspin}
\chi & \equiv \frac{J}{M_{\rm BH}^2}
    = \sqrt{ 1 - \bigg( \frac{2\pi A_{\rm AH}}{C_{\rm e}^2} - 1 \bigg)^2 }
\,,
\end{align}
and mass according to Christodoulou's formula,
\begin{align}
M_{\rm BH}^2 & = M_{\rm irr}^2 + \frac{J^2}{4M_{\rm irr}^2}
\,.
\label{eq:christodouloumass}
\end{align}

\subsection{Initial data solver: \TPBBHSF{}}
\label{ssec:initialdatacode}
We have designed the initial data routine
\TPBBHSF{}~\cite{TPBBHSFGitWeb}
to prepare constraint-satisfying initial configurations of binary \bh{s} surrounded by a massive scalar field.
It is public as a new thorn of the
{\textsc{Scalar}} arrangement~\cite{CanudaGitWeb}.
We base \TPBBHSF{} on the \TP{} pseudo-spectral solver~\cite{Ansorg:2004ds},
for conformally flat binary \bh{} initial data~\cite{Brandt:1997tf}.
In particular, we implement Eqs.~\eqref{eq:CTT_HC_final_SF_rescaled} and~\eqref{eq:CTT_HC_final_SF_rescaled_lin},
while leaving  \TP{'s} underlying infrastructure unchanged.
The solver employs pseudo-spectral methods, and expands the solution in the basis of
Chebyshev  or (for the azimuthal coordinate) Fourier
polynomials.

The accuracy of a solution is determined by
the number of collocation points $\left(N_A,N_B,N_{\varphi}\right)$, i.e., the number of terms in the series expansion.
We validate the code by performing a convergence analysis.
To do so, we fix the number of collocation points in the azimuthal coordinate, $N_\varphi = 24$, and vary $N_A = N_B = N$.
We set the solution $u_{160}$ with $N=160$ as our reference solution, and define the relative error
\begin{align}
\label{eq:ID_convergence_estimator}
\Delta_{N,160} & = \max\left( 1 - \frac{u_N}{u_{160}} \right)
\,,
\end{align}
where $u_N$ denotes solutions constructed with a lower resolution $N < 160$.

\begin{figure}[htp!]
    \centering    \includegraphics[width=.47\textwidth]{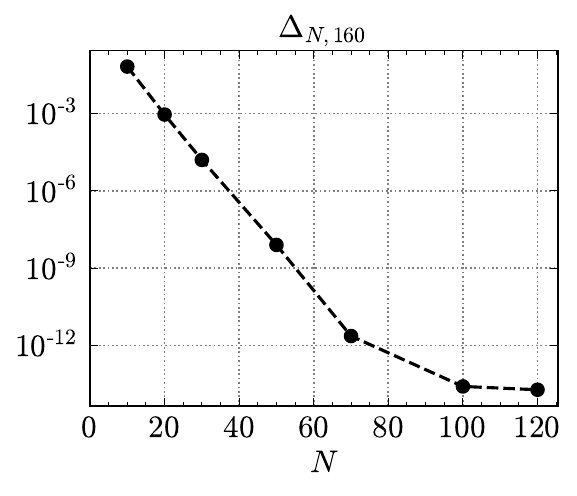}
    \caption{Convergence plot for \TPBBHSF{,} computing the initial solution of a binary \bh{} with mass ratio $q=1/2$ immersed in a scalar cloud with mass parameter $M\muS=0.4$; run \textbf{q12mu04} in Table~\ref{tab:simulations_list}.
    We show the relative error $\Delta_{N,160}$ between a solution $u_{N}$ obtained with $N$ collocation points and the reference solution $u_{160}$,
    as a function of the number of collocation points $N$.
    }
    \label{fig:ID_convergence}
\end{figure}

We perform the convergence test for
run \textbf{q12mu04}, which is
one of the most demanding setups in our pool of simulations listed in Table~\ref{tab:simulations_list}.
The run represents a binary \bh{} with a mass ratio $q=M_1/M_2 = 1/2$ surrounded by a scalar cloud with mass parameter $M\muS=0.4$.
We construct this setup's initial solution, $u_{N}$, with a series of collocation points $N\in\{10,120\}$,
and compute the relative error defined in Eq.~\eqref{eq:ID_convergence_estimator}. Specifically, we compute the relative error along the $x$-axis, and its
maximum typically corresponds to the position of the punctures.
The results are presented in  Fig.~\ref{fig:ID_convergence}.
With increasing number of collocation points $N$, the relative error, $\Delta_{N,160}$,
decays exponentially until about $N\sim100$,
and then saturates.
This cross-over from exponential to algebraic convergence is already present in the original \TP{} solver,
and it is due to
the solution's logarithmic fall-off at infinity~\cite{Ansorg:2004ds}.

For all simulations in this paper, we choose $N=80$
and $N_{\varphi}=24$
for which we still obtain exponential convergence and the relative error is $\Delta_{N=80,160}\lesssim 10^{-12}$.

\subsection{Simulation setup}
\label{ssec:simulationsetup}

We perform a simulation campaign to determine the impact of a scalar condensate on the dynamics of coalescing binary \bh{s} and their gravitational-wave emission,
as sketched in Fig.~\ref{fig:sketch}
and summarized in Table~\ref{tab:simulations_list}.

We set up binaries of quasi-circular, nonspinning \bh{s} with
ADM masses $M_{1,2}$, total mass $M=M_{1}+M_{2} = 1$, and mass ratios $q=M_{1}/M_{2}=1, 1/2$.
The \bh{s} have an initial separation of $d = \left|x_1 - x_2\right| = 10M$.
The binary is initially surrounded by a
spherically symmetric scalar condensate
with a Gaussian radial profile of the conformal scalar field, $\bar{\Phi}$, given by Eq.~\eqref{eq:ID_SF_profile},
and angular profile $Z(\theta,\varphi)=Y_{00}(\theta,\varphi)=1/\sqrt{4\pi}$.
The Gaussian is centered around the origin, $r_{0}=0$,
has a width of $w=10M$, and the amplitude $A_{\rm{SF}}$ is chosen such that the maximum of the energy density (in flat space) is about $\bar{\rho}_{\rm max} M^{2} = 1.6\times10^{-7}$.

\begin{table}[ht!]
    \label{tab:bbh_parameters_Longlist}
    \centering
    \begin{tabular}{ |c|c|c|c|c|c|c|c|c|c|c|c|c|c| }
        \hline
        Name & $q$ & $M\muS$ & $A_{\rm SF}/M$ & $\lambda_r$ & $\lambda_t$ & $M_{\rm ADM}/M$ & $\epsilon_{\Omega}$ \\ \hline
        \textbf{q1vac}  & 1   & NA  & NA     & 1    & 1     & 0.9898 & 0.0041 \\ \hline
        \textbf{q1mu00}    & 1   & 0.0 & 0.0233 & 0.5  & 1     & 0.9906 & 0.0089 \\ \hline
        \textbf{q1mu02}    & 1   & 0.2 & 0.0100 & 0.8  & 1     & 0.9902 & 0.0010 \\ \hline
        \textbf{q1mu04}    & 1   & 0.4 & 0.0050 & 1    & 1     & 0.9901 & 0.0029 \\ \hline
        \textbf{q1mu06}    & 1   & 0.6 & 0.0033 & 1    & 1.001 & 0.9901 & 0.0089 \\ \hline
        \textbf{q1mu08}    & 1   & 0.8 & 0.0025 & 1    & 1.001 & 0.9906 & 0.0091 \\ \hline
        \textbf{q1mu10}    & 1   & 1.0 & 0.0020 & 1    & 1.001 & 0.9901 & 0.0089 \\ \hline
        \textbf{q12vac} & 1/2 & NA  & NA     & 1    & 1     & 0.9909 & 0.0051 \\ \hline
        \textbf{q12mu00}   & 1/2 & 0.0 & 0.0233 & 0.84 & 1.001 & 0.9918 & 0.0013 \\ \hline
        \textbf{q12mu02}   & 1/2 & 0.2 & 0.0100 & 0.90 & 1.001 & 0.9913 & 0.0009 \\ \hline
        \textbf{q12mu04}   & 1/2 & 0.4 & 0.0050 & 1.12 & 1.001 & 0.9912 & 0.0048 \\ \hline
        \textbf{q12mu06}   & 1/2 & 0.6 & 0.0033 & 1.12 & 1.001 & 0.9912 & 0.0045 \\ \hline
        \textbf{q12mu08}   & 1/2 & 0.8 & 0.0025 & 1.12 & 1.001 & 0.9912 & 0.0047 \\ \hline
        \textbf{q12mu10}   & 1/2 & 1.0 & 0.0020 & 1.12 & 1.001 & 0.9912 & 0.0046 \\ \hline
    \end{tabular}
    \caption
    {
    Summary of the initial binary \bh{} and scalar field parameters.
    We report the name of the simulation,
    the \bh{s'} mass ratio $q = M_1/M_2\leq 1$,
    the scalar's (dimensionless) mass parameter $M\muS$ and amplitude $A_{\rm{SF}}$,
    eccentricity reduction parameters $\lambda_r$ and $\lambda_t$,
    ADM energy $M_{\rm ADM}$,
    and eccentricity $\epsilon_\Omega$.
    We choose the scalar's amplitude $A_{\rm SF}$ such that its maximum energy density
    (in the flat space approximation)
    is $\bar{\rho}_{\max} M^{2} = 1.6\times 10^{-7}$.
    The initial momenta
    in vacuum are set to
    $(P_r^{(0)}, P_t^{(0)}) = (-0.101,\, 9.63) \times 10^{-2} M$ for \textbf{q1vac}, and $(P_r^{(0)}, P_t^{(0)}) = (-0.071,\, 8.557) \times 10^{-2} M$ for \textbf{q12vac}.
    In simulations with the scalar field,
    we apply the eccentricity reduction parameters
    and set the initial momenta to
    $(P_r, P_t) = (\lambda_r P_r^{(0)}, \lambda_t P_t^{(0)})$.
    \label{tab:simulations_list}
    }
\end{table}

There are two factors determining our choice of the initial energy density.
The first, perhaps more practical reason, is that the scalar cloud density should be sufficiently large to have a measurable impact on the binary.
On the other hand, our choice has to be astrophysically viable,
i.e., it should be below the maximum density
that can be reached through
known enhancement mechanisms.
To connect with astrophysical estimates, 
we convert 
the energy density between geometric units used in numerical relativity and physical units according to
\begin{align}
\label{eq:RhoConversionNR2Phys}
\rho_{\rm{NR}} M^{2} &
    = 10^{-28} \left(\frac{\rho}{M_{\odot}/\rm{pc}^{3}}\right)
        \left(\frac{M_{\rm{BH}}}{10^{6}M_{\odot}} \right)^2
\,.
\end{align}
The average galactic dark matter abundance is about
$\rho_{\rm{gal}}\sim 0.01 M_{\odot}/ {\rm{pc}}^3$~\cite{Pato:2015dua}.
In other words, the corresponding value in geometric units is $\rho_{\rm{NR,gal}} M^2\sim 10^{-30} (M_{\rm{BH}}/10^6 M_{\odot})^2$, and therefore negligible in our simulations.
However,
enhancement mechanisms such as the superradiant growth of a bosonic cloud can carry as much as $10\%$ of the \bh{} mass~\cite{Brito:2014wla,Ficarra:2018rfu,East:2017ovw}.
Furthermore, when combined with accretion~\cite{Hui:2022sri},
superradiance can
increase the energy density up to
$\rho_{\rm{NR,SR}} M^{2} \sim 10^{-5} (M_{\rm{BH}}/10^6 M_{\odot})^2$ in geometric units~\cite{East:2017ovw}.
With these considerations, the energy density chosen in our simulations,
$\bar{\rho}_{\rm max}M^{2} = 1.6\times10^{-7},$
may therefore represent a superradiant cloud around a supermassive \bh{.}

We perform a series of binary \bh{} simulations, varying the scalar field's dimensionless mass parameter in the range
$M\muS\in\{0,0.2,0.4,0.6,0.8,1\}$
for each mass ratio $q\in\{1,1/2\}$.
The simulation suite is summarized in Table~\ref{tab:simulations_list}.
We list the selection of the initial \bh{} and scalar field parameters, the binaries' ADM mass, initial momenta and eccentricity.

To identify the effect of a
scalar field environment, we need to compare
the evolution of quasi-circular binaries -- of comparable, low eccentricities -- in the presence of scalar matter against those in vacuum.
However, even in a consistent solution of the constraint equations, the scalar cloud can yield different initial \bh{} parameters
and, thus, introduce artificially eccentric orbits that are due to the initial conditions rather than being a signature of dark matter~\cite{Bamber:2022pbs}.
To rule out this effect, we therefore use an iterative procedure to reduce the binaries' eccentricity.
Guided by state-of-the-art in vacuum binary 
\bh{} simulations, we aim at generating initial configurations with an eccentricity $\epsilon_\Omega\lesssim 9\times 10^{-3}$.

In vacuum spacetimes, low eccentricity is achieved by
setting the \bh{s'} momenta from post-Newtonian approximations,
estimating the binary's eccentricity after evolving it for a few orbits
and correcting the holes' momenta~\cite{Ramos-Buades:2018azo}
iteratively until 
the target eccentricity is met.

For binary \bh{s} surrounded by a scalar condensate,
we adjust this procedure as follows:
\begin{enumerate}[noitemsep,leftmargin=*,labelindent=0mm,label=(\roman*)]
    \item We set the initial guess of the \bh{s'}
    linear momenta with the 
    \textsc{NRPyPN} code~\cite{Habib:2020dba},
    based on Post-Newtonian calculations of quasi-circular binaries in vacuum~\cite{Husa:2007rh,Healy:2017zqj,Ramos-Buades:2018azo,Ciarfella:2024clj}.
    We set the tangential and radial components of the initial momenta for an equal-mass binary with separation $d=10M$ to
    $P^{(0)}_t = 9.63\times 10^{-2} M$ and $P^{(0)}_r = 1.01\times 10^{-3}M$.
    For the binary with mass ratio $q=1/2$
    we set $P^{(0)}_t = 8.557\times 10^{-2}M$ and $P^{(0)}_r = 7.1\times 10^{-4}M$.
    \item We initialize the conformal scalar field as
    a Gaussian
    using
    Eq.~\eqref{eq:ID_SF_profile} with
    $Z(\theta,\varphi)=1/\sqrt{4\pi}$,
    $r_{0}=0$, $w=10M$, and $A_{\rm SF}$ and $M\muS$ given in Table~\ref{tab:simulations_list}.
    \item We solve the constraints for the initial metric and extrinsic curvature
    using \TPBBHSF{}  with the initial parameters set in steps~(i) and~(ii); see Sec.~\ref{sec:initialdata}.
    \item We evolve the initial data for several orbits.
    After the gauge adjustment
    (present in the first $\sim 200M$)
    we compute the eccentricity estimator~\cite{Ramos-Buades:2018azo}
    \begin{equation}
    \label{eq:EccentricityEstimator}
    \epsilon_{\Omega} = \frac{\Omega(t) - \Omega(\epsilon = 0)}{2\Omega(\epsilon = 0)}
    \,.
    \end{equation}
    It is a measure for the time dependent oscillations of the orbital frequency $\Omega(t)$ relative to the frequency of a circular orbit $\Omega(\epsilon = 0)$.
    The former is computed from the coordinate separation.
    The latter is obtained by fitting the numerical data assuming a quasi-circular ansatz for $\Omega$ from PN expansions, and averaging any eccentricity or gauge effects. For more details Sec.~IIIB of Ref.~\cite{Ramos-Buades:2018azo}.
    \item Based on the eccentricity estimate in Eq.~\eqref{eq:EccentricityEstimator},
    we correct the initial linear momenta as
    $P_{t,r} \to \lambda_{t,r} P_{t,r}$,
    and then recompute the initial data.
    We repeat steps (iii)~--~(iv), each time choosing a different set of correction factors $\lambda_{t, r}$
    (by trial-and-error),
    until the eccentricity satisfies $\epsilon_{\Omega}\lesssim 9\times 10^{-3}$.
    In the simulations presented here, we typically need 3-5 iterations to reach the eccentricity threshold.
\end{enumerate}

With the initial data carefully set up to represent low eccentricity, quasi-circular binary \bh{s},
we turn our attention to the evolution.
All simulations presented here
employ eighth order finite difference stencils.

The simulation domain consists of a three-dimensional Cartesian grid with an extent of $\sim 512M$ in each dimension, where $M$ is the total mass in the spacetime.
In order to sufficiently resolve the \bh{} horizons, we aim for a grid spacing finer than $M/64$ at each of the \bh{s}, and to make the simulation computationally feasible we employ 7 levels of moving, box-in-box style mesh refinement using the \carpet driver \cite{Schnetter:2003rb}.
The grid resolution on the outermost refinement level is
$\Delta x_{\rm c} = 0.854M$, corresponding to a grid spacing of $0.854M/2^6 \sim 1/75 M$ at the center of the \bh{s}. We provide a full suite of convergence tests in App.~\ref{app:convergenctest}, with relative error of $<4\%$ for the modulus and $<0.3\%$ for the complex phase of $\Psi_{4, 22}$.

\section{Results}
\label{sec:results}

\subsection{Evolution of the binary-cloud system}
\begin{figure*}[htp!]
    \centering
    \includegraphics[width=.29\textwidth]{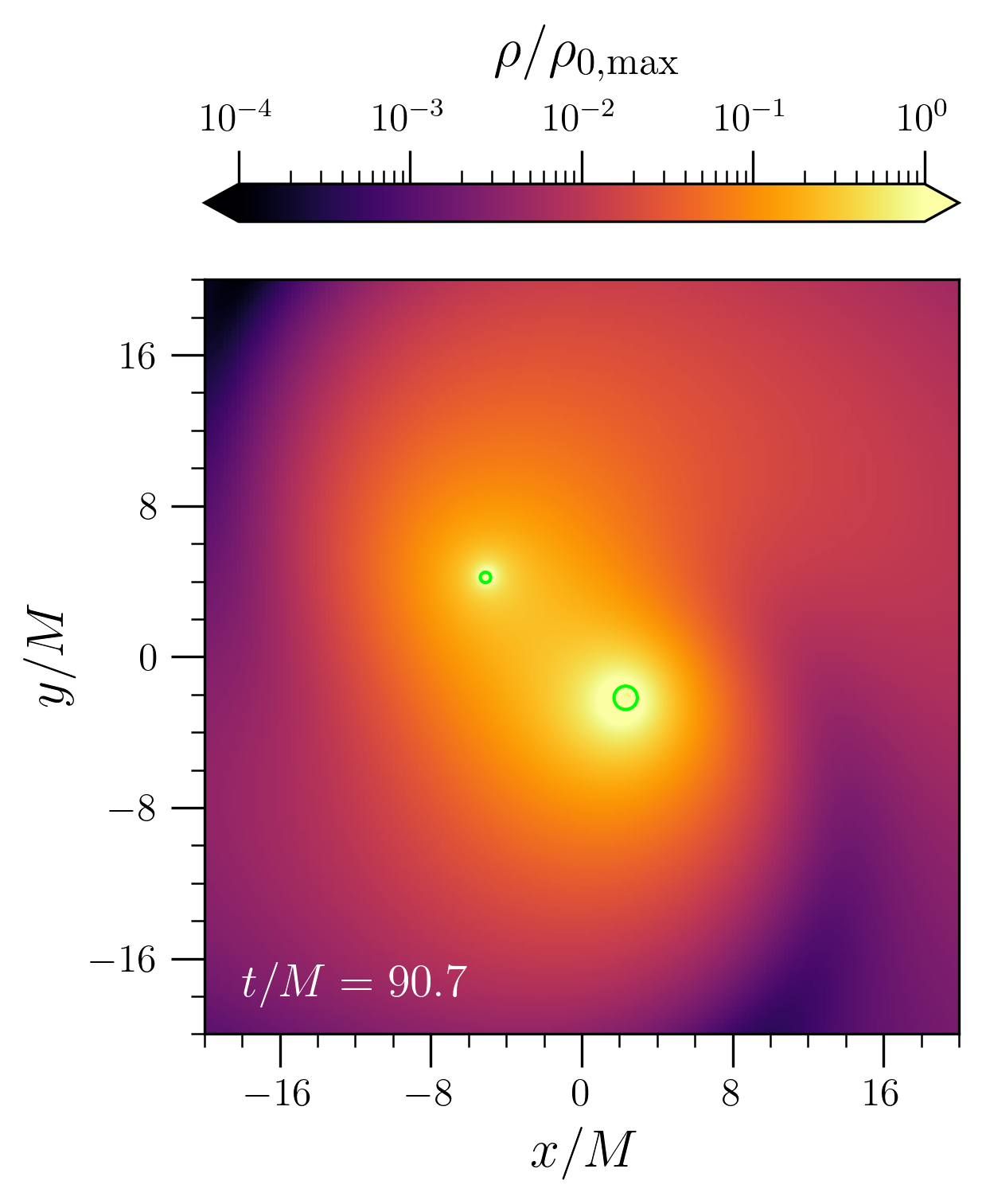}
    \includegraphics[width=.29\textwidth]{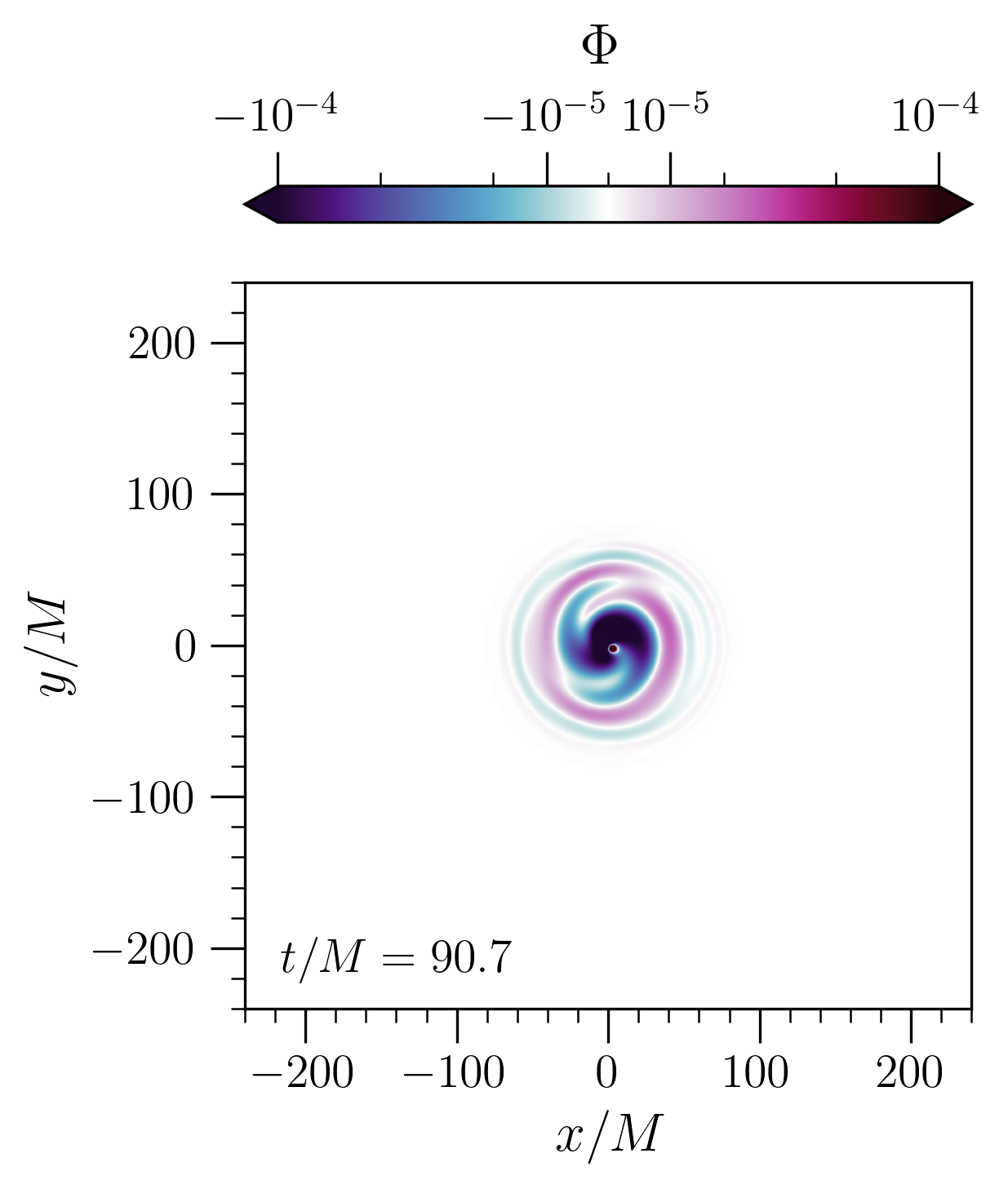}
    \includegraphics[width=.29\textwidth]{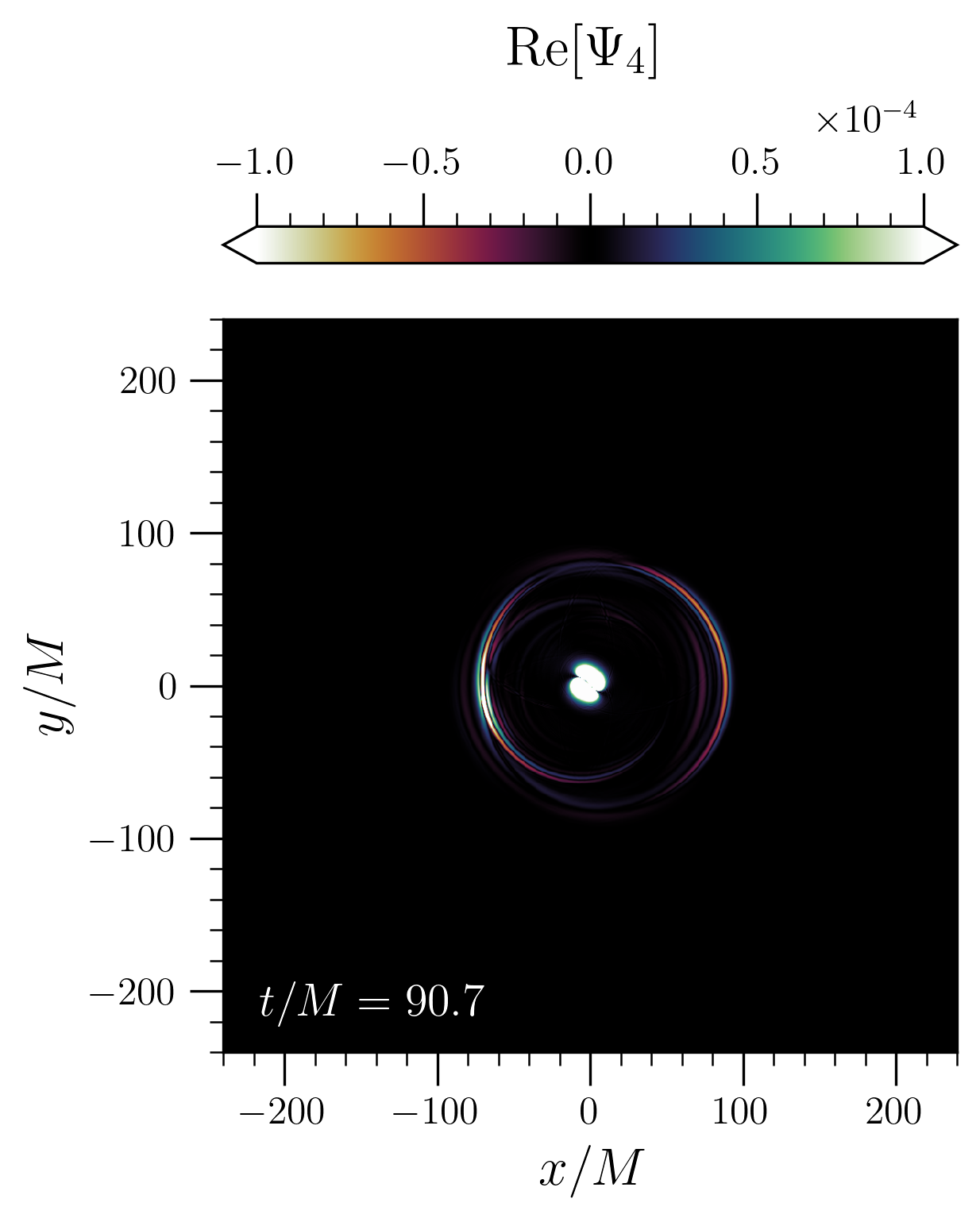}
    \includegraphics[width=.29\textwidth]{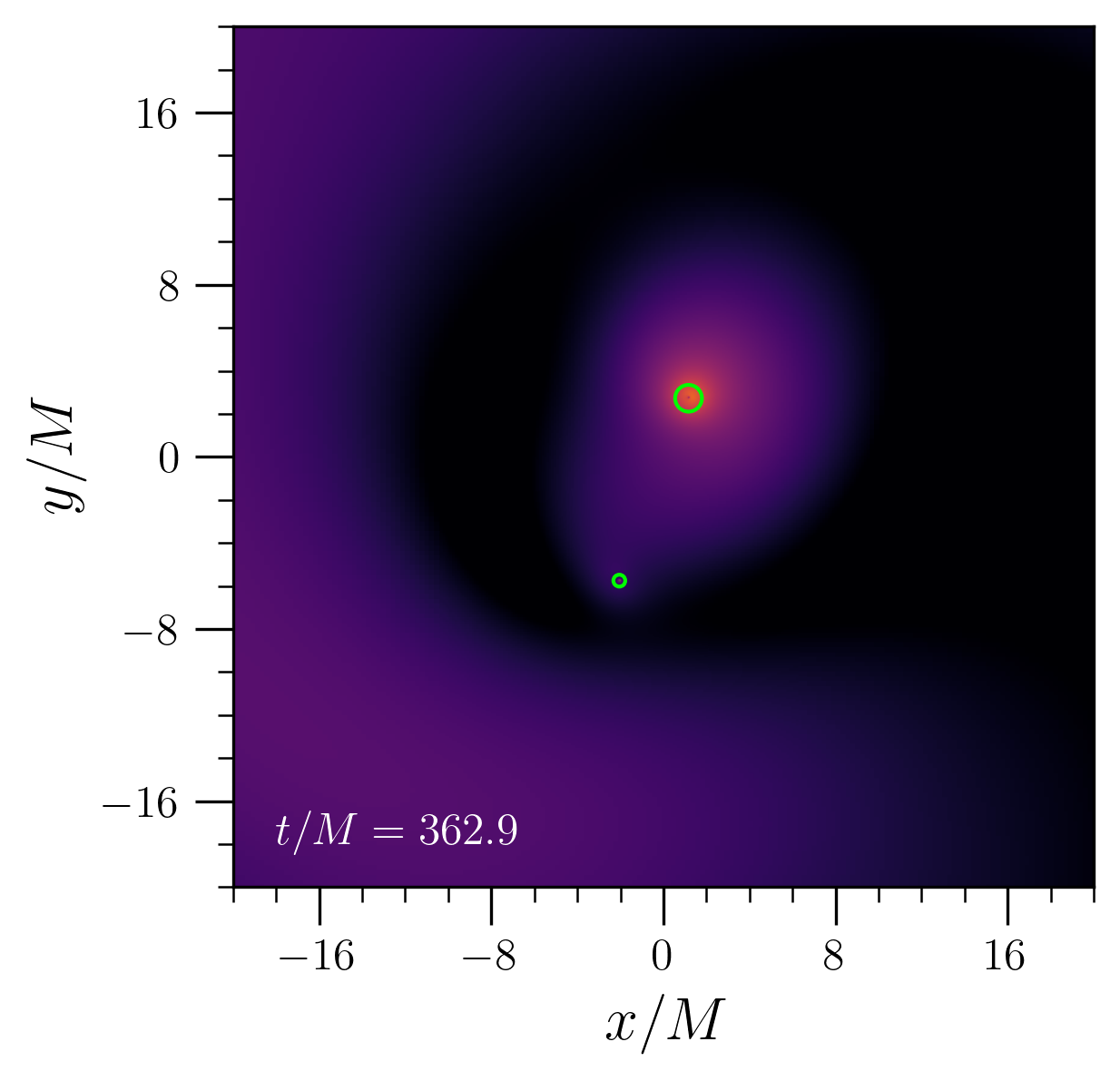}
    \includegraphics[width=.29\textwidth]{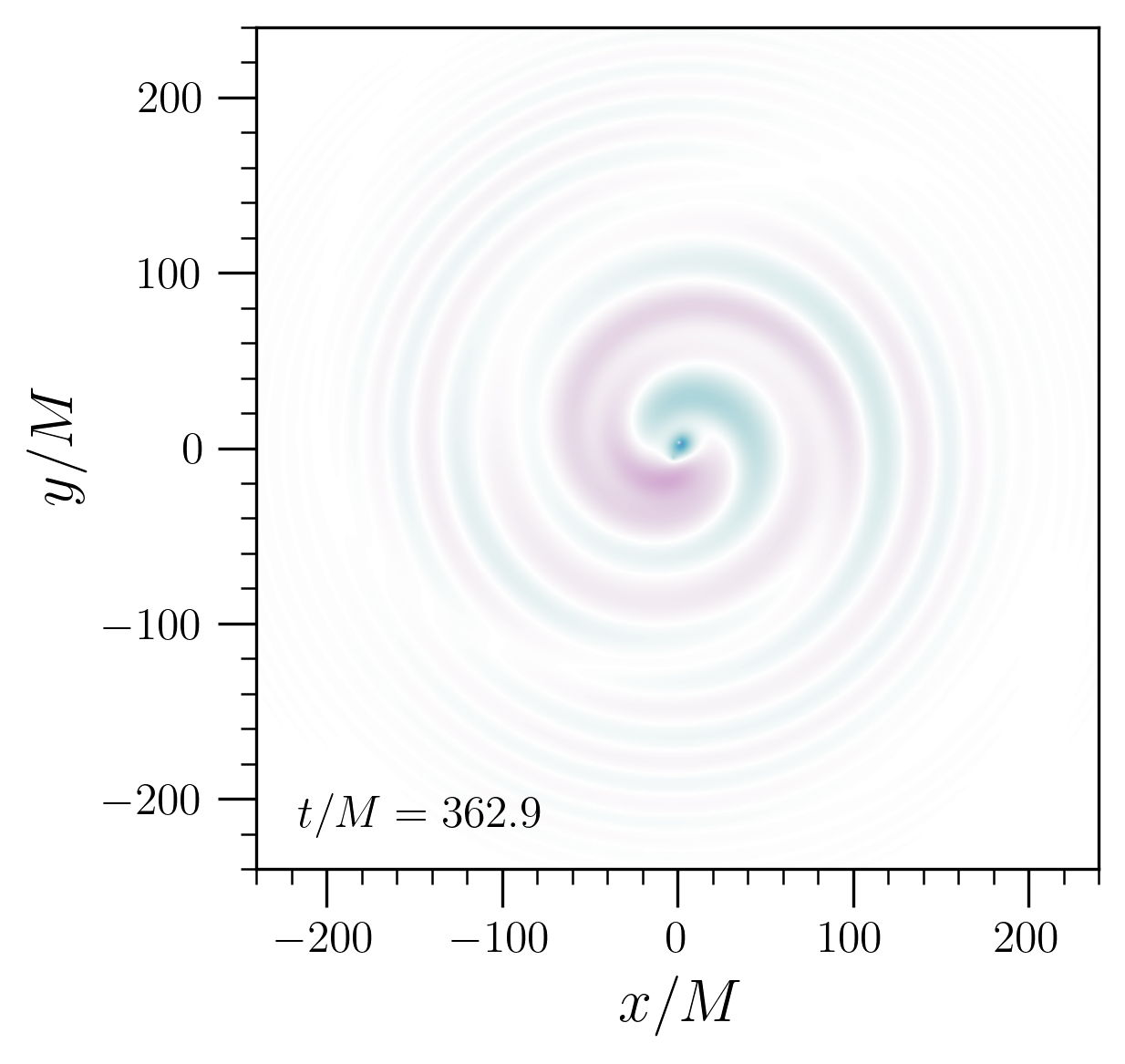}
    \includegraphics[width=.29\textwidth]{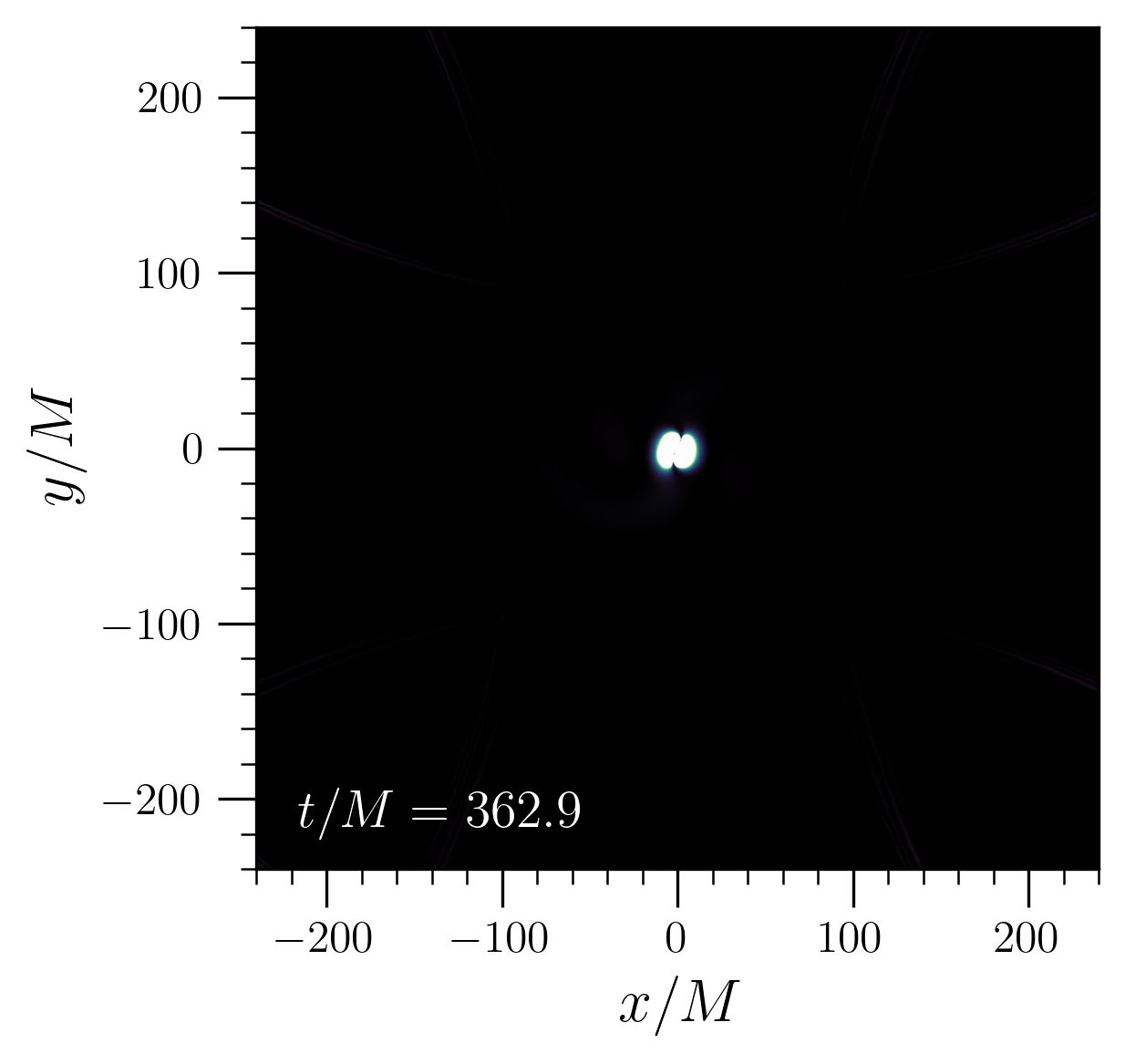}
    \includegraphics[width=.29\textwidth]{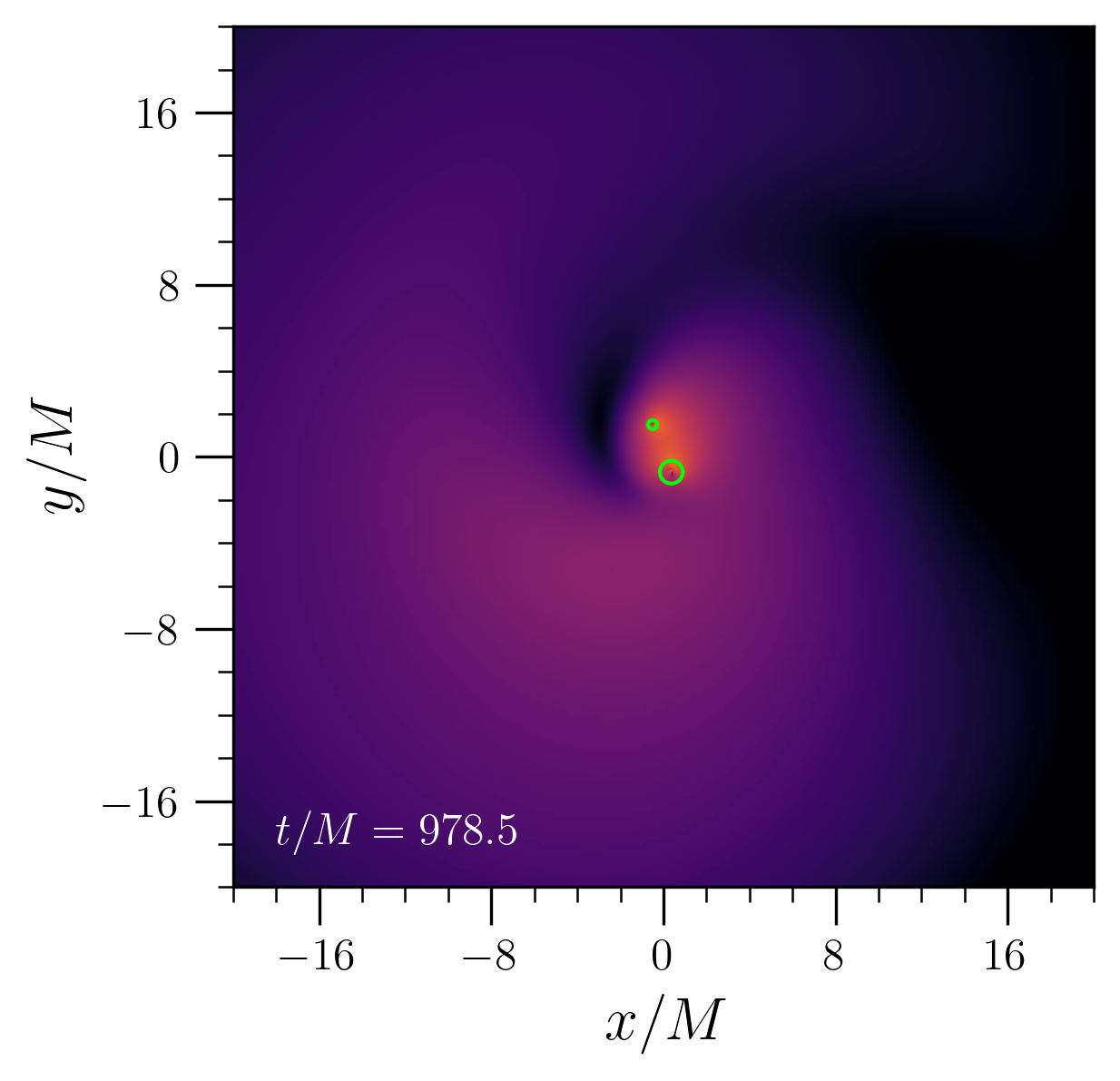}
    \includegraphics[width=.29\textwidth]{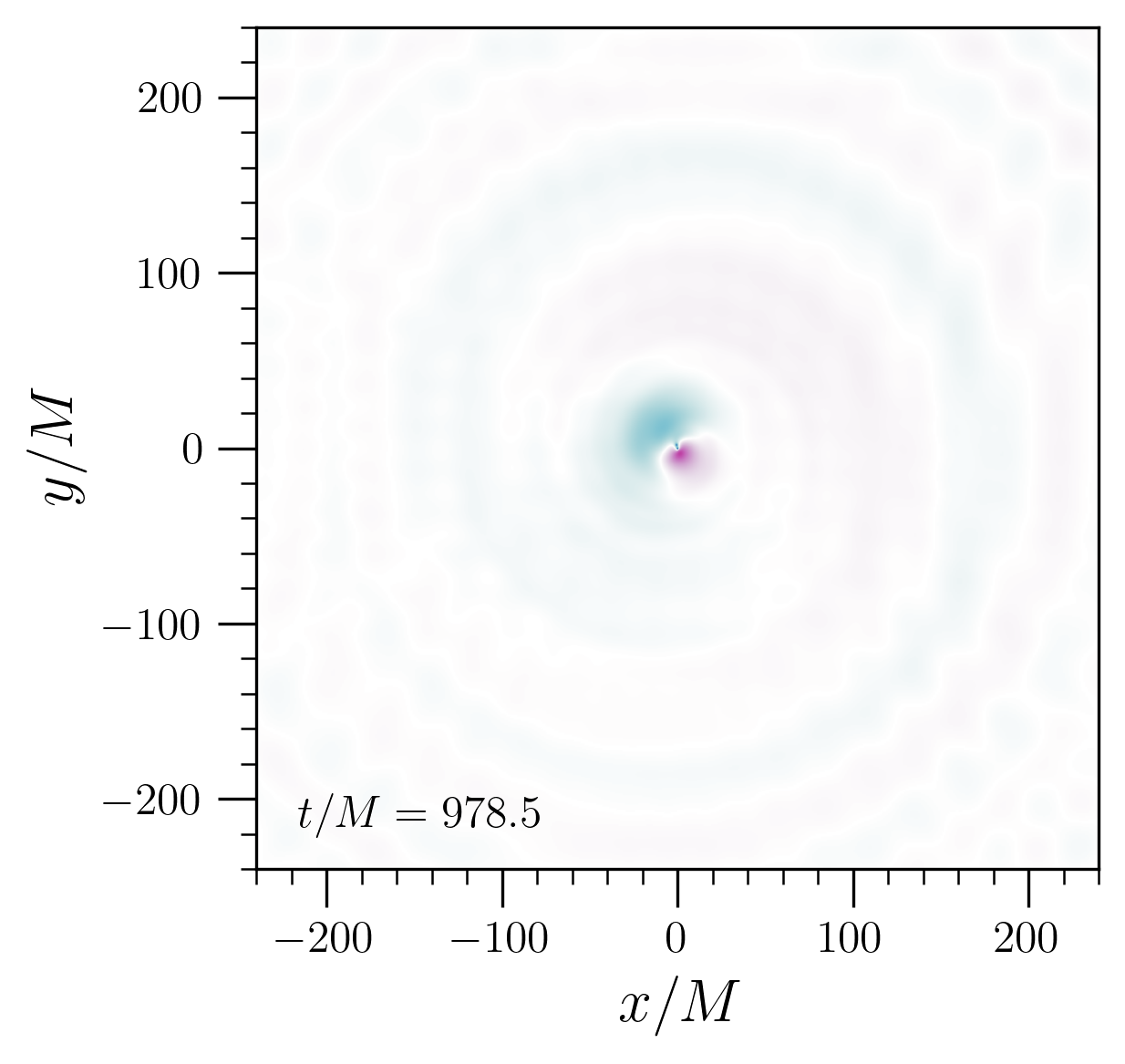}
    \includegraphics[width=.29\textwidth]{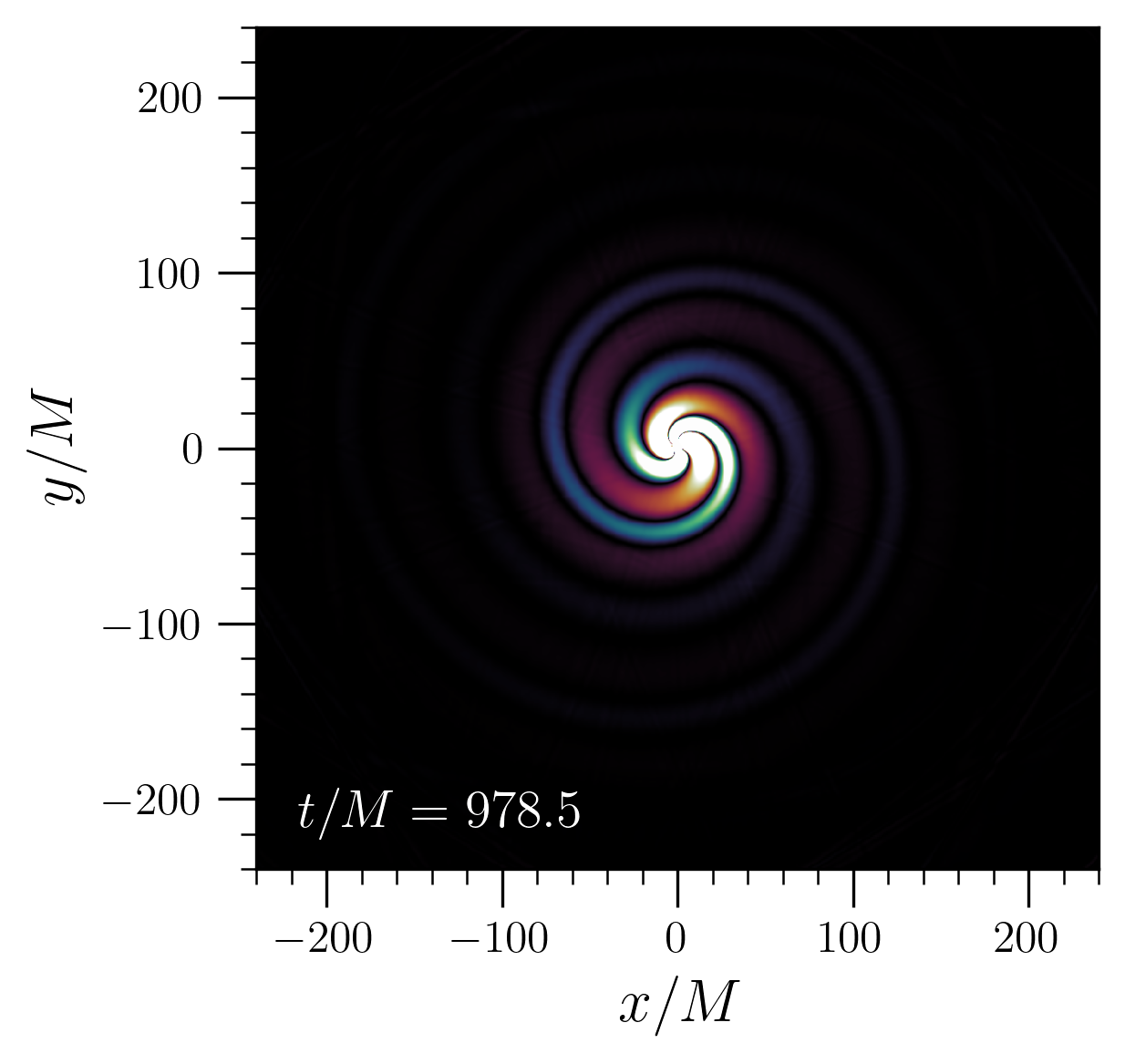}
    \includegraphics[width=.29\textwidth]{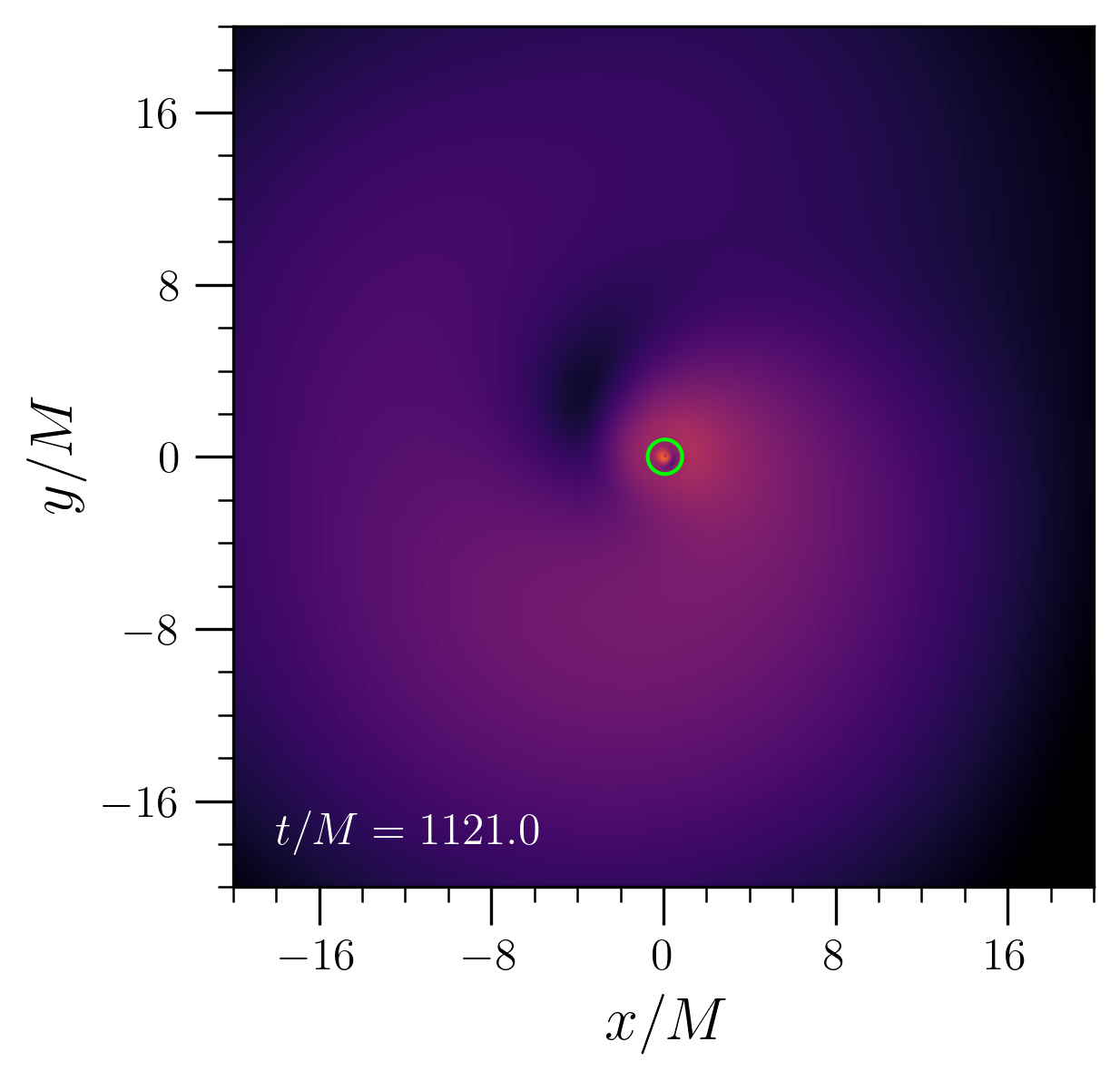}
    \includegraphics[width=.29\textwidth]{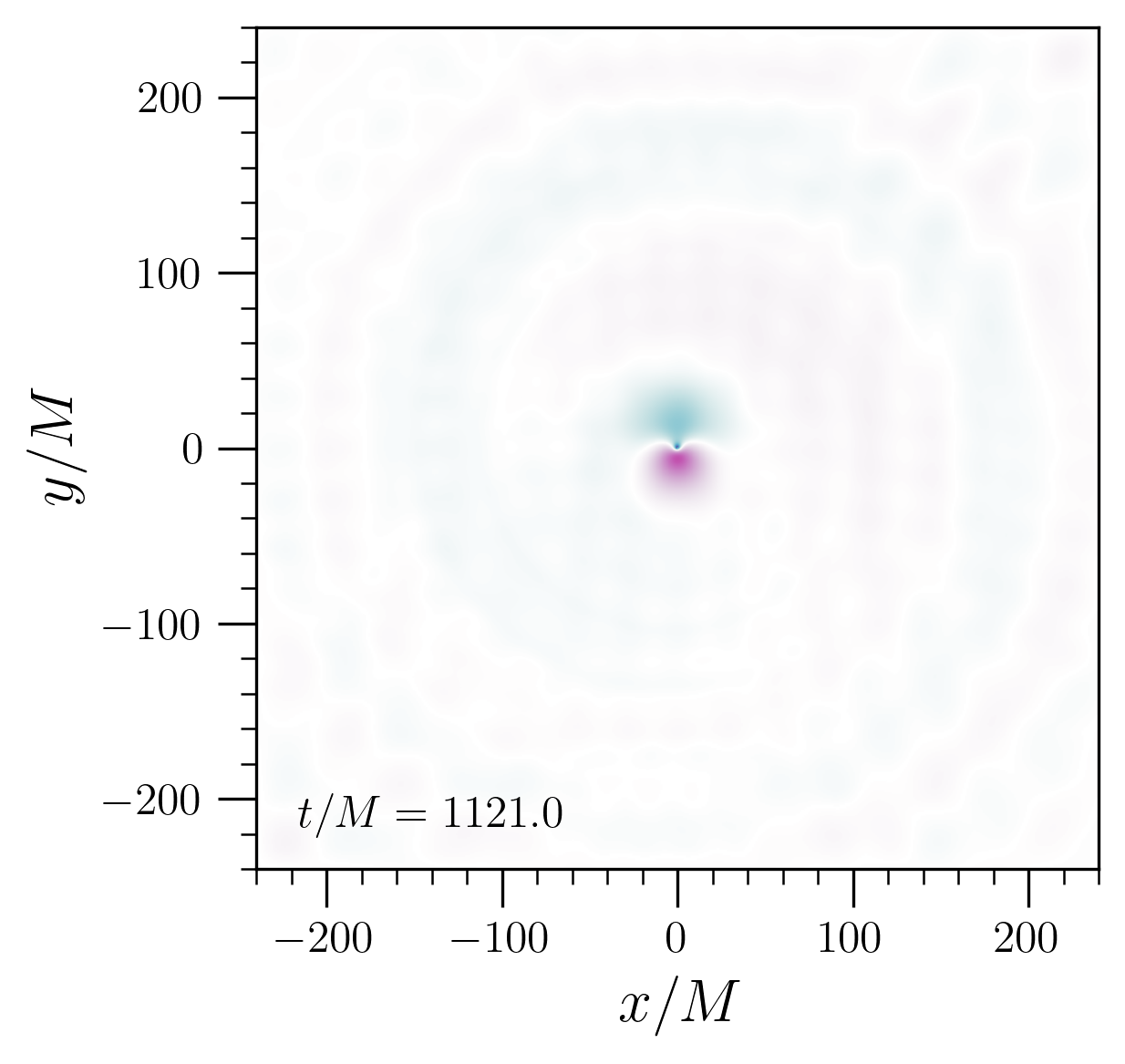}
    \includegraphics[width=.29\textwidth]{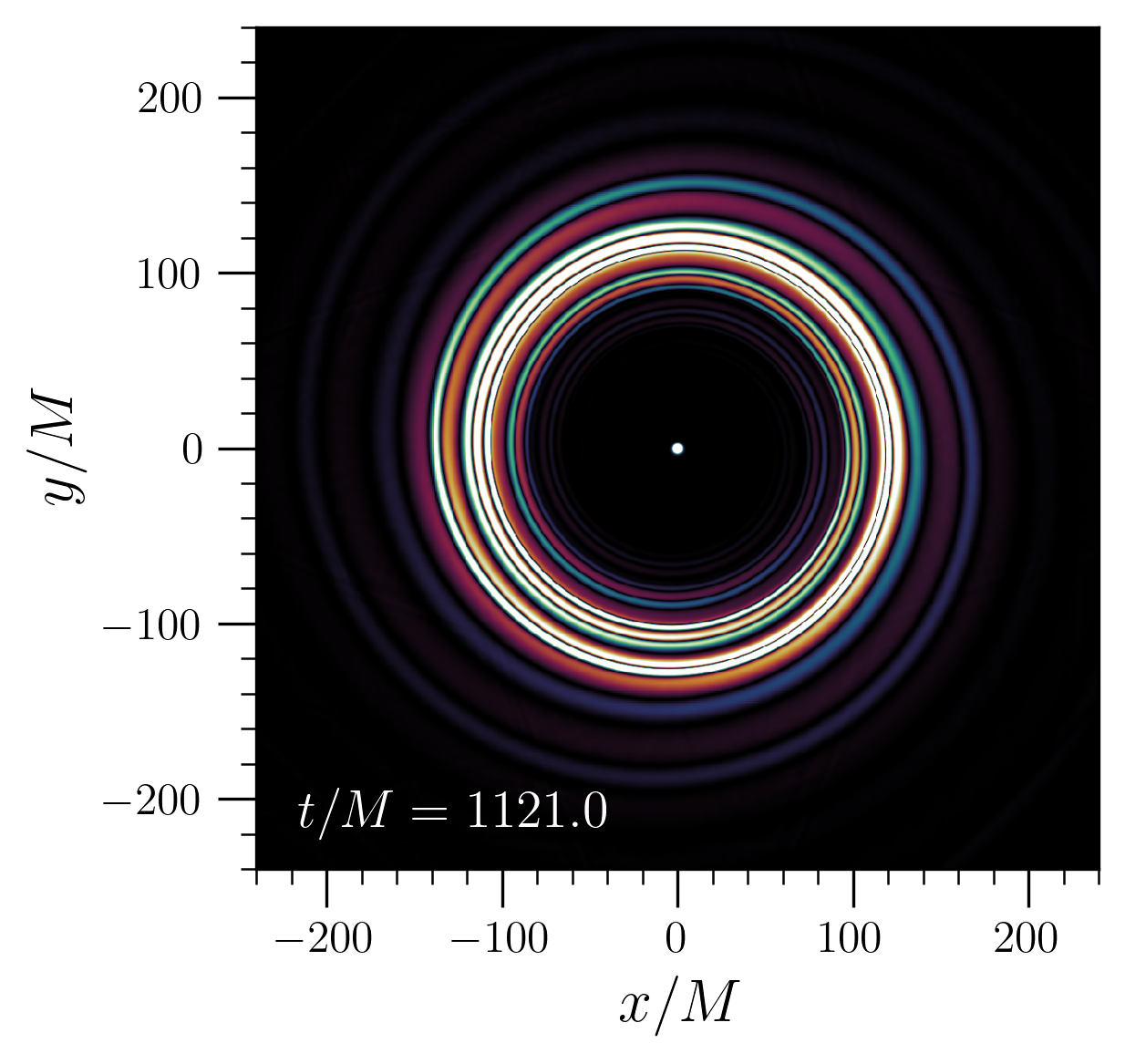}
    \caption{Two-dimensional snapshots of the simulation \textbf{q12mu04}.
    From left to right, the pseudocolor plots show
    the scalar's energy density $\rho$ normalized by the initial maximum value $\rho_{0,\max}=1.6\times 10^{-7} M^{-2}$ (left),
    the scalar field $\Phi$ (middle), and real part of the Newman-Penrose scalar $\Psi_4$ (right).
    The green circles in the left column
    indicate the locations and radii of the \bh{s'} apparent horizons.
    From top to bottom we display
    these quantities at
    different stages of the binary's coalescence:
    (1) shortly after initial data settles and a burst junk radiation is emitted,
    (2) $1.5$ orbits after the beginning of the simulation,
    (3) $0.5$ orbit before merger,
    and (4) $\sim130$M after the merger.
    Note the difference in the displayed spatial domain, chosen to highlight features of the quantities.
    }
    \label{fig:q2_snapshots}
\end{figure*}

We illustrate the evolution of a \bh{} binary immersed in a scalar cloud in Fig.~\ref{fig:q2_snapshots},
exemplarily for the simulation \textbf{q12mu04} in Table~\ref{tab:simulations_list}.
We present a series of two-dimensional
pseudocolor plots
displaying, from left to right,
the scalar energy density, scalar and gravitational radiation in the \bh{s'} orbital plane.
We overlay the scalar's density with circles indicating the position and mean radius of the \bh{s'} apparent horizons.
From top to bottom, the snapshots show
(1) the initial transient,
(2) the binary's evolution after about $1.5$ orbits,
(3) the binary about $0.5$ orbits before the \bh{s} merge,
and (4) about $130$M after the merger.
The complete animation is available on our \canuda{} youtube channel~\cite{YoutubeCanuda}.

As the \bh{} binary begins its quasi-circular inspiral, the initially spherically-symmetric scalar cloud is stirred up and accreted onto the individual \bh{s},
forming a pair of scalar
overdensities
trailing the \bh{s}.
The overdensities are consistent with transient scalar ``charges'' found for single \bh{s}
immersed in a time-dependent scalar field~\cite{Clough:2019jpm,Bamber:2020bpu,Okawa:2014nda}.
We find that in the binary they act like charges
and, through their motion along with the \bh{s}, they generate scalar radiation.
The snapshot of the Newman-Penrose scalar depicts a burst of initial ``junk'' radiation that is a known artifact of puncture initial data~\cite{Habib:2020dba, Habib:2024soh}.
As the \bh{s} progress through their inspiral,
the scalar ``charges'' produce scalar dipole radiation
shown in the top two rows of Fig.~\ref{fig:q2_snapshots}.
The gravitational radiation increases in frequency and amplitude, although too faint to be visible in the snapshot in the second row.
As the \bh{s} merge,
a burst of gravitational radiation is emitted,
after which the final rotating \bh{} rings down.
The scalar forms a slowly decaying cloud anchored around the final \bh{.}
We also note an interference pattern due to 
boundary effects.

Compared to a binary \bh{} coalescence in vacuum,
the presence of a massive scalar condensate may accelerate or slow down the coalescence.
The change in merger time manifests as a phase shift in the gravitational radiation.
The emission of scalar radiation and accretion of the scalar field on to the \bh{} can both accelerate the coalescence.
On the other hand, the \bh{s} may experience a delay in the merger under the effects of dynamical friction from the scalar cloud.

In the following, we focus 
on simulations with scalar mass parameters $M\muS \in \{0.2, 0.4, 0.6\}$ as a representative sample, for both \bh{} mass ratios.

\subsection{Evolution of black hole properties}\label{ssec:blackholeproperties}
The energy density profile shown in Fig.~\ref{fig:q2_snapshots},
a steep gradient peaking at the location of the \bh{s} is formed within the first two orbits.
This sharp profile persists up until just before the merger.
As the scalar overdensities accumulate around each \bh{},
a fraction of the field is accreted onto the holes
and increases their masses.
The increase depends on the gravitational coupling, $M_{(a)}\muS$,
between the scalar and individual \bh{s}
listed in Table~\ref{tab:simulations_properties}.

\begin{figure}[h]
    \centering
    \includegraphics[width=.49\textwidth]{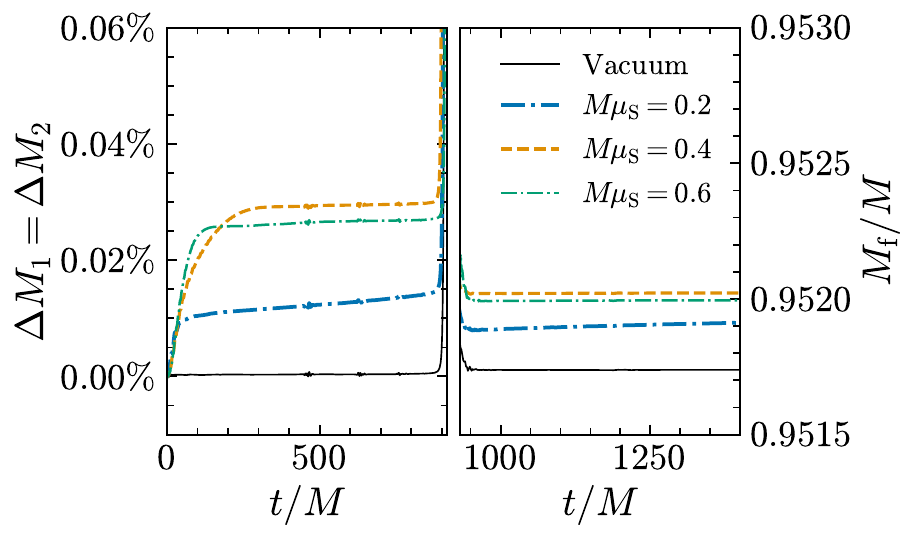}
    \caption{Evolution of the Christodoulou mass, Eq.~\eqref{eq:christodouloumass},
    for a \bh{} binary with $q=1$
    and scalar mass parameters $M\muS$.
    The reference simulation in vacuum is indicated by the solid black line.
    Left: Percent change of the
    individual \bh{s'} masses
    compared to their initial values.
    Right: Mass of the final \bh{.}
    }
    \label{fig:q1_ah_mass_bh}
\end{figure}

\begin{figure}[h]
    \centering
    \includegraphics[width=.49\textwidth]{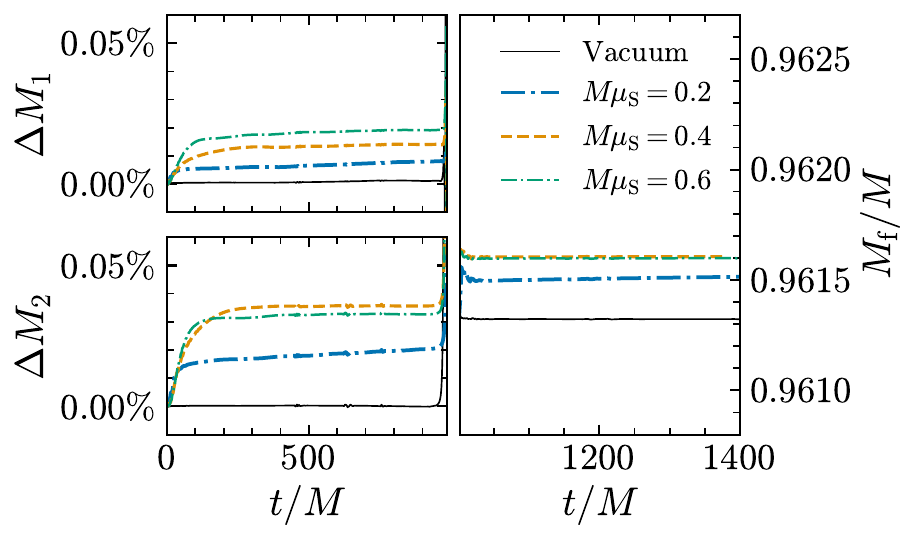}
    \caption{Same as Fig.~\ref{fig:q1_ah_mass_bh} but for
    mass ratio $q=1/2$.
    Left: Percent change of the
    mass of \bh{} 1 (top) and \bh{} 2 (bottom)
    relative to their initial masses.
    Right: Mass of the final \bh{}.
    }
    \label{fig:q2_ah_mass_bh}
\end{figure}

We see the accretion more clearly in Figs.~\ref{fig:q1_ah_mass_bh} and~\ref{fig:q2_ah_mass_bh},
where we show the evolution of the Christodoulou mass,
defined in Eq.~\eqref{eq:christodouloumass},
of the individual and of the final \bh{s}
in binaries with mass ratio $q=1$ and $q=1/2$, respectively.
The masses of the individual \bh{s} grow during their first orbit as the scalar cloud transitions from its initial, spherically symmetric configuration to the scalar ``charges'' around each \bh{.}
Following the early transition, the \bh{s'} masses
increases only very slowly
throughout the inspiral.
This is consistent with the snapshots in Fig.~\ref{fig:q2_snapshots}, where the energy density relaxes to a
nearly constant
profile in the \bh{s'} vicinity.

In the case of the equal-mass binary, the growth is around $0.01\%$ to $0.03\%$ of the initial \bh{} masses depending on the scalar field mass parameter, and this percentage increase is consistent with the mass of the final \bh{}.
Albeit small, the increase is above the numerical error of about $0.001\%$;
see  App.~\ref{app:convergenctest}.
The increase is largest for $M\muS=0.4, 0.6$ which correspond to a gravitational coupling of
$M_{(a)}\muS=0.2,0.3$
between the scalar and individual \bh{s},
as listed in Table~\ref{tab:simulations_properties}.
For the $q=1/2$ binary, the more massive \bh{} experiences a higher increase in its mass of up to $0.04\%$.
Similar to the equal-mass case,
the growth is only prominent during the first orbit, and
it tapers off for the remainder of the inspiral.
The mass of the final \bh{} in the presence of a massive scalar cloud is about $0.02\%$ larger than in the vacuum case.

The amount by which the \bh{} masses increase is linked to the gravitational coupling,
$M_{(a)}\muS$,
between the scalar and the individual \bh{s}.
We find that a larger gravitational coupling
$M_{(a)}\muS$
yields a larger increase in the \bh{s} masses due to partial accretion of the surrounding scalar field.
This is consistent with studies of single \bh{s}, where the
flux of scalar field
into the \bh{} increases with the
scalar's mass parameter~\cite{Bamber:2020bpu}.

\begin{figure}[htp!]
    \centering
    \includegraphics[width=.47\textwidth]{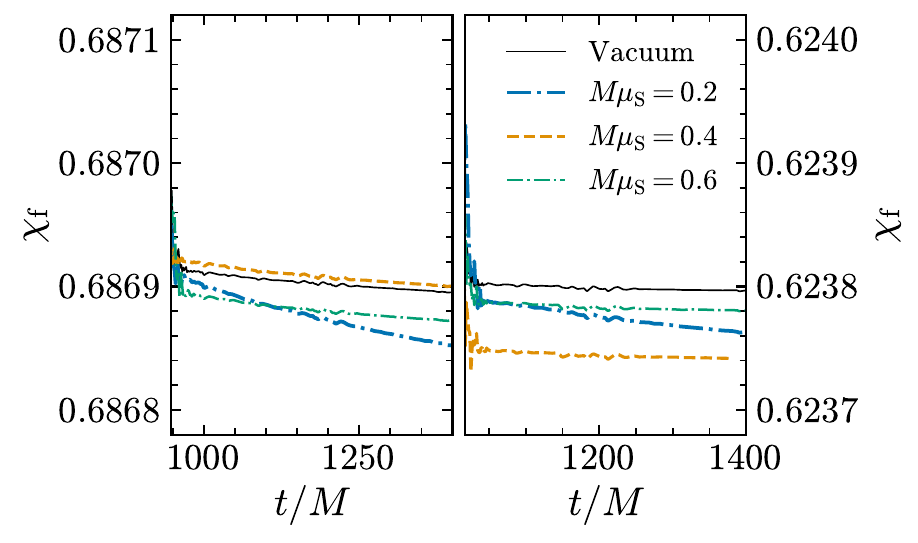}
    \caption{Dimensionless angular momentum $\chi_{\rm{f}} \equiv J/M_{\rm BH}^2$ of the final \bh{} in the equal-mass (left panel) and $q=1/2$ (right panel) simulations.
    }
    \label{fig:q2_ah_angular_momentum}
\end{figure}

The individual \bh{s} were initialized as non-spinning \bh{s},
and their spins remain zero throughout the inspiral.
This is consistent with the scalar field being initialized as spherically symmetric Gaussian with zero angular momentum.
Therefore, we
only present the dimensionless spin
of the final \bh{} in Fig.~\ref{fig:q2_ah_angular_momentum}.
We find that its dimensionless spin is
$\chi_{\rm{f}} = 0.687$ for $q=1$ and
$\chi_{\rm{f}} = 0.6238$ for $q=1/2$.
There is also little change in final spin among the scalar clouds with different mass parameter $M\muS$, which are well comparable with numerical error.

In conclusion,
the initially spherically symmetric scalar field forms overdensities around each of the \bh{s} in the early inspiral,
a small fraction of which is accreted onto the \bh{s} and increases their masses by about $\lesssim 0.05\%$.

\begin{table}[ht!]
    \centering
    \begin{tabular}{ |c|c|c|c|c|c|c|c|c|c|c|c|c|c| }
        \hline
        Name & $M_1\muS$ & $M_2\muS$ & $t_{\rm merger}/M$ & $\Delta\phi_{22}$ & $M\omega_{22}^{peak}$ & $E_{\rm GW} / M$ \\ \hline 
        \textbf{q1vac}    & NA   & NA   & 907.7 &  0.00 & 0.28  & 0.0386 \\ \hline 
        \textbf{q1mu00}   & 0.0  & 0.0  & 905.0 & -1.36 & 0.28  & 0.0386 \\ \hline 
        \textbf{q1mu02}   & 0.1  & 0.1  & 905.0 & -1.35 & 0.28  & 0.0386 \\ \hline 
        \textbf{q1mu04}   & 0.2  & 0.2  & 900.0 & -3.74 & 0.29  & 0.0386 \\ \hline 
        \textbf{q1mu06}   & 0.3  & 0.3  & 911.9 &  2.18 & 0.27  & 0.0386 \\ \hline 
        \textbf{q1mu08}   & 0.4  & 0.4  & 911.6 &  2.03 & 0.28  & 0.0386 \\ \hline 
        \textbf{q1mu10}   & 0.5  & 0.5  & 912.3 &  2.28 & 0.27  & 0.0386 \\ \hline 
        \textbf{q12vac}   & NA   & NA   & 977.3 &  0.00 & 0.30  & 0.0300 \\ \hline 
        \textbf{q12mu00}  & 0.0  & 0.0  & 987.2 &  4.39 & 0.29  & 0.0300 \\ \hline 
        \textbf{q12mu02}  & 0.07 & 0.13 & 988.4 &  4.83 & 0.29  & 0.0300 \\ \hline 
        \textbf{q12mu04}  & 0.13 & 0.27 & 983.8 &  2.95 & 0.29  & 0.0300 \\ \hline 
        \textbf{q12mu06}  & 0.20 & 0.40 & 983.0 &  2.73 & 0.30  & 0.0300 \\ \hline 
        \textbf{q12mu08}  & 0.27 & 0.53 & 983.8 &  2.80 & 0.29  & 0.0300 \\ \hline 
        \textbf{q12mu10}  & 0.33 & 0.67 & 983.4 &  2.75 & 0.29  & 0.0300 \\ \hline 
    \end{tabular}
    \caption{Summary of the results.
    We report the name of the simulation,
    gravitational coupling $M_{\rm (a)}\muS$ between the $a$-th individual \bh{} and the scalar field,
    merger time $t_{\rm merger}$ determined by the formation of the common apparent horizon,
    \gw phase shift $\Delta\phi_{22}$, peak frequency $\omega_{22}^{\rm peak}$,
    and energy radiated gravitationally $E_{\rm GW}$.
    }
    \label{tab:simulations_properties}
\end{table}

\subsection{Scalar radiation}\label{subsec:ScalarRadiation}

We observe that the scalar field builds up overdensities around the \bh{s} in the early inspiral.
Most of the overdensities
remain anchored around each \bh{}
and decay very slowly.
That is, for a transition period longer than the simulations, the field acts like a binary of scalar ``charges'' that follow the \bh{s'}
coalescence.
Through their motion, the ``charges'' generate
scalar radiation as shown in the middle column of Fig.~\ref{fig:q2_snapshots}.

\begin{figure}[htp!]
    \centering
    \includegraphics[width=.46\textwidth]{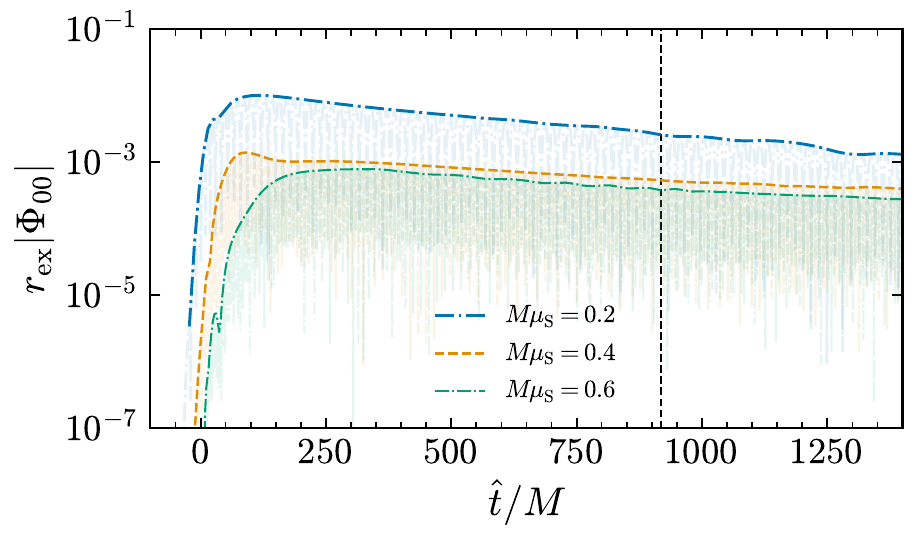}
    \includegraphics[width=.46\textwidth]{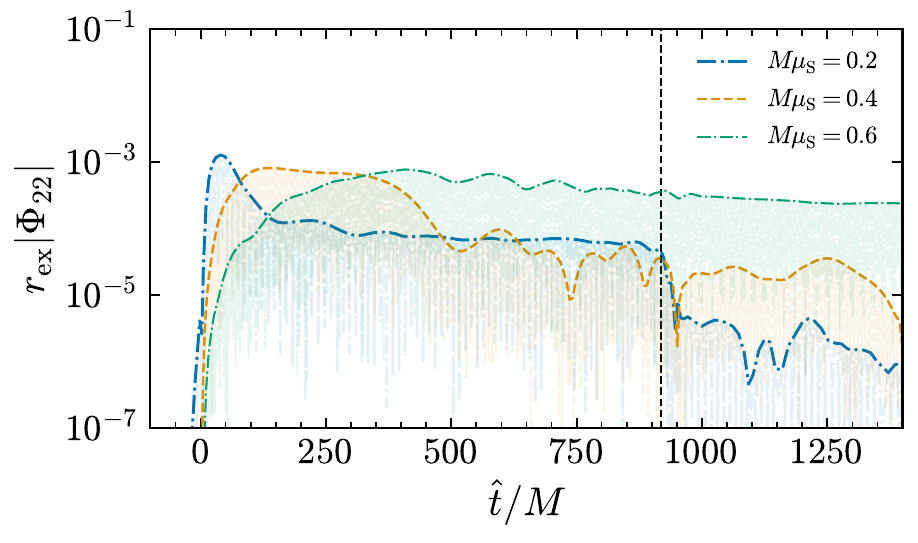}
    \caption{
    Scalar field $\ell=m=0$ (top panel) and $\ell=m=2$ (bottom panel) multipoles, extracted at $\rex=100$M, for equal-mass binaries and different scalar mass parameters.
    The thick lines are the envelopes of $|\Phi_{\ell m}|$.
    The vertical lines indicate the merger time of the reference (vacuum) simulation.}
    \label{fig:q1_mp_phi_time}
\end{figure}

\begin{figure}[htp!]
    \centering
    \includegraphics[width=.46\textwidth]{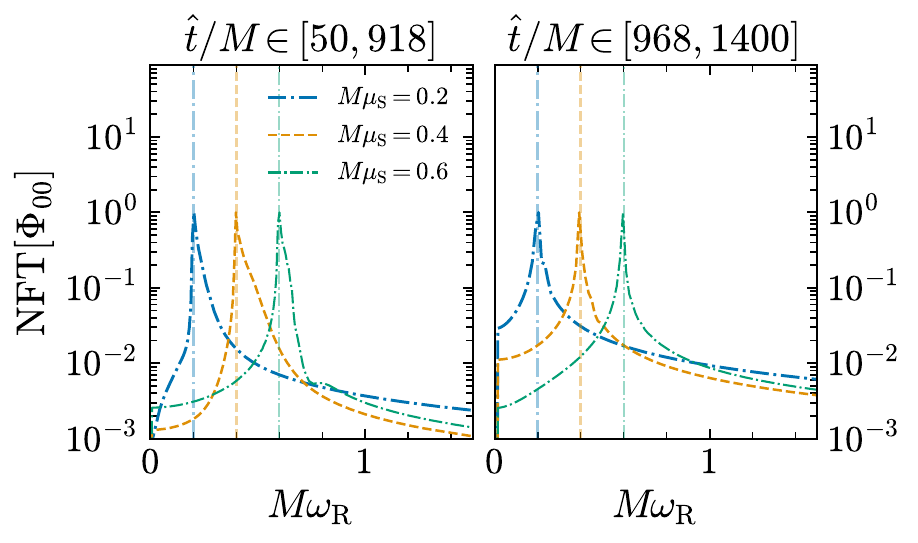}
    \includegraphics[width=.46\textwidth]{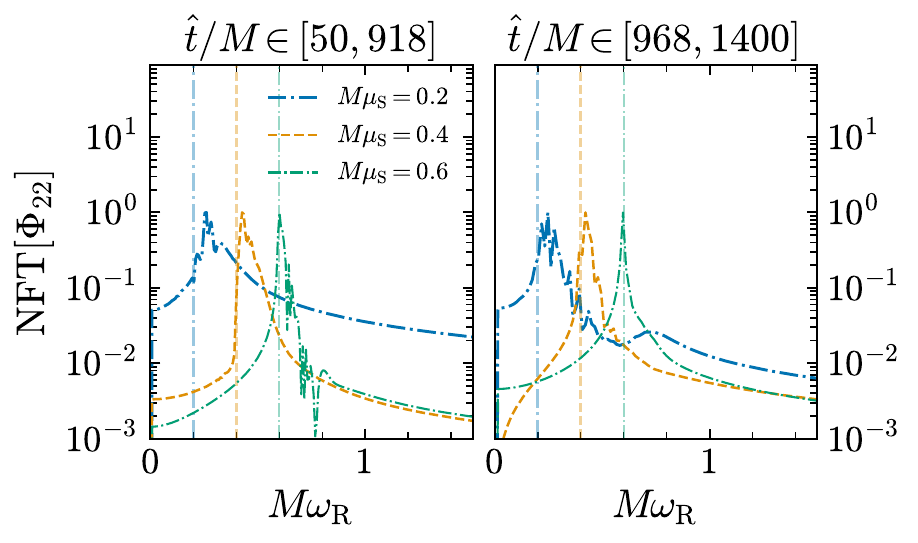}
    \caption{Normalized Fourier transform of the scalar field $\ell = m =0, 2$
    multipoles extracted at $\rex = 100M$
    as function of the oscillation frequency
    for equal-mass binaries.
    We show the spectra
    before (left panels) and after the merger (right panels).
    The faded vertical lines indicate
    $\omega_{\rm R} \sim \muS$.}
    \label{fig:q1_mp_phi_freq}
\end{figure}

\begin{figure}[htp!]
    \centering
    \includegraphics[width=.47\textwidth]{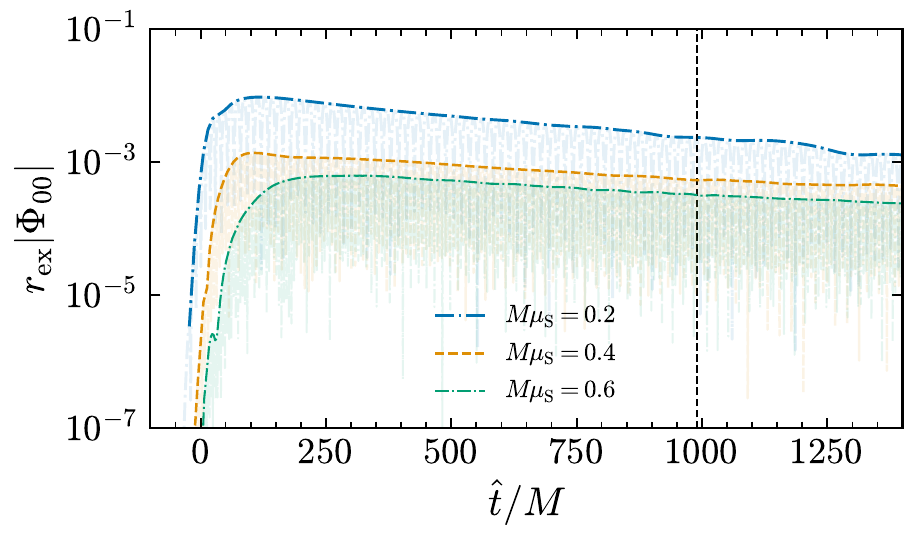}
    \includegraphics[width=.47\textwidth]{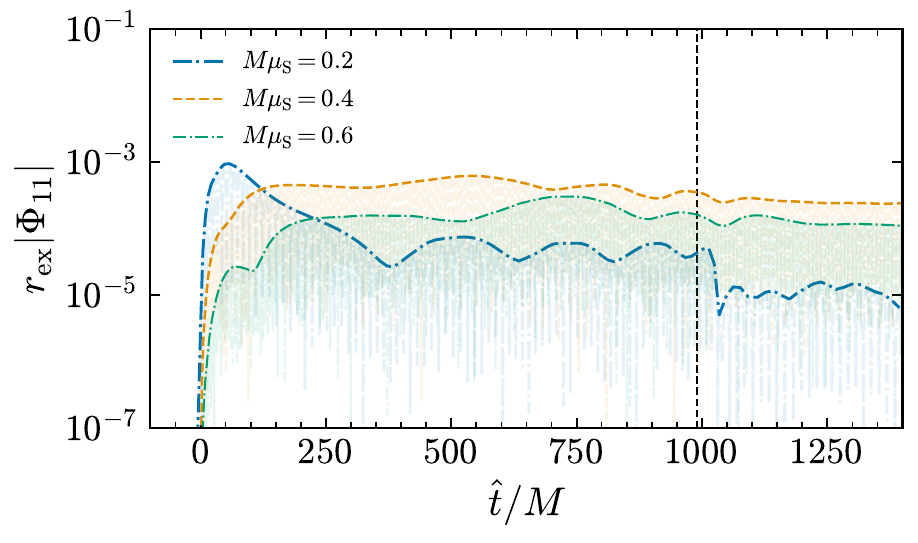}
    \includegraphics[width=.47\textwidth]{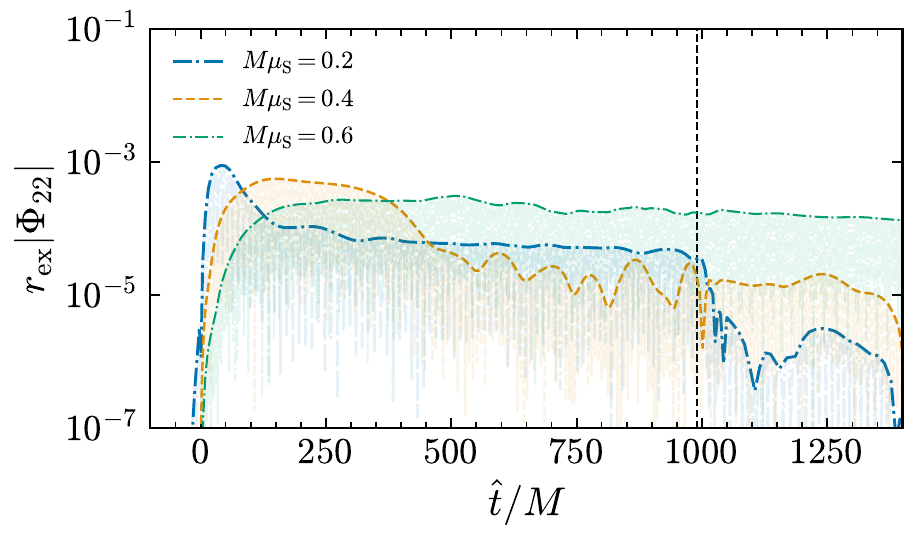}
    \caption{
    Same as Fig.~\ref{fig:q1_mp_phi_time} but for binaries with $q=1/2$.
    We show the scalar's $\ell=m=0$ (top panel), $\ell=m=1$ (middle panel)
    and $\ell=m=2$ (bottom panel)
    multipoles.
    }
    \label{fig:q2_mp_phi_time}
\end{figure}

\begin{figure}[htp!]
    \centering
    \includegraphics[width=.47\textwidth]{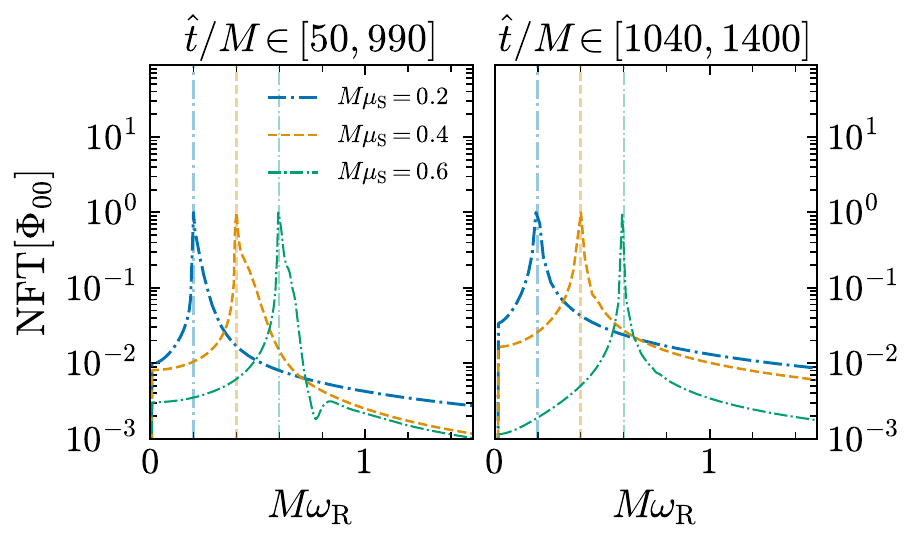}
    \includegraphics[width=.47\textwidth]{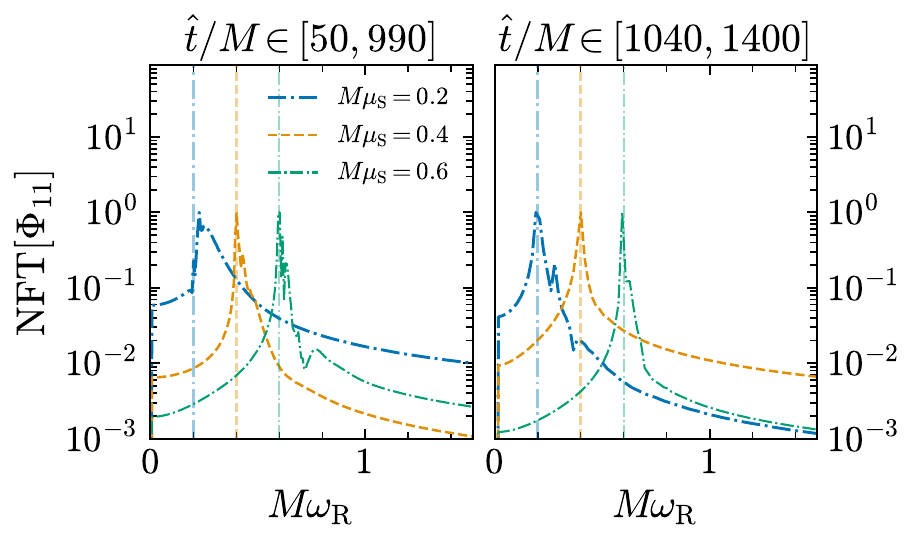}
    \includegraphics[width=.47\textwidth]{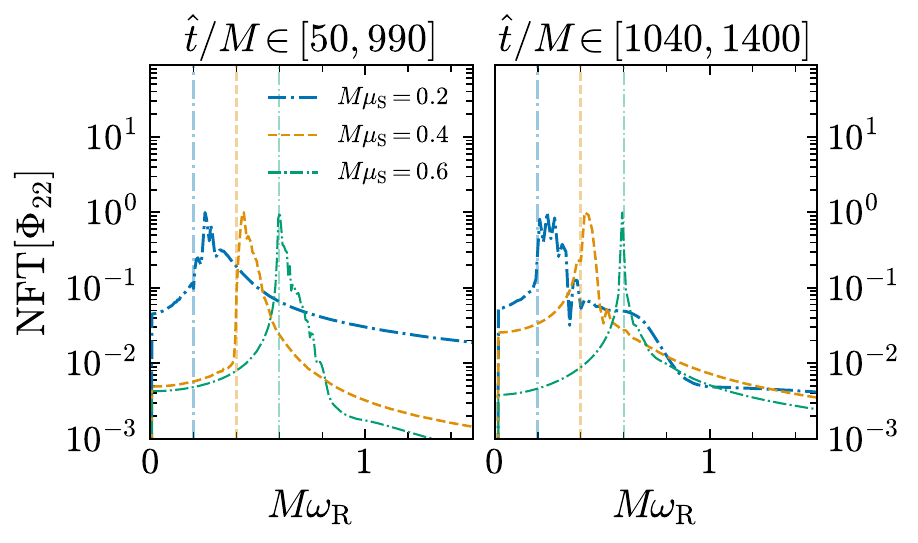}
    \caption{Same as Fig.~\ref{fig:q1_mp_phi_freq},
    but for binaries with $q=1/2$.
    We show the normalized Fourier transforms of the scalar's
    $\ell=m=0$ (top panel),
    $\ell=m=1$ (middle panel)
    and
    $\ell=m=2$ (bottom panel)
    multipoles.
    The faded vertical lines indicate
    $\omega_{\rm R} \sim \muS$.
    }
    \label{fig:q2_mp_phi_freq}
\end{figure}

We present the evolution of the scalar's multipoles up to $\ell=2$ in
Figs.~\ref{fig:q1_mp_phi_time} and~\ref{fig:q2_mp_phi_time},
and their frequency spectra
in Figs.~\ref{fig:q1_mp_phi_freq}
and~\ref{fig:q2_mp_phi_freq},
for the simulations with mass ratios $q=1$ and $q=1/2$, respectively.
We rescale the multipoles
by the extraction radius $r_{\rm ex}=100$M to account
for the $r^{-1}$ radial falloff,
and shift the time coordinate $\hat{t} \equiv t - r_{\rm ex}$
to account for the time to propagate to the extraction sphere.
We indicate the time of merger,
determined
by the formation of the common apparent horizon and listed in Table~\ref{tab:simulations_properties},
for the reference simulation in vacuum.

We first focus on the scalar's monopole that indicates the scalar ``charge'' enclosed in the sphere of radius $\rex$.
In both sets of binary \bh{} simulations and for all scalar field mass parameters, we see that the monopole
oscillates with a frequency $\omega_{\rm{R}}\sim\muS$
(c.f. top panels of Fig.~\ref{fig:q1_mp_phi_freq} and~\ref{fig:q2_mp_phi_freq})
and its amplitude
slowly decreases as a function of time
(c.f. top panels of Fig.~\ref{fig:q1_mp_phi_time} and~\ref{fig:q2_mp_phi_time}).
We also observe that the overall magnitude of the monopole decreases as the scalars' mass parameter increase.
This trend can be understood from the construction of the initial profile:
we fixed the scalars' maximum energy density
(instead of the amplitude),
so scalar fields with a larger mass parameter were initialized with a smaller amplitude and hence smaller initial scalar monopole.

We now turn to the scalar radiation.
The leading contribution in the equal-mass  \bh{} binary is the $\ell=m=2$ quadrupole shown in the bottom panel of Fig.~\ref{fig:q1_mp_phi_time}
while the dipole is absent due to symmetry.
In case of the mass parameter $M\muS=0.2$ we see that, after an initial transient, the amplitude is approximately constant throughout the inspiral and rapidly decays after the \bh{s} have merged.
For $M\muS=0.4$ we observe an almost constant amplitude for the first $t\sim500$M, corresponding to about $2.5$ orbits,
before the quadrupole's magnitude drops by an order of magnitude and slowly decays.
In case of $M\muS=0.6$, the amplitude of the quadrupole remains approximately constant throughout the simulations.

The bottom panel of Fig.~\ref{fig:q1_mp_phi_freq} presents the frequency spectra of the scalar's quadrupole before and after the \bh{s'} merger.
We find that for the mass parameters of $M\muS=0.2, 0.4$, the oscillation frequencies are shifted to values larger than the mass parameter, $\omega_{\rm{R}}>\muS$.
For $M\muS=0.6$, we see that the dominant oscillation frequency is determined by the mass parameter, $\omega_{\rm{R}}=\muS$.

The scalar radiation excited by the \bh{} binary with mass ratio $q=1/2$,
and the scalar's frequency spectra
are shown in Figs.~\ref{fig:q2_mp_phi_time} and~\ref{fig:q2_mp_phi_freq}.
The scalar's quadrupole,
shown in the bottom panels of these figures,
exhibits the same features and dependencies
on the mass parameters as for the equal-mass binary.
Because of the \bh{s'} unequal mass ratio,
the scalar ``charge'' around each \bh{} differs and,
additionally, generates scalar dipole radiation
shown in the middle panel of Fig.~\ref{fig:q2_mp_phi_time}.
For $M\muS=0.2$, the dipole's amplitude remains approximately constant throughout the inspiral, and its frequency is slightly larger than the mass parameter, $\omega_{\rm{R}}>\muS$.
After the \bh{s} have merged, the dipole's amplitude decays and its frequency is 
$\omega_{\rm{R}}\simeq\muS$;
see middle panel of Fig.~\ref{fig:q2_mp_phi_freq}.
For larger mass parameters, $M\muS=0.4,0.6$, the amplitude of the dipole remains approximately constant throughout the binary's coalescence.
Its frequency is determined by the mass parameters, $\omega_{\rm{R}}\simeq\muS$, both before and after the merger.

The relative amount of scalar radiation for different scalar field mass parameters is related to the accretion onto the \bh{s}.
In the case of $M\muS=0.2$, less of the scalar field is accreted onto the individual \bh{s}, and it dissipates away from the binary.
In turn, the amplitudes of the scalar's dipole and quadrupole drop
during the first few orbits,
and they remain approximately constant for the remainder of the inspiral.
On the other hand, for the heavier scalar with $M\muS=0.6$, there is stronger accretion and formation of scalar charges around the \bh{s}.
Consequently, the scalar dipole and quadrupole radiation remain nearly constant throughout the coalescence.

\subsection{Gravitational radiation}\label{subsec:GravitationalRadiation}
\begin{figure}[htp!]
    \centering
    \includegraphics[width=.47\textwidth]{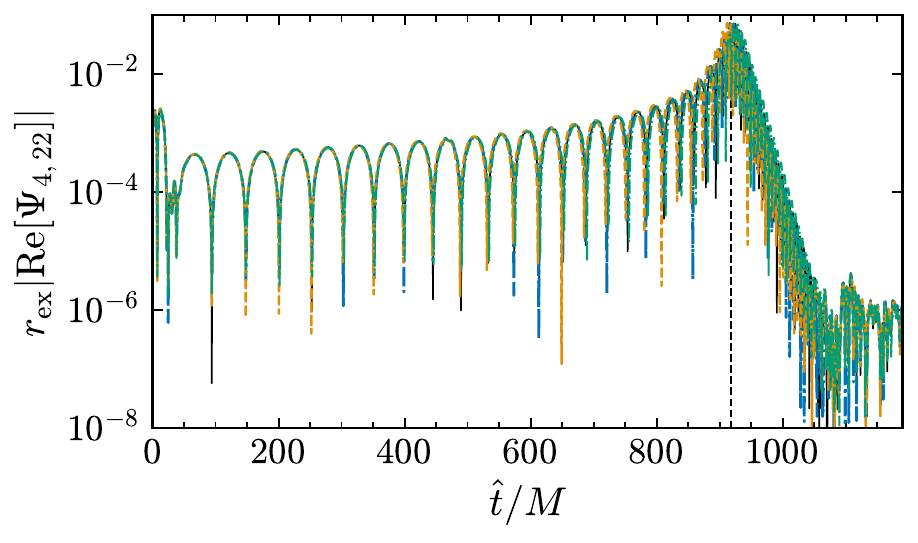}
    \includegraphics[width=.47\textwidth]{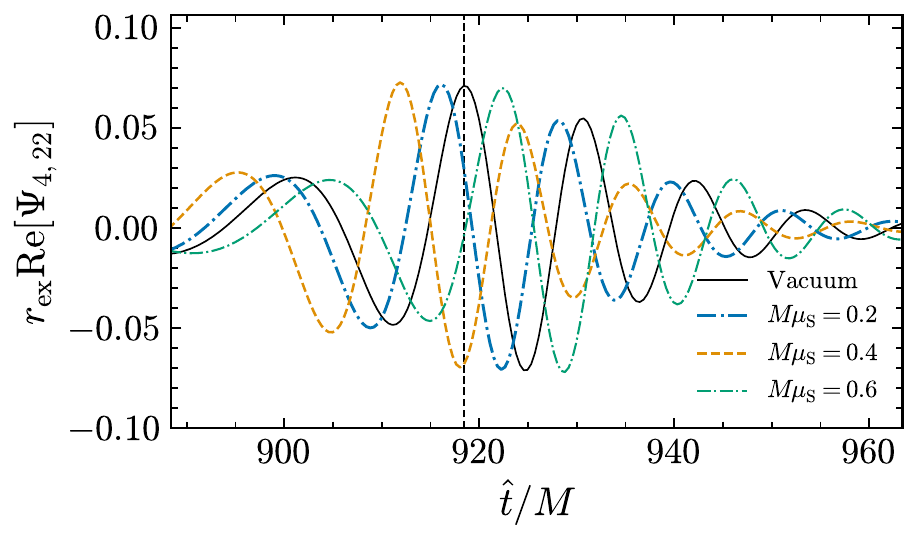}
    \includegraphics[width=.47\textwidth]{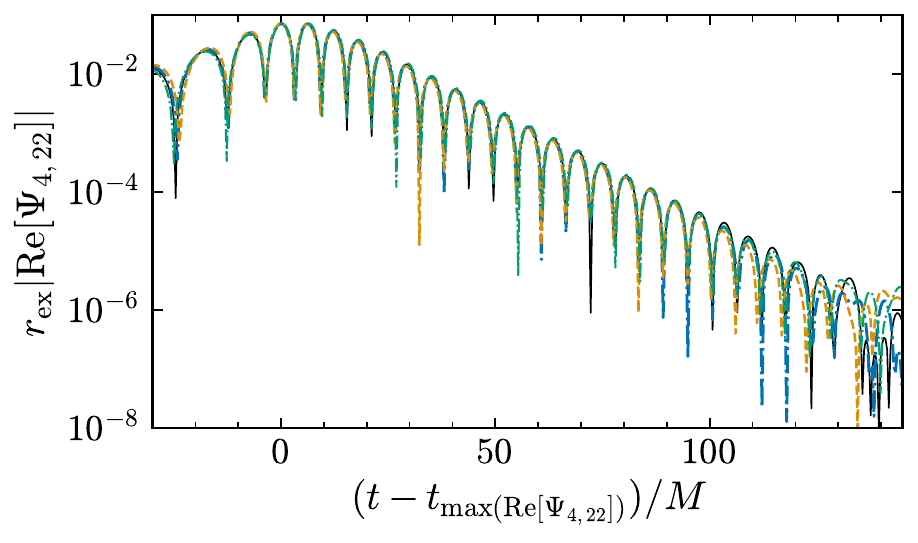}
    \caption{Evolution of the real part of the quadrupole of the Newman-Penrose scalar, indicating outgoing gravitational radiation, for the equal-mass \bh{} binary.
    We display the waveform for a binary in vacuum and binaries interacting with a scalar cloud of mass parameters $M\muS$.
    The waveforms in top and middle panels are shifted in time and rescaled by the extraction radius $\rex=100$M.
    The vertical dashed line indicates the peak in the waveform of the vacuum simulation.
    Top panel: complete waveform.
    Middle panel: zoom-in around the merger.
    Bottom panel: zoom-in of the quasinormal mode ringdown.
    Here, the time is shifted such that the peaks of the waveforms are aligned
    at $t - t_{\max({\rm Re}[\Psi_{4,22}])} = 0$.
    }
\label{fig:q1_mp_psi4_l2_m2}
\end{figure}

\begin{figure}[htp!]
    \centering
    \includegraphics[width=.47\textwidth]{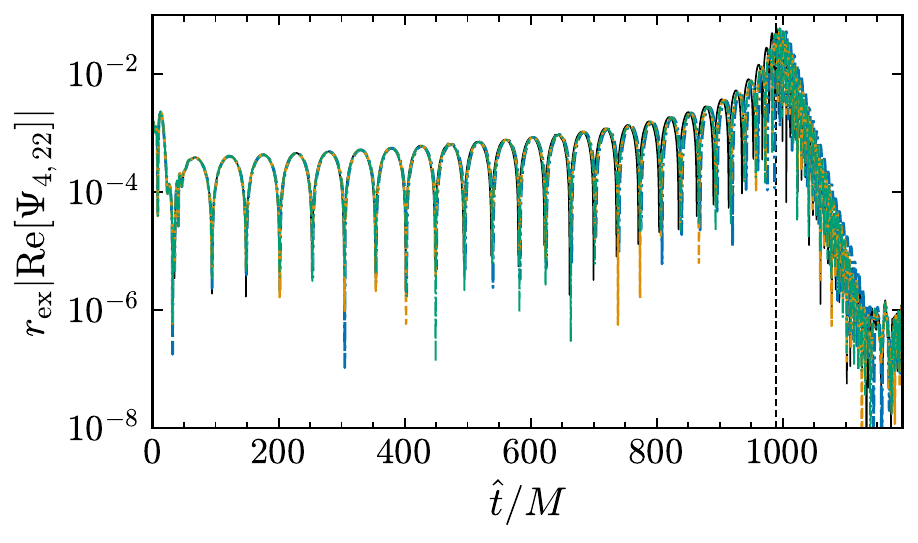}
    \includegraphics[width=.47\textwidth]{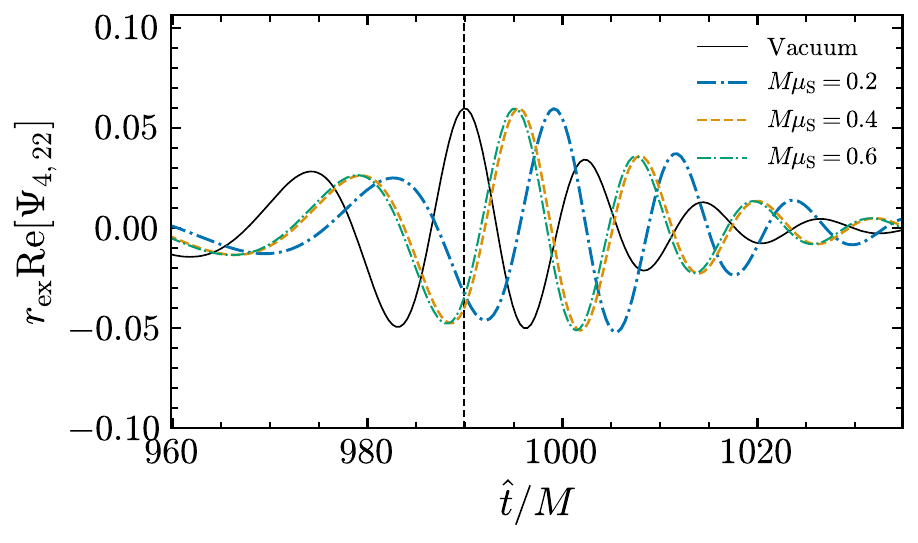}
    \includegraphics[width=.47\textwidth]{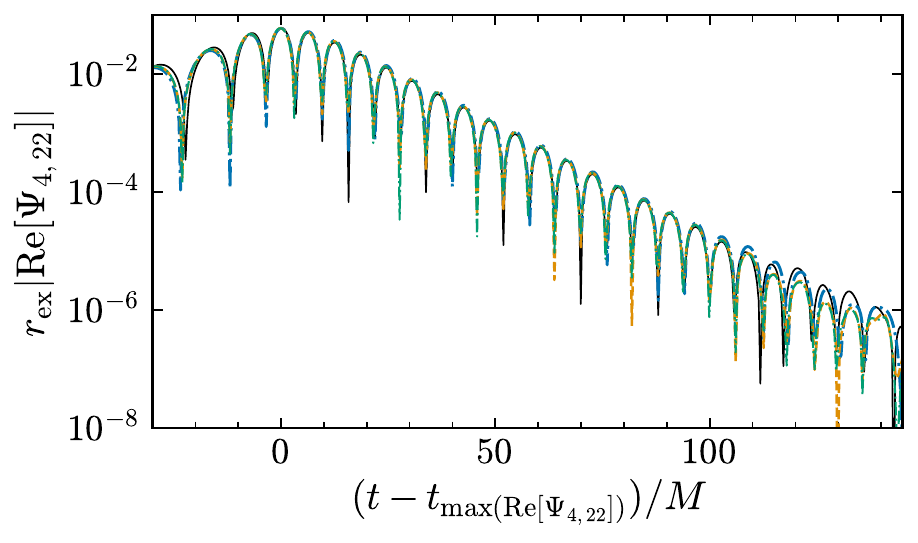}
    \caption{
    Same as Fig.~\ref{fig:q1_mp_psi4_l2_m2}, but for a \bh{} binary with mass ratio $q=1/2$.
}
\label{fig:q2_mp_psi4_l2_m2}
\end{figure}

We now turn our attention to the gravitational radiation.
In Figs.~\ref{fig:q1_mp_psi4_l2_m2} and~\ref{fig:q2_mp_psi4_l2_m2}, we present the quadrupole of the Newman-Penrose scalar $\Psi_{4}$
(determining the outgoing gravitational radiation)
for \bh{} binaries with mass ratio $q=1$ and $q=1/2$, respectively.
The gravitational waveforms are shifted in time by $\rex=100$M to account for their propagation to the extraction sphere,
and
rescaled by the extraction radius to account for their $\sim 1/r$ fall-off.
We present waveforms produced by a \bh{} binary in vacuum and
by binaries interacting with a scalar cloud of mass parameters $M\muS=\{0.2,0.4,0.6\}$.
In all cases the gravitational radiation shows the familiar pattern
of a sinusoid that is increasing in frequency and amplitude during the inspiral
until it peaks as the \bh{s} merge,
and is followed by the
nearly monochromatic, exponentially decaying
quasinormal mode ringdown.

As can be seen in the top panels of Figs.~\ref{fig:q1_mp_psi4_l2_m2} and ~\ref{fig:q2_mp_psi4_l2_m2},
the waveforms for different mass parameters agree in the early inspiral.
This behaviour is consistent with the initial setup of a quasi-circular inspiral
and comparable eccentricity of $\epsilon\lesssim 10^{-3}$
for all scalar mass parameters.

In the late inspiral we observe a dephasing shown more clearly
in the middle panels of Figs.~\ref{fig:q1_mp_psi4_l2_m2} and~\ref{fig:q2_mp_psi4_l2_m2}
where we display a zoom-in near the merger.
The peak in the vacuum waveform is indicated by a vertical dashed line.
In the case of equal-mass \bh{} binaries displayed in Fig.~\ref{fig:q1_mp_psi4_l2_m2},
we find that the peak in the waveform
is shifted to earlier times for small scalar masses $M\muS=0.2,0.4$
while it is shifted to later times for the larger mass parameter $M\muS=0.6$.
We have verified that all phase shifts are larger than numerical error; see App.~\ref{app:convergenctest}.
The former indicates an accelerated merger
(time-to-merger)
due to additional energy dissipated in scalar radiation.
The latter indicates a delayed merger
which we attribute to dynamical friction as the \bh{s} sweep through the massive scalar cloud.
For \bh{} binaries with mass ratio $q=1/2$, displayed in Fig.~\ref{fig:q2_mp_psi4_l2_m2},
we find that the peak in the waveform is shifted to later times for all scalar masses that we consider.
We interpret the delayed merger as a consequence of dynamical friction which competes with an acceleration due to the scalar radiation.

The same trends in the phase shift are found in the $\ell=m=3$ mode in the $q=1/2$ simulations and
in the $\ell=m=4$ modes for both $q=1$ and $q=1/2$.
In Fig.~\ref{fig:observables}
below,
we see that
the phase shifts in the $\ell=m>2$ modes are multiples of the shift in the $\ell=m=2$ mode.

Our finding of an intricate
dependency of the time-to-merger
on the binary's and scalar's parameters is
consistent with and
complementary to those of Ref.~\cite{Aurrekoetxea:2023jwk},
which report an accelerated merger for scalar mass parameters $M\muS \in \{0.0068, 0.86\}$.
The authors focused on an equal-mass \bh{} binary with a larger initial separation
(and no eccentricity reduction)
interacting with an initially homogeneous scalar field distribution of varying mass parameter.
We find consistent results for simulations of  equal-mass binaries with a scalar field of $M\muS=0.2, 0.4$; see Fig.~\ref{fig:q1_mp_psi4_l2_m2}.
We observe a delayed merger in particular for unequal-mass \bh{} binaries that were not studied previously; see Fig.~\ref{fig:q2_mp_psi4_l2_m2}.
Furthermore, we
reduced
the initial eccentricity to $\epsilon\lesssim 9\times10^{-3}$, whereas the simulations presented in
Ref.~\cite{Aurrekoetxea:2023jwk} may contain some residual eccentricity.
The latter typically accelerates the merger~\cite{Ramos-Buades:2022lgf,Ficarra:2024nro,Scheel:2025jct}.

Finally, we analyse the quasinormal mode ringdown shown in the bottom panels of Figs.~\ref{fig:q1_mp_psi4_l2_m2} and~\ref{fig:q2_mp_psi4_l2_m2} for $q=1$ and $q=1/2$, respectively.
To aid the visualization, we shift the waveforms such that their peaks are aligned.
For both series of simulations, we find that the gravitational wave ringdown frequencies appear indistinguishable until about $\sim100$M after the merger.
Only in the late stages of the ringdown ($\gtrsim100$M) do we observe small differences in the quasinormal mode frequencies due to
scalar cloud.
This is consistent with results of Ref.~\cite{Choudhary:2020pxy},
where the authors reported a small dependence of the gravitational wave ringdown frequency on the scalar's mass parameter $M\muS$
that became observable only
for scalar clouds whose density was three times larger than that of our simulations.

In the following subsection, we will discuss the dependence of physical
observables on the properties of the scalar clouds
in more detail.

\subsection{Trends of physical observables}
\label{subsec:discussion}
\begin{figure}[htp!]
    \centering
    \includegraphics[width=.23\textwidth]{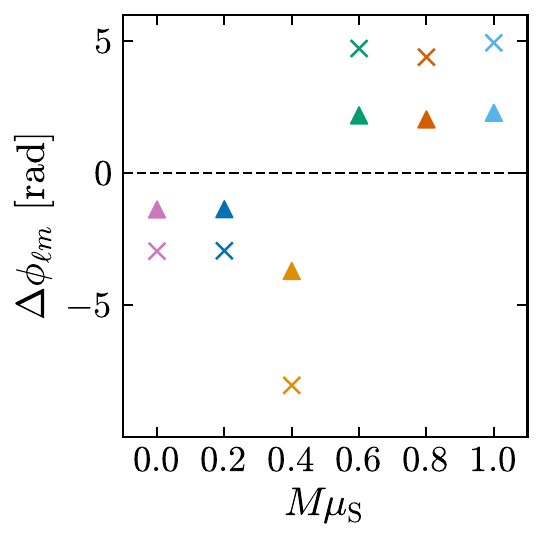}
    \includegraphics[width=.23\textwidth]{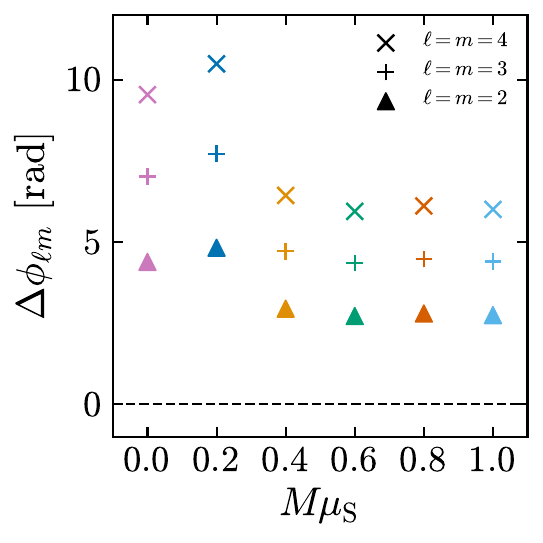}
    \includegraphics[width=.23\textwidth]{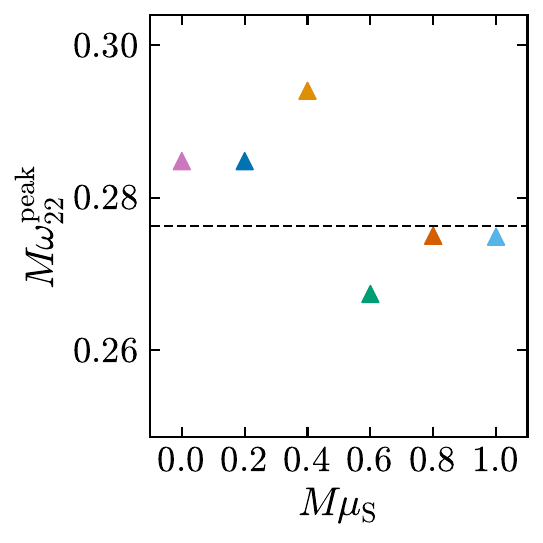}
    \includegraphics[width=.23\textwidth]{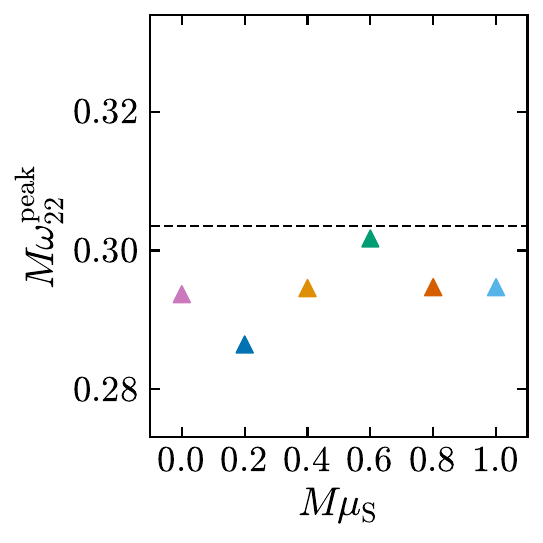}
    \includegraphics[width=.235\textwidth]{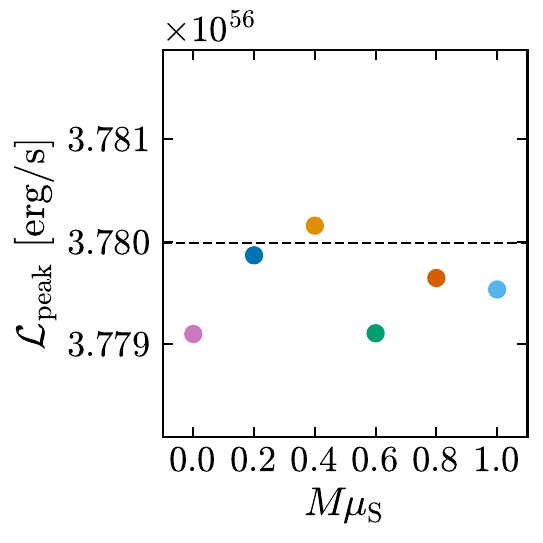}
    \includegraphics[width=.235\textwidth]{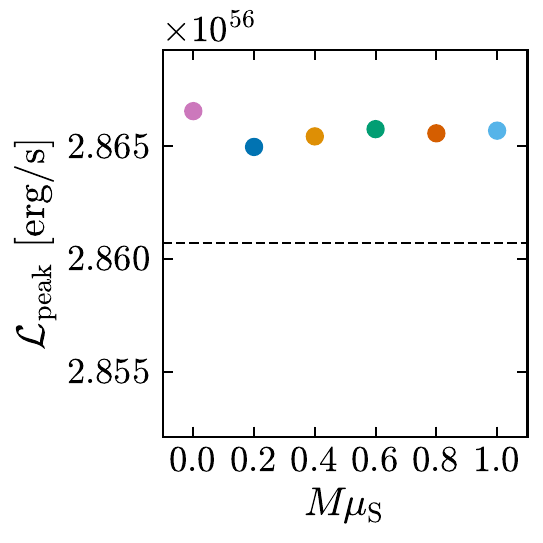}
    \includegraphics[width=.235\textwidth]{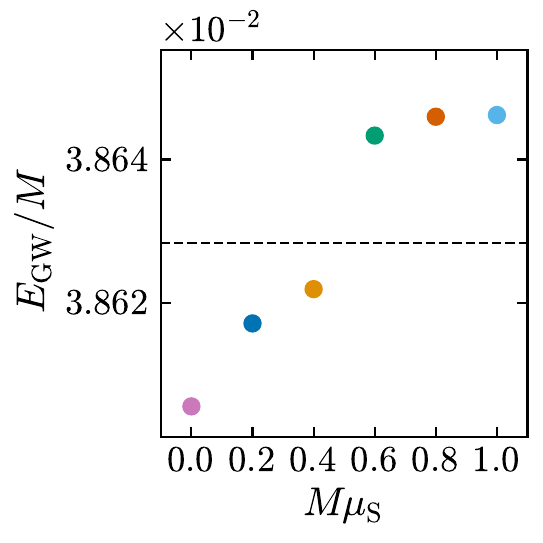}
    \includegraphics[width=.235\textwidth]{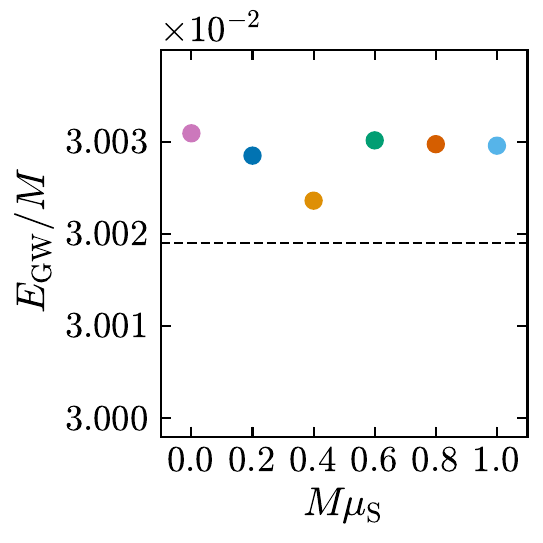}
    \caption{
    Physical observables derived from $\Psi_4$, extracted at $\rex=100M$,
    as a function of the scalar's mass parameter $M\muS$
    for binaries with mass ratios $q=1$ (left panels) and $q=1/2$ (right panels).
    From top to bottom: GW phase shift $\Delta \phi_{\ell m}$ as compared to the vacuum waveform for the $\ell=m=2,3,4$ multipoles;
    peak frequency $M\omega_{22}^{\rm peak}$ of the GW quadrupole;
    peak luminosity $\mathcal{L}_{\rm peak}$;
    and energy radiated in GWs, $E_{\rm GW}/M$.
    The dashed horizontal lines denote the value of the observable in vacuum.
    The color coding of the symbols for each mass parameter is the same as in previous figures.
    }
\label{fig:observables}
\end{figure}

Here, we present the dependence of physical observables
on all scalar field mass parameters $M\muS\in\{0.0, 0.2, \cdots, 1.0\}$
considered in our simulation suites.
The results are summarized in Fig.~\ref{fig:observables},
where we show
from top to bottom
the gravitational wave phase shift $\Delta \phi_{\ell m}$ in the $\ell=m=2,3,4$ modes compared to the vacuum evolution,
the peak frequency of the dominant gravitational waveform $\Psi_{4,22}$ extracted at $\rex=100M$,
the peak luminosity of the gravitational radiation,
and
the total radiated energy in gravitational waves, $E_{\rm{GW}}/M$.

The phase $\phi_{\ell m}$ of the $(\ell, m)$ mode of gravitational waveform is computed as a function of time by taking the complex argument of the Newman-Penrose scalar $\Psi_{4,\ell m}$.
We quantify the \gw phase shift
as the difference between the \gw phase in the presence of the scalar cloud and the phase in vacuum,
\begin{align}
\Delta\phi_{\ell m} \equiv \phi_{\ell m} - \phi_{\ell m}^{\rm Vacuum}
\,.
\end{align}
We align the waveforms at $\hat{t}=t-\rex=100M$, after the initial junk radiation has passed.
During the coalescence the magnitude of the phase difference
accumulates
and it saturates after the \bh{s}' merge.
The phase shift reported in the top panel of Fig.~\ref{fig:observables}
is computed after the saturation
(corresponding to about $30$M after the merger)
when it has reached a constant value.

In the top panels of Fig.~\ref{fig:observables}
we observe an intricate dependence of the \gw phase on the scalar mass parameter and the \bh{s'} mass ratio.
In particular, in simulations of equal-mass binaries (top left panel),
we find that the \bh{s}' merger is accelerated
with a peak (negative) phase shift for $M\muS=0.4$.
As the scalar's mass parameter is increased to $M\muS\geq0.6$, we observe that the difference of the \gw phase becomes positive, indicating a delayed merger, and this difference appears to saturate to a constant that is independent of the mass parameter.
In simulations of \bh{} binaries with mass ratio $q=1/2$ shown in the top right panel of Fig.~\ref{fig:observables},
we find a positive \gw{} phase shift, indicating a delayed merger, for all scalar masses that we considered.
The phase shift peaks around $M\muS=0.2$ and then saturates to a constant value that seems to be independent of the scalar's mass.
For the scalar cloud parameters considered in our simulations,
the phase shift of the quadrupole, accumulated over seven orbits, can exceed half a \gw cycle.
We observe that the higher multipoles, $\ell=m=3,4$, also exhibit a \gw phase shift.
Their qualitative behaviour follows that of the quadrupole,
but with a larger phase difference
as shown in Fig.~\ref{fig:observables}
that corresponds to
$1.2$ \gw cycles for $\ell=m=3$
and
$1.7$ \gw cycles for $\ell=m=4$.

Next, we compute the peak frequency of the \gw quadrupole, $M\omega_{22}^{\rm peak}$,
as the maximum
of the Fourier transformation of ${\rm Re}[\Psi_{4,22}]$
taken in the time interval
$\hat{t}\in(100M, \hat{t}_{\max({\rm Re}[\Psi_{4,22}])})$.
The latter corresponds to the maximum of the time domain waveform, and so this interval captures the frequency evolution during the binaries' inspiral and merger.
The result is shown in the second row of Fig.~\ref{fig:observables}.
In case of equal-mass binaries (left panel),
we find a higher peak frequency than in vacuum for $M\muS\lesssim0.4$, a lower frequency for $M\muS=0.6$, and a value comparable to the vacuum one for $M\muS\gtrsim0.8$.
This dependency on the scalar's mass parameter is consistent with that of the \gw phase shift.
In case of the binaries with mass ratio $q=1/2$ (right panel),
we find a lower peak frequency for $M\muS\lesssim0.2$, and a value comparable to the vacuum simulation for all $M\muS\gtrsim0.4$.
This trend is, again, consistent with the \gw{} phase shift.

The previous discussion focuses on the frequency during the \bh{s}' inspiral and merger.
In the post-merger, the \gw{} exhibits a quasinormal mode ringdown
shown in the bottom panels of Figs.~\ref{fig:q1_mp_psi4_l2_m2} and~\ref{fig:q2_mp_psi4_l2_m2}.
We observe that the \gw{} ringdown in the presence of the scalar field is essentially indistinguishable from that in vacuum.
A frequency analysis indicates
that the ringdown frequencies agree with those in vacuum within numerical error.
This is due to the small density of the scalar cloud.
To produce a difference of the ringdown frequency as reported in Ref.~\cite{Choudhary:2020pxy}, it requires a density that is at least three times as large as the one we employ in our simulations.

We compute the \gw{} luminosity,
$\Lie\sim\dif E_{\rm{GW}}/\dif t$,
by integrating the Newman-Penrose scalar in time
as defined in Eq.~\eqref{eq:RadiatedEnergyGW}.
In practice, we sum up to $\ell=8$ multipoles, and perform the integration at $\rex=100M$.
We display the maximum of the luminosity, $\Lie_{\rm{peak}}$,
in units of erg/s,
in the third row of Fig.~\ref{fig:observables}.
For both the equal-mass (left panel) and unequal-mass (right panel) binary we find that the peak luminosity is comparable to that in vacuum.
We also observe that the luminosity appears to be independent of the scalar's mass parameter.

Finally, we obtain the total energy radiated in gravitational waves,
$E_{\rm{GW}}$,
by integrating the luminosity in time.
The result, in geometric units,
is shown in the bottom panel of Fig.~\ref{fig:observables}.
As for the luminosity,
we find that the radiated energy is largely independent of the scalar's mass parameter
and that it differs
from the vacuum case
only at the sub-percent level.

In summary, we find that
the \gw{} phase and peak frequency depend on the scalar field's mass as well as the \bh{s'} parameters
(such as the mass ratio).
In particular, we find an accelerated merger of equal-mass binaries that is most prominent for $M\muS=0.4$,
and a delayed merger for the unequal-mass binaries that is most prominent for $M\muS=0.2$.
For larger scalar field masses, the phase and frequency shifts seem to become insensitive to them.

There are several mechanisms that can introduce the \gw{} dephasing
in the presence of a  scalar cloud.
These include the growth of the \bh{}
through accretion of the scalar field,
additional channels for energy loss through scalar radiation, and the formation of scalar wakes exerting a dynamical friction force on the \bh{s}.

Let us consider the first mechanism:
in Figs.~\ref{fig:q1_ah_mass_bh} and~\ref{fig:q2_ah_mass_bh} we see that the \bh{s} accrete only a small amount of the scalar field 
during the inspiral,
changing their masses by $\lesssim0.04\%$.
There is also no clear correspondence between the small increase in \bh{} mass and the \gw{} dephasing
for the scalar configurations studied in this work.

Let us now consider the scalar radiation,
discussed in Sec.~\ref{subsec:ScalarRadiation}.
We find that energy is
dissipated in the form of scalar waves with almost constant amplitude throughout the inspiral; see Figs.~\ref{fig:q1_mp_phi_time} and~\ref{fig:q2_mp_phi_time}.
The scalars' amplitude is comparable to that of the \gw{s}
in the early inspiral, and becomes subdominant in the late inspiral and merger.
It may, thus, yield an acceleration of the inspiral.

Another contributor to accelerating the merger is the radial attraction between the \bh{s} that results from the growth of a scalar overdensity in the binary's center of mass from an initially homogeneous distribution~\cite{Aurrekoetxea:2023jwk}.
Such central overdensities appear small in our simulations which
instead use an initial Gaussian profile; see the left panels in Fig.~\ref{fig:q2_snapshots} and
animations in Ref.~\cite{YoutubeCanuda}.

Finally, let us turn to dynamical friction and drag forces
that may cause an additional dephasing and,
thus, compete with energy loss in scalar radiation.
The drag forces depend on the \bh{s'} speed as well as the scalar's mass parameter as was shown for single \bh{s} in Refs.~\cite{Vicente:2022ivh,Traykova:2023qyv},
and, in the test field approximation, for \bh{} binaries moving through a scalar wind in Ref.~\cite{Xin:2025ymm}.
They also show that the drag force
increases for large relative velocities between the \bh{s} and scalar field.
For the simulations presented here,
using the coordinate data from \textsc{PunctureTracker}, we estimate that the \bh{s'} average coordinate speed
during the inspiral is $v\sim 0.15$ for the equal-mass binary
and $v\sim 0.1, 0.2$ for unequal-mass \bh{s} with $q=1/2$.
Invoking the results of Refs.~\cite{Vicente:2022ivh,Traykova:2023qyv,Xin:2025ymm}, we expect the drag forces to be small.
It remains to be explored how such a formalism 
can be applied to fully nonlinear evolutions of the late inspiral,
and if it may account for a delay of the merger.
In summary, there is a clear understanding of an accelerated merger through scalar radiation, while the origin of the delayed mergers is not entirely clear.
Longer simulations, left for future work, may reveal whether the acceleration becomes always dominant or if the delay becomes more prominent.

\section{Conclusions}
\label{sec:conclusions}
In this work, we have investigated the impact of a massive scalar cloud,
that may represent an overdense dark matter environment
or a scalar condensate formed through superradiance,
on coalescing \bh{} binaries and their
\gw{} signal.
We have performed fully general relativistic simulations
that consistently include the
interaction between the scalar field and the spacetime
in the initial data, evolution and wave extraction.
Complementary to existing initial data construction methods for the Einstein-Klein-Gordon system~\cite{East:2012zn,Ruchlin:2014zva,Aurrekoetxea:2022mpw},
we solve the constraints using the puncture approach.
Therefore, we have extended the {\textsc{TwoPunctures}} spectral solver~\cite{Ansorg:2004ds}
to include scalar condensates,
and the resulting code \TPBBHSF{}~\cite{TPBBHSFGitWeb}
is open-source as a thorn of the \ETK{}~\cite{roland_haas_2024_14193969,EinsteinToolkitWeb}.

One of the goals of this work has been the implementation of software capable to produce high quality waveforms
that include (dark matter) environmental effects on \bh{} mergers
and that are en par
with numerical relativity waveform used in
\gw{} template building and data analysis.
Therefore, we have extended our open-source software \canuda{}
by implementing up to eighth order finite difference stencils for spatial derivatives in the scalar and metric evolution thorns, and in the wave extraction.

We have also
carefully
reduced the initial orbital eccentricities
below
$\epsilon_\Omega \lesssim 9\times10^{-3}$ through an iterative procedure~\cite{Ramos-Buades:2018azo,Habib:2020dba}.
Improving this procedure by extending post-Newtonian computations of binaries interacting with massive scalar fields~\cite{Huang:2018pbu}
and extending numerical fitting procedures~\cite{Habib:2024soh}
would be valuable for extended simulation campaigns of quasi-circular \bh{} binaries in such environments.
Here, we have presented our upgraded software and
a series of simulations geared towards high-quality waveform templates for next-generation gravitational wave detectors.

In this paper we presented two simulation series of 
binary \bh{s} with mass ratios $q=1,1/2$.
In each series, we systematically varied the scalar field's mass parameter in the range $M\muS\in\{0.0,1.0\}$,
while keeping the maximum of its initial
(conformal) energy density approximately fixed at
$\bar{\rho}_{\rm max} \approx 1.6\times10^{-7}M^{-2}$.
While this value does represent a large dark matter overdensity, it is still physically viable;
densities up to $\rho\sim10^{-5} M^{-2}$
can be reached by superradiant clouds~\cite{Hui:2022sri} or dark matter surrounding an adiabatically growing \bh{}~\cite{Speeney:2024mas}.
This initial density, $\bar{\rho}_{\rm max} \approx 1.6\times10^{-7}M^{-2}$
in the geometric units of our simulations,
corresponds to a density of
$\rho \sim 10^{21} M_\odot/{\rm pc}^3$
for a supermassive \bh{} of $10^{6}M_{\odot}$; c.f. Eq.~\eqref{eq:RhoConversionNR2Phys}.

Our simulations have shown that
\bh{} binaries surrounded by a scalar cloud, experience a dephasing.
Specifically, we find both an accelerated and a delayed merger
depending on the scalar's initial configuration and properties, and on the \bh{s'} parameters.
In the cases of $M\muS=0.4$ for an equal-mass binary and $M\muS=0.2$ for a binary with mass ratio $q=1/2$,
our simulations have shown that the \gw{} phase shift of the dominant (quadrupole) mode
induced by the scalar cloud
can exceed half a \gw{} cycle as compared to the merger in vacuum.
Moreover, for $M\muS \gtrsim 0.5$, the observables seem to become insensitive to the mass parameter of the scalar.
We find consistent trends in the peak frequency of the \gw{} quadrupole,
and in the phase shifts of the
$\ell=m=3,4$ multipoles.
In our analysis, we have examined the underlying mechanisms of scalar field accretion onto the \bh{s},
scalar radiation, and
drag forces which remain to be quantified in a fully nonlinear evolution.

A natural follow-up question to our study is whether the presence of a scalar cloud can be inferred from \gw{} observables.
Given the scalar cloud density considered in this work, we found that the main feature distinguishing the waveforms from the vacuum case is the \gw{} dephasing and its shift in the peak frequency.
These features appear consistently not only in the dominant \gw{} quadrupole, but also in the higher multipoles.
However, it is unclear whether the dephasing and frequency shift is unique to the presence of a massive scalar cloud,
or if it can be degenerate with
eccentric or spinning \bh{} binaries
that are known to introduce a phase shift~\cite{Ramos-Buades:2022lgf,Ficarra:2024jen,Scheel:2025jct}.

Our work can be extended in several directions.
To quantify potential degeneracies,
it is crucial to compute
the \gw{} signature of spinning or eccentric \bh{} binaries in a scalar cloud environment
by extending 
proof-of-principle simulations~\cite{Zhang:2022rex}, especially as new dynamical effects appear for spinning \bh{s}~\cite{Wang:2024cej,Dyson:2024qrq}.
With our method of constructing initial data for non-homogeneous scalar clouds, it will be interesting to consider more realistic scalar field profiles
representing dark matter halos or superradiant clouds.
Our simulations also form the foundation for first mock parameter estimation studies
to clarify if, and under which conditions, the presence of a scalar cloud
may be detected in a \gw{} signal. 
A very recent work in this direction can be found in Ref.~\cite{Roy:2025qaa}.

Furthermore, the waveforms presented in this paper may inform tests of gravity
and quantify potential degeneracies
between modified gravity and
environmental effects on \gw{} waveforms.
In particular, popular models of modified gravity
such as scalar Gauss-Bonnet gravity
or effective field theories of gravity
yield a similar \gw{} phase shift as observed here~\cite{Cayuso:2023xbc,AresteSalo:2025sxc,Lara:2025kzj}.
Whether or not \gw{} signatures of modified gravity are degenerate with
environmental effects on
\bh{} binaries is an open question.

\section{Acknowledgements}
We thank
J.~Aurrekoetxea, J.~Bamber, K.~Clough,
G.~Pratten,
P.~Schmidt
and H.~O.~Silva
for insightful discussions and comments.
The authors acknowledge support provided by the National Science Foundation under NSF Award No.~OAC-2004879,
No.~PHY-2409726 and
No.~OAC-2411068.
C.-H.~C. acknowledges support provided by the Ministry of Education of Taiwan under the Taiwan-UIUC Scholarship.
G. F. gratefully acknowledges the support of University of Calabria through a research fellowship funded by DR 1688/2023.
We acknowledge the Texas Advanced Computing Center (TACC) at the University of Texas at Austin for providing HPC resources on Frontera via allocation PHY22041.
This work used Delta CPU at NCSA through allocation PHY240117 from the Advanced Cyberinfrastructure Coordination Ecosystem: Services \& Support (ACCESS) program, which is supported by U.S. National Science Foundation grants Nos. 2138259, 2138286, 2138307, 2137603, and 2138296.
This research used resources provided by the Delta research computing project, which is supported by the NSF Award No. OAC-2005572 and the State of Illinois.
This research was supported in part by the Illinois Computes project which is supported by the University of Illinois Urbana-Champaign and the University of Illinois System.

This work used the open-source softwares \textsc{xTensor}~\cite{xActWeb,Brizuela:2008ra},
the \ETK{}~\cite{EinsteinToolkitWeb,roland_haas_2024_14193969},
\textsc{Canuda} \cite{witek_helvi_2023_7791842,CanudaGitWeb},
{\textsc{kuibit}}~\cite{kuibit},
\textsc{NumPy} \cite{harris2020array}, \textsc{MatPlotLib} \cite{Hunter:2007}, \textsc{SciPy} \cite{2020SciPy-NMeth},
\textsc{NRPy+} \cite{Ruchlin:2017com,NRPyWeb}.

The \TPBBHSF{} code developed
to conduct
this work is open source and available in the git repository~\cite{TPBBHSFGitWeb}.
A YouTube playlist with two-dimensional animations is available at Ref.~\cite{YoutubeCanuda}.

\appendix
\section{BSSN formulation of the Einstein--Klein-Gordon equations}
\label{appendix:BSSN}
For completeness, we list the \bssn{} evolution equations as they are implemented in the
{\textsc{Canuda\_LeanBSSNMoL}}
and the
{\textsc{ScalarEvolve}}
thorns of the \canuda{} code~\cite{CanudaGitWeb}.

We evolve the spacetime
using the $W$-version of the \bssn formulation \cite{Marronetti:2007wz}.
Here, $W$ refers to the conformal factor given by
$W=\gamma^{-1/6}$,
and $\gamma$ is the determinant of the $3$-metric.
Other common choices for
the conformal factor are
$\chi$ or $\phi$, which are related by $W^2 = \chi = e^{-4\phi}$~\cite{Baker:2005vv,Campanelli:2005dd}.
The benefits of using
$W$ are
(i) its avoidance of the
$\mathcal{O}(\log r)$ singularity in $\phi$ at the puncture,
and (ii) that the determinant of the 3-metric $\gamma=W^6$ remains positive.

The \bssn{} evolution variables
are the conformal factor $W$ and metric $\tgamma_{ij}$,
the trace $K$ and (conformally rescaled) tracefree part $\tA_{ij}$ of the extrinsic curvature and the conformal connection function $\tGamma^{i}$
defined as
\begin{subequations}
\label{appeq:BSSNVars}
\begin{align}
W &= \gamma^{-1/6}
\,, \quad
\tgamma_{ij} = W^2 \gamma_{ij}
\,,\\
K &= \gamma^{ij} K_{ij}
\,, \quad
\tA_{ij} = W^2  \bracket{K_{ij} - \frac{1}{3} \gamma_{ij} K}
\,,\\
\label{appeq:DefTGamma}
\tGamma^{i} & =     \tgamma^{jk} \tGamma^{i}{}_{jk}
   = - \p_{j} \tgamma^{ij}
\,.
\end{align}
\end{subequations}
Here, $\tGamma^{i}{}_{jk}$ is the Christoffel connection of the conformal metric $\tgamma_{ij}$
and the last relation holds because $\det\tgamma_{ij}=1$.

The time evolution of the \bssn{}
variables is determined by a set of coupled, partial differential equations.
For completeness, we provide the
evolution equations as they are implemented in {\textsc{Canuda\_LeanBSSNMoL}},

\begin{subequations}
\label{appeq:BSSNEvolutionEquations}
\begin{align}
\label{appeq:BSSN_W}
\p_t W & =
  \beta^i \p_i W
- \frac{1}{3} W \p_i \beta^i
+ \frac{1}{3} \alpha W K
\,,\\
\label{appeq:BSSN_tildegamma}
\p_t \tgamma_{ij} & =
\beta^k \partial_k \tgamma_{ij}
+ \tgamma_{ik} \partial_j \beta^k
+ \tgamma_{jk} \partial_i \beta^k
- \frac{2}{3} \tgamma_{ij} \partial_k \beta^k
\nonumber \\ & \quad
- 2 \alpha \tA_{ij}
\,,\\
\label{appeq:BSSN_K}
\p_t K & =
    \beta^k \partial_k K
    - \gamma^{ij} D_i D_j \alpha
\nonumber\\ & \quad
    + \alpha \Bracket{\tilde{A}^{ij}\tilde{A}_{ij} + \frac{1}{3} K^2 + 4\pi (\rho + S)}
\,, \\
\label{appeq:BSSN_tildeA}
\p_t \tA_{ij} & =
    \beta^k \partial_k \tA_{ij}
    + \tA_{ik} \partial_j \beta^k
    + \tA_{jk} \partial_i \beta^k
    - \frac{2}{3} \tA_{ij} \partial_k \beta^k
\nonumber\\ & \quad
    + W^2 \tracefree{\alpha R_{ij} - D_i D_j \alpha }
    + \alpha \bracket{K \tA_{ij}
    - 2 {\tilde{A}_i}^{\ k} \tA_{jk}}
\nonumber\\ & \quad
    - 8 \pi \alpha W^2 \tracefree{S_{ij} }
\,,\\
\label{appeq:BSSN_tildeGamma}
\partial_t \tGamma^{i} & =
    \beta^k \partial_k \tGamma^{i}
    + \frac{2}{3} \tGamma^{i} \partial_k \beta^k
    - \tGamma^{k} \partial_k \beta^i
\nonumber\\ & \quad
    - \frac{4}{3} \alpha \tilde{\gamma}^{ik} \tilde{D}_k K
    + \frac{1}{3} \tilde{\gamma}^{ik} \partial_k \partial_j \beta^j
    + \tilde{\gamma}^{kj} \partial_j \partial_k \beta^i
\nonumber\\ & \quad
    - 2 \alpha {\tilde{\Gamma}^i}_{jk} \tilde{A}^{jk}
    - 2 \tilde{A}^{ik} \bracket{ \tilde{D}_k \alpha
    + 3  \alpha \frac{\tilde{D}_k W}{W}}
    - 16\pi \alpha j^i
\nonumber\\ & \quad
- \bracket{\sigma + \frac{2}{3}} \bracket{\tGamma^{i} - \tilde{\gamma}^{jk} \tilde{\Gamma}^i_{jk}} \partial_l \beta^l
\,.
\end{align}
\end{subequations}
Here, $\tilde{D}_{i}$ indicates the covariant derivative associated with the conformal metric $\tgamma_{ij}$,
$\tracefree{\ldots}$ denotes the tracefree part,
and we use as short-hand notation
\begin{subequations}
\begin{align}
D_i D_j \alpha & =
    \frac{1}{W} \Big(\tilde{D}_i \alpha \tilde{D}_j W + \tilde{D}_j \alpha \tilde{D}_i W - \tgamma_{ij} \tilde{D}^k\alpha \tilde{D}_k W\Big)
\nonumber\\ & \quad
    + \tilde{D}_i \tilde{D}_j \alpha
\,,\\
R_{ij} & =
    \tilde{R}_{ij}
    + \frac{1}{W} \Big( \tilde{D}_i \tilde{D}_j W + \tgamma_{ij} \tilde{D}^k \tilde{D}_k W \Big)
\nonumber\\ & \quad
    - \frac{2}{W^2} \tgamma_{ij} \tilde{D}^k W \tilde{D}_k W
\,.
\end{align}
\end{subequations}
The energy density $\rho$, energy-momentum flux $j_{i}$, and stress tensor $S_{ij}$ are read in from the {\textsc{TmunuBase}} thorn.

During the numerical evolution, we enforce the algebraic constraints
that  $\tA_{ij}$ is trace-free, i.e.,
$\tgamma^{ij} \tA_{ij} = 0$.

The definition of the conformal connection function in Eq.~\eqref{appeq:DefTGamma} implies a differential constraint.
As in the \lean code~\cite{Sperhake:2006cy},
we augment the evolution equation of the conformal connection function
with a term $\sim (\tGamma^{i} - \tgamma^{jk} \tGamma^{i}{}_{jk})$ in Eq.~\eqref{appeq:BSSN_tildeGamma}.
This modification was introduced by Yo et al~\cite{Yo:2002bm} to mitigate the growth of any numerical error in $\tGamma^{i}$ which could result in numerical instabilities.
This extra term vanishes in the continuum limit by construction of $\tGamma^{i}$.
We typically set the free parameter $\sigma=0$ by default.

The spacetime evolution equations are closed by the moving puncture gauge for the lapse function and shift vector ~\cite{Campanelli:2005dd,Baker:2005vv,vanMeter:2006vi}, in the form given in Eq.~\eqref{eq:movingpuncturegauge}.

The scalar field's evolution equations are implemented in
{\textsc{Canuda's}}
{\textsc{ScalarEvolve}} thorn using the \bssn variables.
Their explicit form is given by
\begin{subequations}
\begin{align}
\label{appeq:BSSN_Phi}
\partial_t \Phi & =
    - 2 \alpha \Pi + \beta^k \tilde{D}_k \Phi
\,, \\
\label{appeq:BSSN_Pi}
\partial_t \Pi & =
    \frac{\alpha}{2} \bracket{ - W^2 \tilde{D}^k \tilde{D}_k \Phi + \tilde{D}^k \Phi \tilde{D}_k W + 2 K\Pi + \muS^2 \Phi }
\nonumber \\ & \quad
    + \beta^k \tilde{D}_k \Pi
    - \frac{1}{2} W^2 \tilde{D}^i \alpha \tilde{D}_i \Phi
\,.
\end{align}
\end{subequations}

\section{Convergence test and error estimate} \label{app:convergenctest}
We perform convergence tests
to verify that the simulations are run at a sufficiently high resolution to be in the convergent regime,
and to assess their numerical error.
Therefore, we run simulations at three different spatial resolutions $\Delta x_{f} < \Delta x_{m} < \Delta x_{c}$, and compute the residuals
of a quantity $h$
between the fine and medium resolution results $|h_{f} - h_{m}|$, and the medium and coarse resolution results $|h_{m} - h_{c}|$.
If these results are convergent at $n$-th order, then the residuals are
proportional to the numerical error
and satisfy~\cite{alcubierre2008introduction}
\begin{align}
\label{eq:convergence_factor}
\frac{|h_{m} - h_{c}|}{|h_{f} - h_{m}|}
& = Q_n \equiv
\frac{\Delta x_{m}^n - \Delta x_{c}^n}{\Delta x_{f}^n - \Delta x_{m}^n}
\,,
\end{align}
where $Q_n$ is the $n$-th order convergence factor for the resolutions considered.

We perform the convergence analysis using the most demanding simulation setup,
\textbf{q12mu04} in Table~\ref{tab:simulations_list},
i.e.,
a binary \bh{} with mass ratio $q=1/2$ and a scalar field with mass parameter $M\muS=0.4$.
We run setup \textbf{q12mu04} at three different resolutions $\Delta x_{f} = 0.729$, $\Delta x_{m} = 0.81$, $\Delta x_{c} = 0.854$
defined on the outermost refinement level,
and examine if the criterion in Eq.~\eqref{eq:convergence_factor} is satisfied.
The corresponding convergence factors are $Q_4 = 0.685, Q_6 = 0.797$, and $Q_8 = 0.925$ for fourth, sixth and eighth order convergence, respectively.
In all production simulations, we employed eighth order finite difference stencils for spatial derivatives of the metric, scalar field and for the Newman-Penrose scalars
The time integration employs a fourth order Runge-Kutta integrator,
while the interpolator at refinement boundaries is of second order in time and fifth order in space.
Therefore, we may expect a mixed convergence order in our simulations.

\begin{figure}[htp!]
    \centering
    \includegraphics[width=.47\textwidth]{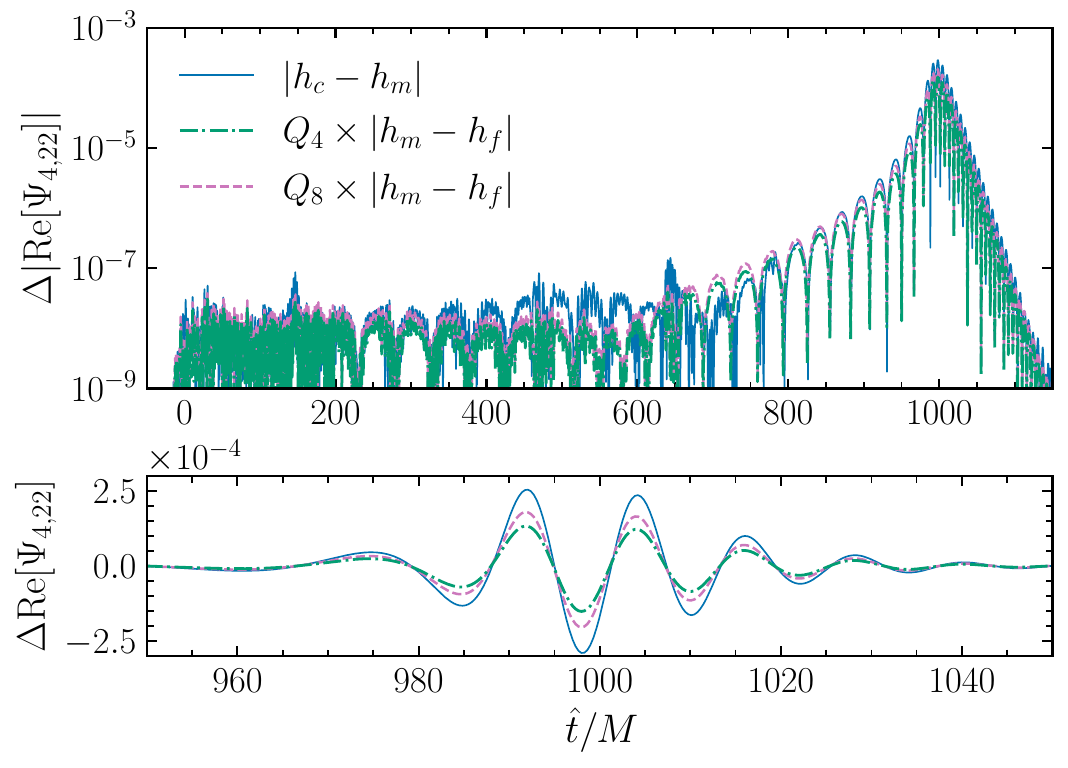}
    \caption{Convergence plot for the evolution of the quadrupole gravitational wave mode,
    $h = {\textrm{Re}} [\Psi_{4,22}]$,
    extracted at $\rex = 100M$. We plot the residual $|h_f - h_m|$ between the fine $(\Delta x_{f} = 0.729)$ and medium resolution $(\Delta x_{m} = 0.81)$ simulations,
    and the residual $|h_m - h_c|$ between the medium and coarse resolution ($\Delta x_{c} = 0.854$) simulations multiplied by the convergence factors $Q_4$ and $Q_8$.
    The bottom panel displays a zoom in around the merger.
    }
    \label{fig:convtest_psi4re_residuals}
\end{figure}

In Fig.~\ref{fig:convtest_psi4re_residuals}, we present the
convergence test
of the real part of the $(\ell, m) = (2, 2)$ mode of the Newman-Penrose scalar $\Psi_4$ extracted on coordinate spheres of radius $\rex = 100M$.
We find that
the waveforms are eighth order convergent in the early part of the simulation before $\hat{t}\lesssim600M$,
over-convergent during the merger and ringdown
$\hat{t}\gtrsim 900M$, until the
waves amplitude becomes comparable to numerical noise at $\hat{t}\gtrsim 1100M$.
However, the convergence order drops below fourth order in the interval $600M \lesssim \hat{t} \lesssim 900M$ right before the merger.
This is likely due to the different discretization orders used in spatial derivatives, time integrator and interpolator as outlined above.

\begin{figure}[htp!]
    \centering
    \includegraphics[width=.47\textwidth]{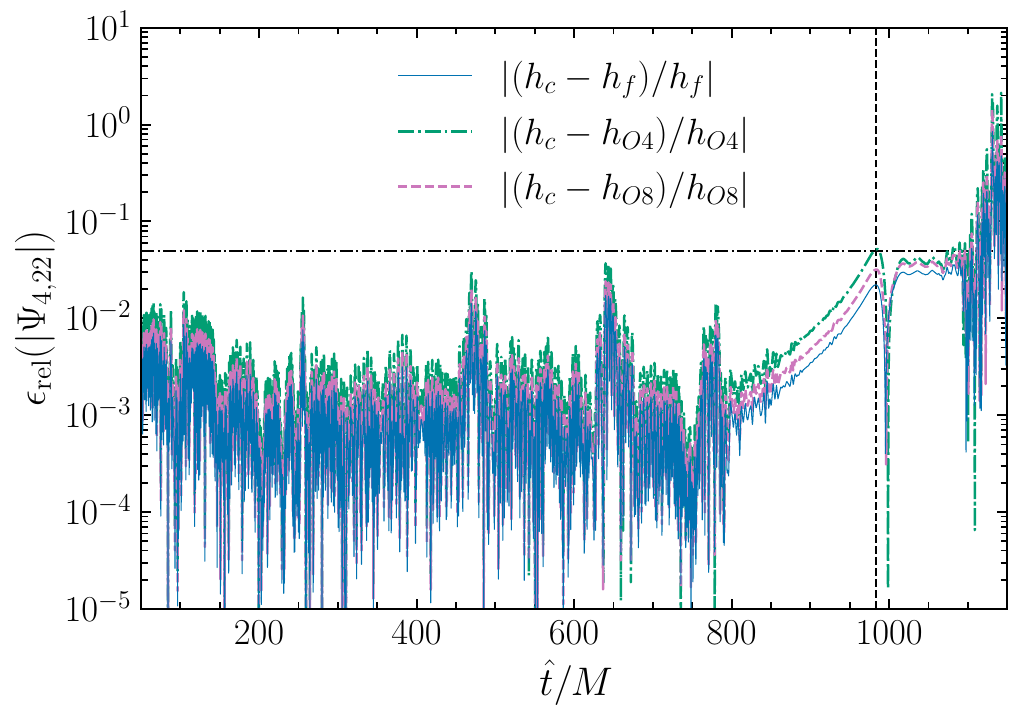}
    \caption{Relative error of
    $|\Psi_{4,22}|$,
    computed by the coarse resolution simulation
    and compared to
    the fine resolution simulation,
    the fourth and eighth order Richardson extrapolated waveforms
    as reference solutions.
    The black horizontal dotted-dashed line indicates a relative error of $5\%$, and the black vertical dashed line indicates the merger time
    determined by the formation of the common apparent horizon.}
    \label{fig:convtest_psi4abs_relerror}
\end{figure}

\begin{figure}[htp!]
     \centering
     \includegraphics[width=.30\textwidth]{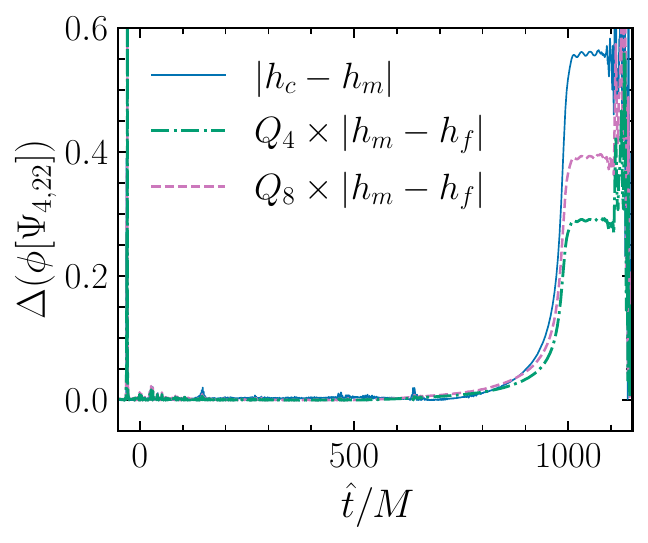}
     \caption{Convergence plot for the complex phase $h = \phi[\Psi_{4,22}]$ using the waveform extracted at $\rex = 100M$. The phase is measured from the time $t - \rex = 100M$ to define the zero point of the phase. We plot the residual $h_f - h_m$ between the fine $(\Delta x_{f} = 0.729)$ and medium resolution $(\Delta x_{m} = 0.81)$ simulations, and the residual $h_m - h_c$ between the medium and coarse resolution ($\Delta x_{c} = 0.854$) simulations scaled by the convergence factors $Q_6$ and $Q_8$.
     }
     \label{fig:convtest_phase_residuals}
\end{figure}

\begin{figure}[htp!]
    \centering
    \includegraphics[width=.47\textwidth]{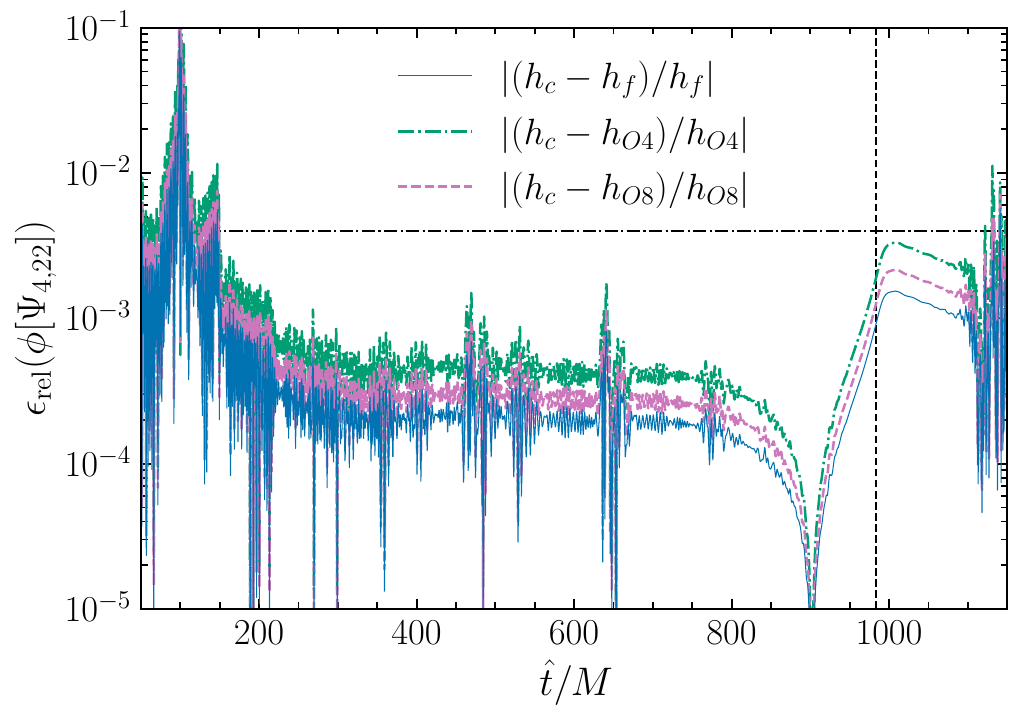}
    \caption{Same as Fig.~\ref{fig:convtest_psi4abs_relerror} but for the phase of $\Psi_{4,22}$.
    The black horizontal dotted-dashed line indicates a relative error of 0.4\%, and the black vertical dashed line indicates the merger time.}
    \label{fig:convtest_phase_relerror}
\end{figure}

The simulations presented in this paper have been produced with the coarse resolution ($\Delta x_{c} = 0.854$) run.
Having determined the convergence order of the waveform,
we now estimate the relative error of the coarse resolution run with respect to
(i) the finest resolution ($\Delta x_{f} = 0.729$),
and (ii) the fourth and eighth order Richardson extrapolation of the waveform
as benchmark.

More precisely, the $n$-th order Richardson extrapolation,
$h_{On}$,
using the fine and coarse numerical solutions $h_c$ and $h_f$
is obtained from~\cite{alcubierre2008introduction}
\begin{align}
\label{appeq:RichardsonExtrapolation}
h_{On} = h_f - \frac{1}{(\Delta x_c/\Delta x_f)^n - 1} (h_c - h_f) + \mathcal{O}(\Delta x_{f}^{n+1})
\,,
\end{align}
where
$\Delta x_{c}/\Delta x_{f} = 0.854/0.729$.
The extrapolated solution in Eq.~\eqref{appeq:RichardsonExtrapolation}
is accurate to order $n+1$ in the spatial resolution if the numerical solutions are in the convergent regime.
Having verified that the gravitational waveform are 
sixth to eighth order convergent,
we can employ the Richardson extrapolated solution as benchmark for the error analysis.

Next, we estimate the relative error
\begin{align}
\epsilon_{\rm rel} & = \left| \frac{(h - h_{\rm reference})}{h_{\rm reference}} \right|
\,,
\end{align}
of a function $h$ at a given resolution
compared to the reference solution of that function.
In the following, we compute the relative error using as reference solution
the fine resolution data, and the fourth or eighth order Richardson extrapolated quantities.
We plot the relative errors of the modulus of $\Psi_{4,22}$ in Fig.~\ref{fig:convtest_psi4abs_relerror},
and find that
it is below $\sim1\%$ throughout the early inspiral ($\hat{t}\lesssim900M$),
and below $\sim4\%$ in the late inspiral, merger and ringdown ($900M\lesssim \hat{t} \lesssim 1100M$).
The relative error increases after the ringdown when the waveform has such a small amplitude that it is dominated by numerical noise.

Next, we test the convergence of the
phase of the $(\ell, m) = (2, 2)$ mode of $\Psi_4$ extracted on $\rex = 100M$.
The phase is computed by measuring the complex argument of $\Psi_{4, 22}$ with respect to a starting point, which we choose to be $\hat{t} = 100M$, corresponding to the first cycle in the gravitational wave.
The result is shown in Fig.~\ref{fig:convtest_phase_residuals}.
Similar to the real part of the Newman-Penrose scalar (c.f.  Fig.~\ref{fig:convtest_psi4re_residuals}),
we find eighth order convergence in the early inspiral ($\hat{t}\lesssim600M$), fourth order convergence in the late inspiral ($600M\lesssim \hat{t}\lesssim900M$), and overconvergence in the merger and ringdown ($\hat{t}\gtrsim900M$).

We compute the fourth and eighth order Richardson extrapolation of the phase $h=\phi[\Psi_{4,22}]$.
We calculate the relative error of the coarse resolution simulation using the fine resolution data $h_f$ and the Richardson extrapolated data.
In Fig.~\ref{fig:convtest_phase_relerror} we show that the relative error in the phase stays below
$1\%$ percent for the duration of the simulation
until the end of the ringdown at
$t-\rex = 1100M$, where the waveforms are dominated by numerical noise.

\begin{figure}[htp!]
    \centering
    \includegraphics[width=.47\textwidth]{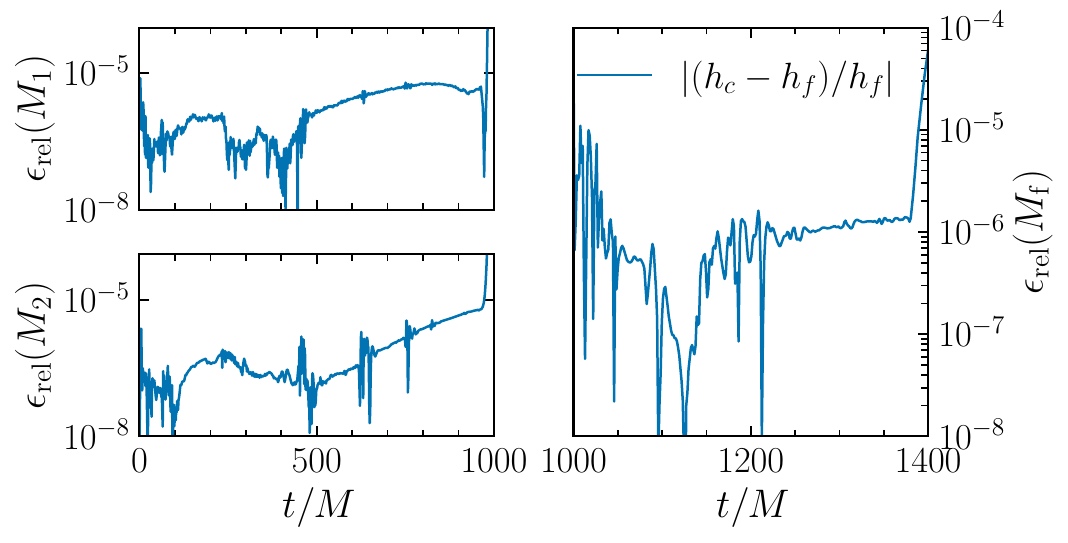}
    \caption{Relative error of the Christodoulou mass
    of the individual \bh{s} before the merger (left panels) and the final \bh{} (right panel).
    }
    \label{fig:convtest_mass_bh_relerror}
\end{figure}

\begin{figure}[htp!]
    \centering
    \includegraphics[width=.30\textwidth]{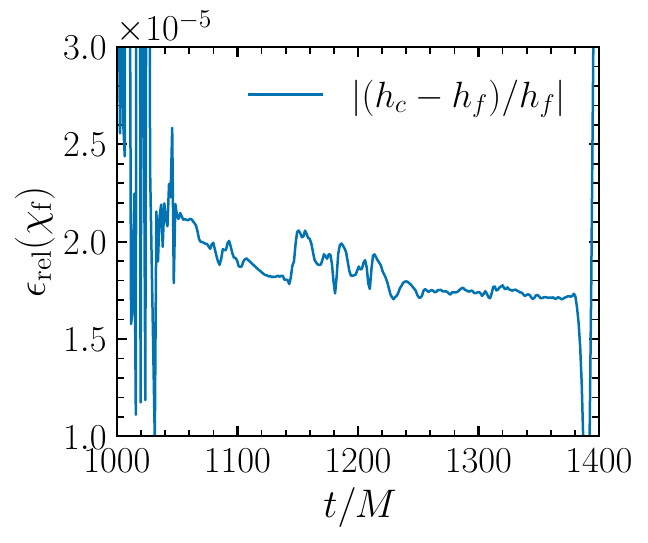}
    \caption{Relative error of the
    dimensionless spin $\chi_f\equiv J_f/M_f^2$ of the final \bh{.}
    }
    \label{fig:convtest_angular_momentum_bh_relerror}
\end{figure}

Finally, in Figs.~\ref{fig:convtest_mass_bh_relerror} and~\ref{fig:convtest_angular_momentum_bh_relerror} we present the relative error of the Christodoulou masses of the individual and final \bh{s},
and the dimensionless spin of the final \bh{,}
extracted by~\textsc{QuasiLocalMeasures}.
These quantities are defined in
Eqs.~\eqref{eq:AHspin} and \eqref{eq:christodouloumass}, and we only show the comparison between low- and high-resolution data.
Prior to the merger, the Christodoulou mass of each \bh{} has relative error $\epsilon_{\rm rel}(M_{(a)})\lesssim10^{-5}$.
After the merger, the final \bh{} has a relative error of $\epsilon_{\rm rel}(M_{\rm f})\lesssim 2\times10^{-6}$.

The final \bh{} acquires a dimensionless spin of $\chi_{\rm f} \sim 0.62$ after the merger.
Its relative error is
$\epsilon_{\rm rel}(\chi_{\rm f})\lesssim10^{-4}$,
which stays approximately constant throughout the ringdown.

\clearpage
\bibliographystyle{apsrev4-2}
\bibliography{BBHSFbib}

\end{document}